\documentclass[aps,preprint]{revtex4}%
\usepackage{amsfonts}
\usepackage{amsmath}
\usepackage{amssymb}
\usepackage{graphicx}
\usepackage{float}%
\renewcommand{\theequation}{\arabic{section}.\arabic{equation}}
\makeatletter
\@addtoreset{equation}{section}
\makeatother
\makeatletter
\def\bibsection{{\let\@hangfroms@section\@hang@froms
\section*{\refname}\@nobreaktrue
}\makeatother
}
\begin{document}
\title{Two-Body Dirac Equations for Nucleon-Nucleon Scattering\\ }
\author{Bin Liu and Horace Crater}
\affiliation{The University of Tennessee Space Institute Tullahoma, Tennessee
37388\footnote{hcrater@utsi.edu} }

\begin{abstract}
We investigate the nucleon-nucleon interaction by using the meson exchange
model and the two-body Dirac equations of constraint dynamics. This approach
to the two-body problem has been successfully tested for QED and QCD
relativistic bound states. An important question we wish to address is whether
or not the two-body nucleon-nucleon scattering problem can be reasonably
described in this approach as well. This test involves a number of related
problems. First we must reduce our two-body Dirac equations exactly to a
Schr\"{o}dinger-like equation in such a way that allows us to use techniques
to solve them already developed for Schr\"{o}dinger-like systems in
nonrelativistic quantum mechanics. Related to this we present a new derivation
of Calogero's variable phase shift differential equation for coupled
Schr\"{o}dinger-like equations. Then we determine if the use of \ nine meson
exchanges in our equations give a reasonable fit to the experimental
scattering phase shifts for $n-p$ scattering. The data involves seven angular
momentum states including the singlet states $^{1}S_{0}$, $^{1}P_{1}$,
$^{1}D_{2}$ and the triplet states $^{3}P_{0}$, $^{3}P_{1}$, $^{3}S_{1}$,
$^{3}D_{1}$. Two models that we have tested give us a fairly good fit. The
parameters obtained by fitting the $n-p$ experimental scattering phase shift
give a fairly good prediction for most of the $p-p$ experimental scattering
phase shifts examined (for the singlet states $^{1}S_{0}$, $^{1}D_{2}$ and
triplet states $^{3}P_{0}$, $^{3}P_{1}$). Thus the two-body Dirac equations of
constraint dynamics present us with a fit that encourages the exploration of a
more realistic model. We outline generalizations of the meson exchange model
for invariant potentials that may possibly improve the fit.

\end{abstract}
\maketitle

\section{\bigskip Introduction}

In this paper \cite{liu}, \ we obtain a semi-phenomenological relativistic
potential model for nucleon-nucleon interactions by using two-body Dirac
equations of \ constraint dynamics \cite{van1,c1,c2,c3,c4} , \cite{Saz1} and
Yukawa's theory of meson exchange. \ In previous work Long and Crater
\cite{long1} have derived the two-body Dirac equations for all non-derivative
Lorentz invariant interactions acting together or in any combinations. They
also reduced the two-body Dirac equations to coupled Schr\"{o}dinger-like
equations in which the potentials appear as covariant generalizations of the
standard spin dependent interactions appearing in the early phenomenological
works in this area \cite{1,2,3,4,5,6,7} based on the nonrelativistic
Schr\"{o}dinger equation. This allows us to take advantage of earlier work
done by other people on the nonrelativistic Schr\"{o}dinger equation. In
particular we use the variable phase method developed by Calogero and
Degasparis \cite{cal,dega} for computation of the phase shift from the
nonrelativistic Schr\"{o}dinger equation, presenting a new derivation for the
case of coupled equations. Our potentials for different angular momentum
states are constructed from combinations of several different meson exchanges.
Furthermore, our potentials, as well as the whole equations, are local, yet at
the same time covariant. This contrasts our approach with other relativistic
schemes such as those by Gross and others \cite{gross1,gross2,stoks,mac}. It
is the aim of this paper to see if the meson exchanges we use are adequate to
describe the elastic nucleon-nucleon interactions from low energy to high
energy (%
$<$%
350 \textrm{MeV}) when using them together with two-body Dirac equations of
constraint dynamics.

Although numerous relativistic approaches have been used in the
nucleon-nucleon scattering problem, none of the other approaches have been
tested nonperturbatively in both QED and QCD as they have been with the
two-body Dirac equations of \ constraint dynamics \cite{c2,c4,c6,c7,c15}.
Unlike the earlier local two-body approaches of Breit
\cite{Breit1,Breit2,Breit3}, the relativistic spin corrections need not be
treated only perturbatively. This means that we can use nonperturbative
methods (numerical methods) to solve the two-body Dirac equations. This is a
very important advantage of the constraint two-body Dirac equations (CTBDE).
The successful numerical tests in QED and QCD gives us confidence that they
may be appropriate relativistic equations for phase shift analysis of
nucleon-nucleon scattering.

In section II\ we introduce the two-body Dirac equations of constraint
dynamics. In section III we obtain the Pauli reduction of the two-body Dirac
equations to coupled Schr\"{o}dinger-like equations. We go a step further than
that achieved in the paper of Long and Crater in that we eliminate the first
derivative terms that appear in the Schr\"{o}dinger-like equation. \ This is
relatively simple for the case of uncoupled equations but not so for the case
of coupled Schr\"{o}dinger-like equations. The reason we perform this extra
reduction is that the formulae we use for the phase shift analysis, the
variable phase method developed by Calogero, has been worked out already for
coupled equations, but ones in which the first derivative terms are absent.
\ This step then becomes an important part of the formalism, allowing us take
advantage of previous work. In section IV, we discuss the phase shift methods
used in our numerical calculations, which include phase shift equations for
uncoupled and coupled states and the phase shift equations with Coulomb
potential. In section V we present the models used in our calculations,
including the expressions for the scalar, vector and pseudoscalar
interactions, and the way they enter into our two-body Dirac equations with
the mesons used in our fits. In section VI we present the results we have
acheived and in section VII are the summaries and conclusions of our work.

\section{\bigskip Review of Constraint two-body Dirac Equations}

The two-body Dirac equations that we will use for studying nucleon-nucleon
interaction bear a close relation to the single particle equation proposed by
Dirac in 1928 \cite{Dirac}$.$%

\begin{equation}
\left[  \mathbf{\alpha\cdot p}+\beta m+V(r)\right]  \psi=E\psi. \label{a}%
\end{equation}
For interactions that transforms as a time-component of a four vector and
world scalar we have $V(r)=A(r)+\beta S(r)$. Of course, the single particle
Dirac equation is not suitable to describe systems such as the mesons,
(quarkonium), muonium, positronium, the deuteron and nucleon-nucleon
scattering because the particles may have equal or near equal mass.

The earliest attempt at putting both particles on an equal footing was in 1929
by G. Breit \cite{Breit1}\cite{Breit2}\cite{Breit3}$.~~$However, the Breit
equations do not retain manifest covariant form and in QED the equations
cannot be treated nonperturbatively beyond the Coulomb term \cite{Breit2,bse}.
\ There have been many attempts to by pass the problems of the Breit equation
and also of the full four dimensional Bethe-Salpeter equation. \ These are
discussed in a number of different contexts in \cite{c1,c2,c3,c4}. \ The
approach of the CTBDE provides a manifestly covariant yet three dimensional
detour around many of the problems that hamper the implementation and
application of Breit's two-body Dirac equations as well as the full four
dimensional Bethe-Salpeter equation (see also \cite{c10}). \ In addition, as
mentioned above, the approach can by a Pauli reduction give us a local
Schr\"{o}dinger-like equation.

The CTBDE make use of Dirac's relativistic Hamiltonian formalism. In a series
of papers (in addition to those cited above see also \cite{c9,c8}) Crater and
Van Alstine have incorporated Todorov's effective particle idea developed in
his quasipotential approach \cite{Tod1} into the framework of Dirac's
Hamiltonian constraint mechanics \cite{dirac2} for a description of two body
systems. Their approach yields manifestly covariant coupled Dirac equations.
The standard reduction of the Breit equation to a Schr\"{o}dinger-like
equation for QED yields highly singular operators (like $\delta$ functions and
attractive $1/r^{3}$ potentials) that can only be treated perturbatively. In
the treatment of the CTBDE for QED \cite{c8,c6}, for example, one finds that
all the operators are quantum mechanically well defined so that one can
therefore use nonperturbative techniques (analytic as well as numerical) to
obtain solutions of bound state problems and scattering. (A quantum
mechanically well defined potential is one no more singular than $-1/4r^{2}$.
If it is not quantum mechanically well defined, it can only be treated
perturbatively.) Although it is encouraging that good results have been
obtained for QED and QCD meson spectroscopy, that is no guarantee that the
formalism so developed will lead to effective potentials in the case of
nucleon-nucleon scattering that render reasonable fits to the phase shift data.

\smallskip Using techniques developed by Dirac to handle constraints in
quantum mechanics and the method developed by Crater and Van Alstine, one can
derive the two-body Dirac equations for eight nonderivative Lorentz invariant
interactions acting separately or together \cite{c12},\cite{long1}. These
include world scalar, four vector and pseudoscalar interactions among others.
We can also reduce the two-body Dirac equations to coupled
Schr\"{o}dinger-like equations even with all these interactions acting
together. Before we test this method in nuclear physics in the phase shift
analysis of the nucleon-nucleon scattering problems we review highlights of
the constraint formalism and the form of the two-body Dirac equations.

\subsection{Hamiltonian Formulation Of The Two-Body Problem From Constraint
Dynamics}

Dirac \cite{dirac2} extended Hamiltonian mechanics to include conjugate
variables related by constraints of the form $\phi(q,p)=0$. For $N$
constraints, we may write%

\begin{equation}
\phi_{n}(q,p)\approx0,\,\,\,\,\,\,n=1,2,3,\cdot\cdot,N
\end{equation}
With these constraints the Hamiltonian of the system (with sum over repeated indices)%

\begin{equation}
H=\dot{q}_{n}p_{n}-L
\end{equation}
is not unique. \ The Dirac Hamiltonian $\mathcal{H~}$includes the constraints%

\begin{equation}
\mathcal{H=}H+\lambda_{n}\phi_{n}.
\end{equation}
in which $H$ is the Legendre Hamiltonian obtained from the Lagrangian by means
of a Legendre transformation. The $\lambda_{n}$ may be functions of conjugate
variables $q^{\prime}s$ and $p^{\prime}s$. The equation of motion for any
arbitrary function $g$ (without explicit time dependence) of the conjugate
variables $q^{\prime}s$ and $p^{\prime}s$ is then%

\begin{equation}
\overset{\cdot}{g}=[g,\mathcal{H}].
\end{equation}
Dirac called the conditional equality, $\approx$ a \textquotedblleft\ weak
\textquotedblright\ equality meaning the constraints $\phi_{n}\approx0$ must
not be applied before working out the Poisson brackets. Dirac called $=$ a
nonconditional equality or a \textquotedblleft\ strong \textquotedblright%
\ equality. The equations of motion are%

\begin{equation}
\overset{\cdot}{g}=[g,\mathcal{H}]=[g,H+\lambda_{n}\phi_{n}]=[g,H]+\lambda
_{n}[g,\phi_{n}]+[g,\lambda_{n}]\phi_{n}\approx\lbrack g,H]+\lambda_{n}%
[g,\phi_{n}]
\end{equation}
for $\phi_{n}\approx0$.

In the two body system, we have two constraints $\phi_{n}(q,p)\approx
0,~n=1,2$. \ For spinless particles they are taken to be the generalized mass
shell constraints of the two particles \cite{c8},\cite{Wong1}, namely
\[
\mathcal{H}_{1}=p_{1}^{2}+m_{1}^{2}+\Phi_{1}(x,p_{1},p_{2})\approx0,
\]%
\begin{equation}
\mathcal{H}_{2}=p_{2}^{2}+m_{2}^{2}+\Phi_{2}(x,p_{1},p_{2})\approx0,
\end{equation}
where
\begin{equation}
x=x_{1}-x_{2}.
\end{equation}

Dirac extended his idea of handling constraints in classical mechanics to
quantum mechanics by replacing the classical constraints $\phi_{n}%
(q,p)\approx0$ with quantum wave equations $\phi_{n}(q,p)\mid\psi\rangle=0$,
where $q$ and $p$ are conjugate variables. Thus the quantum forms for each
individual particle constraint become Schr\"{o}dinger-type equations
\cite{saz4}
\begin{equation}
\mathcal{H}_{i}|\psi\rangle=0{~~~~~~\mathrm{for~~~~~~~~~}}i=1,2. \label{1}%
\end{equation}
\ The total Hamiltonian $\mathcal{H}$ from these constraints alone is
\begin{equation}
\mathcal{H=\lambda}_{1}\mathcal{H}_{1}+\lambda_{2}\mathcal{H}_{2}, \label{2.5}%
\end{equation}
(with $\lambda_{i}$ as Lagrange multipliers). \ In order that each of these
constraints be conserved in time we must have
\begin{equation}
\lbrack\mathcal{H}_{i},\mathcal{H}]|\psi\rangle=i\frac{d\mathcal{H}_{i}}%
{d\tau}|\psi\rangle=0.
\end{equation}
so that
\[
\lbrack\mathcal{H}_{i},\mathcal{\lambda}_{1}\mathcal{H}_{1}+\lambda
_{2}\mathcal{H}_{2}]|\psi\rangle=
\]%
\begin{equation}
\{[\mathcal{H}_{i},\lambda_{1}]\mathcal{H}_{1}|\psi\rangle+\lambda
_{1}[\mathcal{H}_{i},\mathcal{H}_{1}]|\psi\rangle+[\mathcal{H}_{i},\lambda
_{2}]\mathcal{H}_{2}|\psi\rangle+\lambda_{2}[\mathcal{H}_{i},\mathcal{H}%
_{2}]\}|\psi\rangle=0.
\end{equation}

Using Eq.(\ref{1}), the above equation leads to this compatibility condition
between the two constraints
\begin{equation}
\lbrack\mathcal{H}_{1},\mathcal{H}_{2}]|\psi\rangle=0. \label{3}%
\end{equation}
This condition guarantees that with the Dirac Hamiltonian, the system evolves
such that the \textquotedblleft\ motion \textquotedblright\ is constrained to
the surface of the mass shell described by the constraints of $\mathcal{H}%
_{1}$ and $\mathcal{H}_{2}$ $($\cite{Wong1},\cite{c8,c9}$).$ As described most
recently in \cite{Wong1} this requires that
\begin{equation}
\Phi_{1}=\Phi_{2}=\Phi(x_{\perp}) \label{8}%
\end{equation}
(a kind of relativistic Newton's third law) with the transverse coordinate
defined by
\begin{equation}
x_{\nu\perp}=x_{12}^{\mu}(\eta_{\mu\nu}-P_{\mu}P_{\nu}/P^{2}), \label{9}%
\end{equation}
and total momentum by%
\begin{equation}
P=p_{1}+p_{2}.
\end{equation}

To complete our review of the spinless case (\cite{Wong1}) and establish
notation we introduce the transverse relative momentum%
\begin{equation}
p=\frac{\varepsilon_{2}}{w}p_{1}-\frac{\varepsilon_{1}}{w}p_{2} \label{qrel}%
\end{equation}%
\begin{equation}
P\cdot p=0, \label{pq}%
\end{equation}
where the CM energy eigenvalue $w$ is defined from
\begin{equation}
\left\{  P^{2}+w^{2}\right\}  |\psi\rangle=0. \label{pm}%
\end{equation}
Taking the difference of the two constraints,
\begin{equation}
(p_{1}^{2}-p_{2}^{2})|\psi\rangle=-(m_{1}^{2}-m_{2}^{2})|\psi\rangle,
\label{cmpt}%
\end{equation}
we can show that the longitudinal or time-like components of the momenta in
the CM system have the invariant forms%

\begin{align}
\varepsilon_{1}  &  =\frac{w^{2}+m_{1}^{2}-m_{2}^{2}}{2w}\nonumber\\
\varepsilon_{2}  &  =\frac{w^{2}+m_{2}^{2}-m_{1}^{2}}{2w}. \label{eps}%
\end{align}
Thus on these states $|\psi\rangle$ \ we obtain
\begin{equation}
\left\{  p^{2}+\Phi(x_{\perp})-b^{2}(w^{2},m_{1}^{2},m_{2}^{2})\right\}
|\psi\rangle=0,
\end{equation}
where
\begin{equation}
b^{2}(w^{2},m_{1}^{2},m_{2}^{2})=\varepsilon_{1}^{2}-m_{1}^{2}=\varepsilon
_{2}^{2}-m_{2}^{2}\ =\frac{1}{4w^{2}}\left\{  w^{4}-2w^{2}(m_{1}^{2}+m_{2}%
^{2})+(m_{1}^{2}-m_{2}^{2})^{2}\right\}  .
\end{equation}
and indicates the presence of exact relativistic two-body kinematics. (By this
statement we mean that classically $p^{2}-b^{2}=0$, would imply $w=\sqrt
{p^{2}+m_{1}^{2}}+\sqrt{p^{2}+m_{2}^{2}})$. \ Note that both of the
constituent invariant CM energies $\varepsilon_{1}$ and $\varepsilon_{2}$ are
positive for positive total CM energy $w$ greater than the square root of
$\left\vert m_{1}^{2}-m_{2}^{2}\right\vert $ . \ This is a direct consequence
of the Eq.(\ref{cmpt}) which in turn depends on the \textquotedblleft third
law\textquotedblright condition necessary for compatibility . In our
scattering applications below this guarantees that nucleons cannot scatter
into a final state having an overall positive energy but with constituent
positive and negative energy nucleons.

In the center-of-momentum system, $p=p_{\perp}=(0,\mathbf{p)}$, $x_{\perp
}=(0,\mathbf{r)}$ and the relative energy and time are removed from the
problem. The equation for the relative motion is then
\begin{equation}
\left\{  \mathbf{p}^{2}+\Phi(\mathbf{r})-b^{2}\right\}  |\psi\rangle=0,
\label{eq:eig}%
\end{equation}
which is in the form of a non-relativistic Schr\"{o}dinger equation (with
$2mV\rightarrow\Phi,~~2mE_{NR}\rightarrow b^{2}$). Thus the relativistic
treatment of the two-body problem for spinless particles gives a form that has
the simplicity of the ordinary non-relativistic two-body Schr\"{o}dinger
equation and yet maintains relativistic covariance. Spin and different types
of interactions can be included in a more complete framework
\cite{c12,Saz2,long1,c10} and will be reviewed later in this section.

In addition to exact relativistic kinematical corrections, Eq.(\ref{eq:eig})
displays through the potential $\Phi$ relativistic dynamical corrections .
These corrections include dependence of the potential on the CM energy $w$ and
on the nature of the interaction. For spinless particles interacting by way of
a world scalar interaction $S$, one finds \cite{c9,c8,c13,c14}
\begin{equation}
\Phi=2m_{w}S+S^{2} \label{scl}%
\end{equation}
where
\begin{equation}
m_{w}=\frac{m_{1}m_{2}}{w}, \label{mw}%
\end{equation}
while for (time-like) vector interactions (described by $\mathcal{A}$), one
finds \cite{c9,c13,Tod1,c14}
\begin{equation}
\Phi=2\varepsilon_{w}\mathcal{A}-\mathcal{A}^{2}, \label{faiaa}%
\end{equation}
where
\begin{equation}
\varepsilon_{w}=\frac{w^{2}-m_{1}^{2}-m_{2}^{2}}{2w}. \label{ew}%
\end{equation}
For combined space-like and time-like vector interactions (that reproduce the
correct energy spectrum for scalar QED \cite{c8} )
\begin{equation}
\Phi=2\varepsilon_{w}\mathcal{A}-\mathcal{A}^{2}+\frac{1}{2}\mathbf{\nabla
}^{2}\log(1-2\mathcal{A}/w)+\frac{1}{4}[\mathbf{\nabla}\log(1-2\mathcal{A}%
/w)]^{2}. \label{faiaa1}%
\end{equation}
The variables $m_{w}$ and $\varepsilon_{w}$ \ (which both approach the reduced
mass $\mu=m_{1}m_{2}/(m_{1}+m_{2})$ in the nonrelativistic limit) are called
the relativistic reduced mass and energy of the fictitious particle of
relative motion. These were first introduced by Todorov \cite{Tod1} in his
quasipotential approach. \ Thus, in the nonrelativistic limit, $\Phi$
approaches $2\mu(S+\mathcal{A})$ for combined interactions$.$ In the
relativistic case, the dynamical corrections to $\Phi$ referred to above
include both quadratic additions to $S$ and $\mathcal{A}$ as well as CM energy
dependence through $m_{w}$ and $\varepsilon_{w}.$ The two logarithm terms at
the end of Eq.(\ref{faiaa1}) are due to the transverse or space-like part of
the potential. \ Without those terms, spectral results would not agree with
the standard (but more complex) spinless Breit and Darwin approaches (see
references in \cite{c8} including \cite{Tod1}). \ 

Eq.(\ref{eq:eig}) provides a useful way to obtain the solution of the
relativistic two-body problem for spinless particles in scalar and vector
interactions and as reviewed below has been extended to include spin. \ In
that case they have been found to give a very good account of the bound state
spectrum of both light and heavy mesons using reasonable input quark potentials.

These ways of putting the invariant potential functions for scalar $S$ and
vector $\mathcal{A}$ interactions into $\Phi$ will be used in this paper for
the case of two spin one-half particles (see (\ref{A-a} ) to ( \ref{S1-2}) and
(\ref{E1-b}) to (\ref{M22}) below). These exact forms are not unique but were
motivated by work of Crater and Van Alstine in classical field theory and
Sazdjian in quantum field theory \cite{c14,saz3}. Other, closely related
structures will also be used. These structures play a crucial role in this
paper since they give us a nonperturbative framework in which $S$ and
$\mathcal{A}$ appear in the equations we use. This structure has been
successfully tested (numerically) in QED (positronium and muonium bound
states) and is found to give excellent results when applied to the highly
relativistic circumstances of QCD (quark model for mesons). An important
question we wish to answer in this paper is whether such structures are also
valid in the two-body nucleon-nucleon problem. \ This is an important test
since the quadratic forms (see e.g. Eqs.(\ref{scl}), (\ref{faiaa})) that
appear could very well distort possible fits based on Yukawa-type potentials
with strong couplings.

Before going on to describe the constraints for two spin-one-half particles we
mention an important but often overlooked aspect of the foundations of the
generalized mass shell contraint equations given in (\ref{1}). It involves
their derivation from an alternative starting point. In addition to the
connection with the Bethe-Salpeter equation described in \cite{saz4}, \ there
exists a connection between constraint dynamics and Wigner's early formulation
of relativistic quantum mechanics \cite{wig}. In particular,\ \ Polyzou
\cite{plyz}\ has demonstrated that the assumption of both Poincar\'{e}
invariance and manifest Lorentz covariance forces the scalar product for
quantum mechanical state vectors to be interaction dependent. \ So, whereas
for a free particle the kernal involved in the scalar product has the form
$\delta(p^{2}+m^{2})\theta(p^{0}),$ in cases of interactions the self-adjoint
nature of the kernal demands the forms $\delta(\mathcal{H}_{1})\delta
(\mathcal{H}_{2})$ \ with compatible constraints (a related use of such delta
functions to construct the state vectors themselves is discussed in
\cite{Wong1}).

\subsection{Two spin-one-half particles}

\smallskip We continue our review in this section by introducing the two-body
Dirac equations of constraint dynamics. The Dirac equations for two free
spin-one-half particles are%

\begin{align}
\mathcal{S}_{10}\psi &  =(\theta_{1}\cdot p_{1}+m_{1}\theta_{51})|\psi
\rangle=0\nonumber\\
\mathcal{S}_{20}\psi &  =(\theta_{2}\cdot p_{2}+m_{2}\theta_{52})|\psi
\rangle=0
\end{align}
where $\psi$ is the product of the two single-particle Dirac wave functions
(these equations are equivalent to the free one-body Dirac equation). The
\textquotedblleft theta\textquotedblright\ matrices are related to the
ordinary gamma matrices by
\begin{align}
\theta_{i}^{\mu}  &  =i\sqrt{\frac{1}{2}}\gamma_{5i}\gamma_{i}^{\mu},\qquad
\mu=0,1,2,3,\qquad i=1,2\nonumber\\
\theta_{5i}  &  =i\sqrt{\frac{1}{2}}\gamma_{5i}%
\end{align}
and satisfy the fundamental anticommutation relations
\begin{align}
\left[  \theta_{i}^{\mu},\theta_{i}^{\nu}\right]  _{+}  &  =-\eta^{\mu\nu
},\nonumber\\
\left[  \theta_{5i},\theta_{i}^{\mu}\right]  _{+}  &  =0,\nonumber\\
\left[  \theta_{5i},\theta_{5i}\right]  _{+}  &  =-1.
\end{align}
It is much more convenient to use the \textquotedblleft
theta\textquotedblright\ matrices instead of the Dirac gamma matrices for
working out the compatibility conditions. \ In the reduction of complicated
commutators to simpler form one uses reduction brackets that involve
anticommutators for odd numbers of \textquotedblleft theta\textquotedblright%
\ matrices and commutators for even numbers of \ \textquotedblleft
theta\textquotedblright\ matrices and coordinate and momentum operators.
\ This property follows from the relation of the \textquotedblleft
theta\textquotedblright\ matrices to the Grassmann variables used in the
pseudoclassical form of the constraints (see \cite{van1,c1,c3}). These
fundamental anticommutation relations guarantee that the Dirac operators
$\mathcal{S}_{10}$ and $\mathcal{S}_{20}$ are the square root of the mass
shell operators $-\frac{1}{2}(p_{1}^{2}+m_{1}^{2})$ and $-\frac{1}{2}%
(p_{2}^{2}+m_{2}^{2})$. \ Differencing these implies that the relative
momentum $p$ in Eq.(\ref{pq}) satisfies $P\cdot p|\psi\rangle=0.$

Writing $p_{1}$ and $p_{2}$ in terms of the total and relative momenta we
obtain
\begin{align}
\mathcal{S}_{10}\psi &  =(\theta_{1\perp}\cdot p+\epsilon_{1}\theta_{1}%
\cdot\hat{P}+m_{1}\theta_{51})|\psi\rangle=0\nonumber\\
\mathcal{S}_{20}\psi &  =(-\theta_{2\perp}\cdot p+\epsilon_{2}\theta_{2}%
\cdot\hat{P}+m_{2}\theta_{52})|\psi\rangle=0.
\end{align}
The projected \textquotedblleft theta \textquotedblright\ matrices then
satisfy
\begin{align}
\left[  \theta_{i}\cdot\hat{P},\theta_{i}\cdot\hat{P}\right]  _{+}  &
=1,\nonumber\\
\left[  \theta_{i}\cdot\hat{P},\theta_{i\bot}^{\mu}\right]  _{+}  &  =0,
\end{align}
where
\begin{equation}
\theta_{\nu\perp}^{\mu}=\theta_{i\nu}(\eta^{\mu\nu}+\hat{P}^{\mu}\hat{P}^{\nu
}).
\end{equation}
Defining $\alpha_{i\perp}^{\mu}=2\theta_{i}\cdot\overset{\wedge}{P}%
\theta_{i\perp}^{\mu},$ and $\beta_{i}=2\theta_{i}\cdot\overset{\wedge}%
{P}\theta_{5i},$ the above two-body Dirac equations become
\begin{align}
(\alpha_{1}\cdot p+\beta_{1}m_{1})\psi &  =\epsilon_{1}\psi\nonumber\\
(-\alpha_{2}\cdot p+\beta_{2}m_{2})\psi &  =\epsilon_{2}\psi
\end{align}
which have the form of single free particle Dirac equations.

Recall that in the spinless case we had the compatibility condition%

\begin{equation}
\lbrack\mathcal{H}_{1},\mathcal{H}_{2}]|\psi\rangle=0.
\end{equation}
It was a requirement that followed in the classical case (or the Heisenberg
picture in the quantum case) from the individual constraints $\mathcal{H}_{i}$
being conserved in time. Similarly here with $\mathcal{S}_{i}$ designating the
form of the Dirac constraint with intereactions present, the commutator
condition guaranteeing that the Dirac equations for two spin $\frac{1}{2}$
particles form a compatible set is%

\begin{equation}
\lbrack\mathcal{S}_{1},\mathcal{S}_{2}]|\psi\rangle=0.
\end{equation}
(It would follow from an $\mathcal{H}$ as in Eq.(\ref{2.5}) composed of a sum
of the $\mathcal{S}_{i}.$)

We found that even for the simplest interaction, a Lorentz scalar, the naive
replacement such as making the minimal substitutions (corresponding in the
case of the single particle Dirac equation to Eq.(\ref{a}) with $V(r)=\beta
S(r)$ ),%

\begin{equation}
m_{i}\rightarrow M_{i}(r)=m_{i}+S_{i}\hspace{0.2in}i=1,2
\end{equation}
doe not lead to compatible constraints. Rather than detail here the earlier
work steps that were taken to make the interactions meet the compatibility
condition for scalar interactions \cite{van1,c1,c12} we present here the form
of the compatible constraints for general covariant interactions%

\[
\mathcal{S}_{1}|\psi\rangle=(\cosh(\Delta)\mathbf{S}_{1}+\sinh(\Delta
)\mathbf{S}_{2})|\psi\rangle=0
\]

\begin{equation}
\mathcal{S}_{2}|\psi\rangle=(\cosh(\Delta)\mathbf{S}_{2}+\sinh(\Delta
)\mathbf{S}_{1})|\psi\rangle=0 \label{d2}%
\end{equation}
where the operators $\mathbf{S_{1}}$ and $\mathbf{S_{2}}$ are auxiliary
constraints of the form
\[
\mathbf{S}_{1}|\psi\rangle=(\mathcal{S}_{10}\cosh(\Delta)+\mathcal{S}%
_{20}\sinh(\Delta))|\psi\rangle=0
\]%
\begin{equation}
\mathbf{S}_{2}|\psi\rangle=(\mathcal{S}_{20}\cosh(\Delta)+\mathcal{S}%
_{10}\sinh(\Delta))|\psi\rangle=0. \label{d4}%
\end{equation}
Both of these sets of constraints \cite{Saz1}\cite{c10,c12} are compatible%

\begin{align}
\lbrack\mathcal{S}_{1},\mathcal{S}_{2}]|\psi\rangle &  =0\label{S1S2}\\
\lbrack\mathbf{S}_{1},\mathbf{S}_{2}]|\psi\rangle &  =0.
\end{align}
provided only that%
\begin{equation}
\Delta(x)=\Delta(x_{\bot}).
\end{equation}
Furthremore
\begin{equation}
P\cdot p|\psi\rangle=0,
\end{equation}
the same constraint equation on the relative momentum $p\,$as in the spinless case.

The covariant potentials are divided into two categories, four
\textquotedblleft polar\textquotedblright\ and four \textquotedblleft axial
\textquotedblright\ interactions. The four polar interactions (or tensors of
rank 0,1,2) are

\noindent scalar
\begin{equation}
\Delta_{L}=-L\theta_{51}\theta_{52}=-\frac{L}{2}\mathcal{O}_{1},\mathcal{O}%
_{1}=-\gamma_{51}\gamma_{52}, \label{interL}%
\end{equation}

\noindent time-like vector
\begin{equation}
\Delta_{J}=J\theta_{1}\cdot\hat{P}\theta_{2}\cdot\hat{P}\equiv\mathcal{O}%
_{2}\frac{J}{2}=\beta_{1}\beta_{2}\frac{J}{2}\mathcal{O}_{1}, \label{interJ}%
\end{equation}

\noindent space-like vector
\begin{equation}
\Delta_{\mathcal{G}}=\mathcal{G}\theta_{1\perp}\cdot\theta_{2\perp}%
\equiv\mathcal{O}_{3}\frac{\mathcal{G}}{2}=\gamma_{1\perp}\cdot\gamma_{2\perp
}\frac{\mathcal{G}}{2}\mathcal{O}_{1}, \label{interG}%
\end{equation}

\noindent and tensor(polar)
\begin{equation}
\Delta_{\mathcal{F}}=4\mathcal{F}\theta_{1\perp}\cdot\theta_{2\perp}%
\theta_{52}\theta_{51}\theta_{1}\cdot\hat{P}\theta_{2}\cdot\hat{P}%
\equiv\mathcal{O}_{4}\frac{\mathcal{F}}{2}=\alpha_{1}\cdot\alpha_{2}%
\frac{\mathcal{F}}{2}\mathcal{O}_{1}.
\end{equation}
We may use each in Eqs.(\ref{d2}) and (\ref{d4}) separately or as a sum
\begin{equation}
\Delta_{p}=\Delta_{L}+\Delta_{J}+\Delta_{\mathcal{G}}+\Delta_{\mathcal{F}}%
\end{equation}
to generate the sets of two-body Dirac equations with corresponding
interactions. A particularly important combination occurs for electromagnetic
inteactions. \ While time - and space- like vector interactions are
characterized by the respective matrices $\beta_{1}\beta_{2}$ and
$\gamma_{1\perp}\cdot\gamma_{2\perp}$, a potential proportional to $\gamma
_{1}\cdot\gamma_{2}$ \ would correspond to an electromagnetic-like interaction
and would require that $J=-\mathcal{G}$.
\begin{equation}
\Delta_{\mathcal{EM}}=\frac{(\mathcal{O}_{3}-\mathcal{O}_{2})\mathcal{G}%
(x_{\bot})}{2}=\frac{\gamma_{1}\cdot\gamma_{2}\mathcal{G}(x_{\bot})}%
{2}\mathcal{O}_{1}.
\end{equation}

The four \textquotedblleft axial\textquotedblright\ interactions (or
pseudotensors of rank 0,1,2) are

\noindent pseudoscalar
\begin{equation}
\Delta_{C}=\frac{C}{2}\equiv\mathcal{E}_{1}\frac{C}{2}=-\gamma_{51}\gamma
_{52}\frac{C}{2}\mathcal{O}_{1},
\end{equation}

\noindent time-like pseudovector
\begin{equation}
\Delta_{H}=-2H\theta_{1}\cdot\hat{P}\theta_{2}\cdot\hat{P}\theta_{51}%
\theta_{52}\equiv-\mathcal{E}_{2}\frac{H}{2}=\beta_{1}\gamma_{51}\beta
_{2}\gamma_{52}\frac{H}{2}\mathcal{O}_{1},
\end{equation}

\noindent space-like pseudovector
\begin{equation}
\Delta_{I}=-2I\theta_{1\perp}\cdot\theta_{2\perp}\theta_{51}\theta_{52}%
\equiv-\mathcal{E}_{3}\frac{I}{2}=-\gamma_{51}\gamma_{1\perp}\cdot\gamma
_{52}\gamma_{2\perp}\frac{I}{2}\mathcal{O}_{1},
\end{equation}

\noindent and tensor(axial)
\begin{equation}
\Delta_{Y}=-2Y\theta_{1\perp}\cdot\theta_{2\perp}\theta_{1}\cdot\hat{P}%
\theta_{2}\cdot\hat{P}\equiv-\mathcal{E}_{4}\frac{Y}{2}=-{\sigma}_{1}%
\cdot{\sigma}_{2}\frac{Y}{2}\mathcal{O}_{1}.
\end{equation}

Crater and Van Alstine found \cite{c12} that these and their sum
\begin{equation}
\Delta_{a}=\Delta_{C}+\Delta_{H}+\Delta_{I}+\Delta_{Y} \label{interA}%
\end{equation}
would be used in Eqs.(\ref{d2}) and Eqs.(\ref{d4}) but with the $\sinh
(\Delta_{a})$ terms in Eqs.(\ref{d2}) appearing with a negative sign instead
of the plus sign as is the case polar interactions . There is no sign change
in Eqs.(\ref{d4} ) for $\Delta_{a}.$

For systems with both polar and axial interactions \cite{c12}, one uses
$\Delta_{p}-\Delta_{a}$ to replace $\Delta$ in Eqs.(\ref{d2}), and $\Delta
_{p}+\Delta_{a}$ to replace the $\Delta$ in Eqs.(\ref{d4}). $L$, $J$,
$\mathcal{G}$, $\mathcal{F}$, $C$, $H$, $I$, $Y$ are arbitrary invariant
functions of $x_{\bot}.$ In this paper, we include only mesons corresponding
to the interactions $L$, $J$, $\mathcal{G}(J=-\mathcal{G)}$, and $C$. Thus we
are ignoring tensor and pseudovector mesons, limiting ourselves to vector,
scalar and pseudoscalar mesons, all having masses less than or about 1000
$\mathrm{MeV}$. We are also ignoring possible pseudovector couplings of the
pseudoscalar mesons.

For computational convenience we have found it necessary to transform the
Dirac equations to \textquotedblleft external potential\textquotedblright%
\ form. We obtain these forms by combining the two sets of equations
\[
\mathcal{S}_{1}|\psi\rangle=[\cosh(\Delta)(\mathcal{S}_{10}\cosh
(\Delta)+\mathcal{S}_{20}\sinh(\Delta))+\sinh(\Delta)(\mathcal{S}_{20}%
\cosh(\Delta)+\mathcal{S}_{10}\sinh(\Delta))]\psi\rangle=0
\]

\begin{equation}
\mathcal{S}_{2}|\psi\rangle=[\cosh(\Delta)(\mathcal{S}_{20}\cosh
(\Delta)+\mathcal{S}_{10}\sinh(\Delta))+\sinh(\Delta)(\mathcal{S}_{10}%
\cosh(\Delta)+\mathcal{S}_{20}\sinh(\Delta))]\psi\rangle=0.
\end{equation}
and bringing the $\mathcal{S}_{i0}$ operators through to the right. References
\cite{c12,long1} gives the \textquotedblleft external potential
\textquotedblright\ forms of the constraint two-body Dirac equations for each
of the eight interaction matrices, $\Delta_{L}$, $\Delta_{J}$, $\Delta
_{\mathcal{G}}$, $\Delta_{\mathcal{F}}$, $\Delta_{C}$, $\Delta_{H}$,
$\Delta_{I}$, $\Delta_{Y}$ acting alone. These forms are similar in appearance
to individual Dirac equations for each of the particles in an external
potential. In \cite{long1} appeared also the form with all eight interactions
acting simultaneously
\[
\mathcal{S}_{1}|\psi\rangle=
\]%
\[
\{\exp(\mathcal{G}+\mathcal{F}\mathcal{E}_{2}+I\mathcal{O}_{1}+Y\mathcal{O}%
_{2})[\theta_{1}\cdot p-{\frac{i}{2}}\theta_{2}\cdot\partial(L\mathcal{O}%
_{1}-J\mathcal{O}_{2}-\mathcal{G}\mathcal{O}_{3}-\mathcal{F}\mathcal{O}%
_{4}-C\mathcal{E}_{1}+H\mathcal{E}_{2}+I\mathcal{E}_{3}+Y\mathcal{E}_{4})]
\]%
\[
+\epsilon_{1}\cosh(J\mathcal{O}_{2}+\mathcal{F}\mathcal{O}_{4}+H\mathcal{E}%
_{2}+Y\mathcal{E}_{4})\theta_{1}\cdot\hat{P}+\epsilon_{2}\sinh(J\mathcal{O}%
_{2}+\mathcal{F}\mathcal{O}_{4}+H\mathcal{E}_{2}+Y\mathcal{E}_{4})\theta
_{2}\cdot\hat{P}%
\]%
\begin{equation}
+m_{1}\cosh(-L\mathcal{O}_{1}+\mathcal{F}\mathcal{O}_{4}+H\mathcal{E}%
_{2}+I\mathcal{E}_{3})\theta_{51}+m_{2}\sinh(-L\mathcal{O}_{1}+\mathcal{F}%
\mathcal{O}_{4}+H\mathcal{E}_{2}+I\mathcal{E}_{3})\theta_{52}\}|\psi\rangle=0,
\label{e1}%
\end{equation}%
\[
\mathcal{S}_{2}|\psi\rangle=
\]%
\[
\{-\exp(\mathcal{G}+\mathcal{F}\mathcal{E}_{2}+I\mathcal{O}_{1}+Y\mathcal{O}%
_{2})[\theta_{2}\cdot p-{\frac{i}{2}}\theta_{1}\cdot\partial(L\mathcal{O}%
_{1}-J\mathcal{O}_{2}-\mathcal{G}\mathcal{O}_{3}-\mathcal{F}\mathcal{O}%
_{4}-C\mathcal{E}_{1}+H\mathcal{E}_{2}+I\mathcal{E}_{3}+Y\mathcal{E}_{4})]
\]%
\[
+\epsilon_{1}\sinh(J\mathcal{O}_{2}+\mathcal{F}\mathcal{O}_{4}+H\mathcal{E}%
_{2}+Y\mathcal{E}_{4})\theta_{1}\cdot\hat{P}+\epsilon_{2}\cosh(J\mathcal{O}%
_{2}+\mathcal{F}\mathcal{O}_{4}+H\mathcal{E}_{2}+Y\mathcal{E}_{4})\theta
_{2}\cdot\hat{P}%
\]%
\begin{equation}
+m_{1}\sinh(-L\mathcal{O}_{1}+\mathcal{F}\mathcal{O}_{4}+H\mathcal{E}%
_{2}+I\mathcal{E}_{3})\theta_{51}+m_{2}\cosh(-L\mathcal{O}_{1}+\mathcal{F}%
\mathcal{O}_{4}+H\mathcal{E}_{2}+I\mathcal{E}_{3})\theta_{52}\}|\psi\rangle=0.
\label{e2}%
\end{equation}
What is remarkable is that the above hyperbolic and exponential structures
account for all of the \textquotedblleft interference\textquotedblright\ terms
between the various interactions. The interactions acting separately or in
subgroupings are simple reductions of the the above. \ For example, \ in the
case of the combined scalar, time-like, space-like and pseudoscalar
interactions used in this paper,%

\begin{equation}
\Delta=\Delta_{J}+\Delta_{L}+\Delta_{\mathcal{G}}+\Delta_{C}%
\end{equation}
and the two-body Dirac equations (\ref{e1},\ref{e2}) reduce to
\begin{align}
&  \mathcal{S}_{1}|\psi\rangle\label{f1}\\
&  =(\exp(\mathcal{G}\mathtt{)}\theta_{1}\cdot p+E_{1}\theta_{1}\cdot\hat
{P}+M_{1}\theta_{51}+i{\frac{\exp(\mathcal{G}\mathtt{)}}{2}}\theta_{2}%
\cdot\partial(\mathcal{G}\mathcal{O}_{3}+J\mathcal{O}_{2}-L\mathcal{O}%
_{1}+C\mathcal{E}_{1}))|\psi\rangle=0\nonumber
\end{align}%
\begin{align}
&  \mathcal{S}_{2}|\psi\rangle\label{f2}\\
&  =(-\exp(\mathcal{G}\mathtt{)}\theta_{2}\cdot p+E_{2}\theta_{2}\cdot\hat
{P}+M_{2}\theta_{52}-i{\frac{\exp(\mathcal{G}\mathtt{)}}{2}}\theta_{1}%
\cdot\partial(\mathcal{G}\mathcal{O}_{3}+J\mathcal{O}_{2}-L\mathcal{O}%
_{1}+C\mathcal{E}_{1}))|\psi\rangle=0.\nonumber
\end{align}
where
\[
M_{1}=m_{1}\cosh(L)+m_{2}\sinh(L),
\]%
\begin{equation}
M_{2}=m_{2}\cosh(L)+m_{1}\sinh(L), \label{m}%
\end{equation}%
\[
E_{1}=\epsilon_{1}\cosh(J)+\epsilon_{2}\sinh(J),
\]%
\begin{equation}
E_{2}=\epsilon_{2}\cosh(J)+\epsilon_{1}\sinh(J). \label{e}%
\end{equation}
In the limit $m_{1}\rightarrow\infty$ (or $m_{2}\rightarrow\infty$ ), (when
one of the particles become infinitely massive), the extra terms
$\partial\mathcal{G}$, $\partial J,\partial L$ and $\partial C$ in
Eqs.(\ref{f1},\ref{f2}) vanish, and one recovers the expected one-body Dirac
equation in an external potential. The above two-body Dirac equations (without
pseudoscalar interactions ) have been tested successfully in quark model
calculations of the meson spectra \cite{c2}\cite{c3}\cite{c7,c15}. \ 

We may rewrite the \textquotedblleft external potential form\textquotedblright%
\ of the CTBDE for two relativistic spin-one-half particle interacting through
scalar and vector potentials as (see Eqs.(\ref{f1},\ref{f2}) without the
pseudoscalar interaction)%

\begin{equation}
\mathcal{S}_{1}|\psi\rangle\equiv\gamma_{51}(\gamma_{1}\cdot(p_{1}%
-A_{1})+m_{1}+S_{1})|\psi\rangle=0 \label{D-a}%
\end{equation}

\begin{equation}
\mathcal{S}_{2}|\psi\rangle\equiv\gamma_{52}(\gamma_{2}\cdot(p_{2}%
-A_{2})+m_{2}+S_{2})|\psi\rangle=0.\label{D-b}%
\end{equation}
$A_{i}^{\mu}$ and $S_{i}$ introduce the interactions that the $i$ th particle
experience due to the presence of the other particle and are both
spin-dependent \cite{van1}\cite{c1}\cite{c2}\cite{c3}\cite{c4}. In order to
identify these potentials we use Eqs.(\ref{f1},\ref{f2}), and (\ref{m}%
,\ref{e}). \ Then we find that the momentum dependent vector potentials
$A_{i}^{\mu}$ are given in terms of three invariant functions \cite{c3}%
\cite{c4} $G$, $E_{1},E_{2}$
\begin{equation}
A_{1}^{\mu}=((\epsilon_{1}-E_{1})-i\frac{G}{2}\gamma_{2}\cdot\frac{\partial
E_{1}}{E_{2}}\gamma_{2}\cdot\hat{P})\hat{P}+(1-G)p^{\mu}-\frac{i}{2}\partial
G\mathcal{\cdot}\gamma_{2\perp}\gamma_{2\perp}^{\mu}\label{A-a}%
\end{equation}%
\begin{equation}
A_{2}^{\mu}=((\epsilon_{2}-E_{2})-i\frac{G}{2}\gamma_{1}\cdot\frac{\partial
E_{2}}{E1}\gamma_{1}\cdot\hat{P})\hat{P}+(1-G)p^{\mu}-\frac{i}{2}\partial
G\mathcal{\cdot}\gamma_{1\perp}\gamma_{1\perp}^{\mu},\label{A-b}%
\end{equation}
where
\begin{equation}
G=\exp(\mathcal{G}),\label{Ge}%
\end{equation}
(with $\hat{P}^{2}=-1$, where $\hat{P}\equiv P/w$) while the scalar potentials
$S_{i}$ are given in terms of three invariant functions \cite{c1,c3}\cite{c4}
$G$, $M_{1},M_{2}$
\begin{equation}
S_{1}=M_{1}-m_{1}-\frac{i}{2}G\gamma_{2}\cdot\frac{\partial M_{1}}{M_{2}%
}\label{S1-1}%
\end{equation}%
\begin{equation}
S_{2}=M_{2}-m_{2}-\frac{i}{2}G\gamma_{1}\cdot\frac{\partial M_{2}}{M_{1}%
}.\label{S1-2}%
\end{equation}
\ In QCD, the scalar potentials $S_{i}$ are semi-phenomenological long range
interactions. The vector potentials $A_{i}^{\mu}$ are semi-phenomenological in
the long range while in the short range are closely related to perturbative
quantum field theory \cite{c16}. $\ $Of course this rewrite does not change
the fact that $\mathcal{S}_{1}$ and $\mathcal{S}_{2}$ still satisfy the
compatibility condition Eq.(\ref{S1S2})

\section{Pauli Reduction}

Now one can use the complete hyperbolic constraint two-body Dirac equations
Eqs.(\ref{e1},{\ref{e2}}), to derive the Schr\"{o}dinger-like eigenvalue
equation for the combined interactions: $L(x_{\perp}),J(x_{\perp}),H(x_{\perp
}),C(x_{\perp}),\mathcal{G}(x_{\perp}),\mathcal{F}(x_{\perp}),{I}(x_{\perp
}),{Y}(x_{\perp})$ \cite{long1}. In this paper, however, we include only
mesons corresponding to the interactions $L$, $J$, $\mathcal{G}%
(J=-\mathcal{G)}$, $C$, thus limiting ourselves to vector, scalar and
pseudoscalar interactions. The basic method we use here has some similarities
to the reduction of the single particle Dirac equation to a
Schr\"{o}dinger-like form (the Pauli-reduction) and to related work by
Sazdjian \cite{Saz1}\cite{Saz2}.

The state vector $|\psi\rangle$ appearing in the two-body Dirac equations
(\ref{e1},{\ref{e2}}) is a Dirac spinor written as%

\begin{equation}
|\psi\rangle=\left[
\begin{array}
[c]{l}%
|\psi\rangle_{1}\\
|\psi\rangle_{2}\\
|\psi\rangle_{3}\\
|\psi\rangle_{4}%
\end{array}
\right]
\end{equation}
where each $|\psi\rangle_{i}$ is itself a four component spinor. $|\psi
\rangle$ has a total of sixteen components and the matrices $\mathcal{O}_{i}%
$'s, $\mathcal{E}_{i}$'s are all sixteen by sixteen. We use the block forms of
the gamma matrices given by Eq.(4.2) in Ref.\cite{long1} and
\begin{equation}
\Sigma_{i}^{\mu}=\gamma_{5i}\beta_{i}\gamma_{\perp i}^{\mu}%
,\ i=1,2.\label{sig}%
\end{equation}
The $\Sigma_{i}^{\mu}$ are four-vector generalizations of the Pauli matrices
of particles one and two. In the CM frame, the time component is zero and the
spatial components are the usual Pauli matrices for each particle. Appendix A
details the procedure that leads to a second-order Schr\"{o}dinger-like
eigenvalue equation for the four component wavefunction $|\phi_{+}\rangle=$
$|\psi\rangle_{1}+|\psi\rangle_{4}~$in the general form
\begin{equation}
(\mathbf{p}^{2}+\Phi(\mathbf{r},\mathbf{p,}{\mbox{\boldmath $\sigma$}}%
_{1},{\mbox{\boldmath $\sigma$}}_{2},w))|\phi_{+}\rangle=b^{2}(w)|\phi
_{+}\rangle.\label{slik}%
\end{equation}
Below we display all the general spin dependent structures in $\Phi
(\mathbf{r},\mathbf{p},\mbox{\boldmath $\sigma$}_{1}%
,\mbox{\boldmath $\sigma$}_{2},w)$ explicitly, ones very similar to those
appearing in nonrelativistic formalisms such as seen in the older
Hamada-Johnson and Yale group models (as well as the nonrelativistic limit of
Gross's equation). Simplification of the final result in Appendix A by using
identities involving $\mbox{\boldmath $\sigma$}_{1}$ and
$\mbox{\boldmath $\sigma$}_{2}$ and grouping by the $\mathbf{p}^{2}$ term ,
Darwin term $(\hat{\mathbf{r}}\cdot\mathbf{p)}$, spin-orbit angular momentum
term $\mathbf{L}\cdot(\mbox{\boldmath $\sigma$}_{1}%
+\mbox{\boldmath $\sigma$}_{2})$, spin-orbit angular momentum difference term
$\mathbf{L}\cdot(\mbox{\boldmath $\sigma$}_{1}-\mbox{\boldmath $\sigma$}_{2}%
)$, spin-spin term $(\mbox{\boldmath $\sigma$}_{1}\cdot
\mbox{\boldmath $\sigma$}_{2})$, tensor term $(\mbox{\boldmath $\sigma$}_{1}%
\cdot\hat{\mathbf{r}})(\mbox{\boldmath $\sigma$}_{2}\cdot\hat{\mathbf{r}})$,
additional spin dependent terms $\mathbf{L}\cdot(\mbox{\boldmath $\sigma$}_{1}%
\times\mbox{\boldmath $\sigma$}_{2})$ and $(\mbox{\boldmath $\sigma$}_{1}%
\cdot\hat{\mathbf{r}})(\mbox{\boldmath $\sigma$}_{2}\cdot\mathbf{p}%
)+({\mbox{\boldmath $\sigma$}}_{2}\cdot\hat{\mathbf{r}}%
)(\mbox{\boldmath $\sigma$}_{1}\cdot\mathbf{p})$ and spin independent terms we obtain%

\[
\{\mathbf{p}^{2}-i[2\mathcal{G}^{\prime}-\frac{E_{2}M_{2}+M_{1}E_{1}%
}{\mathcal{D}}(J+L)^{\prime}]\hat{\mathbf{r}}\cdot\mathbf{p}-\frac{1}{2}%
\nabla^{2}\mathcal{G-}\frac{1}{4}\mathcal{G}^{\prime2}%
\]

\[
-\frac{1}{4}(C+J-L)^{\prime}(-C+J-L)^{\prime}+\frac{1}{2}\frac{E_{2}%
M_{2}+M_{1}E_{1}}{\mathcal{D}}\mathcal{G}^{\prime}(J+L)^{\prime}%
\]

\[
+\frac{\mathbf{L}\cdot(\mbox{\boldmath $\sigma$}_{1}%
+\mbox{\boldmath $\sigma$}_{2})}{r}[\mathcal{G}^{\prime}-\frac{1}{2}%
\frac{E_{2}M_{2}+M_{1}E_{1}}{\mathcal{D}}(J+L)^{\prime}]-\frac{\mathbf{L}%
\cdot(\mbox{\boldmath $\sigma$}_{1}-\mbox{\boldmath $\sigma$}_{2})}{r}\frac
{1}{2}\frac{E_{2}M_{2}-M_{1}E_{1}}{\mathcal{D}}(J+L)^{\prime}%
\]

\[
+(\mbox{\boldmath $\sigma$}_{1}\cdot\mbox{\boldmath $\sigma$}_{2})[\frac{1}%
{2}\nabla^{2}\mathcal{G+}\frac{1}{2}\mathcal{G}^{\prime2}-\frac{1}{2}%
\frac{E_{2}M_{2}+M_{1}E_{1}}{\mathcal{D}}\mathcal{G}^{\prime}(J+L)^{\prime
}-\frac{1}{2}\mathcal{G}^{\prime}C^{\prime}-\frac{1}{2}\frac{\mathcal{G}%
^{\prime}}{r}-\frac{1}{2}\frac{(-C+J-L)^{\prime}}{r}]
\]

\begin{align*}
&  +(\mbox{\boldmath $\sigma$}_{1}\cdot\hat{\mathbf{r}}%
)(\mbox{\boldmath $\sigma$}_{2}\cdot\hat{\mathbf{r}})[-\frac{1}{2}\nabla
^{2}(-C+J-L)-\frac{1}{2}\nabla^{2}\mathcal{G-G}^{\prime}(-C+J-L)^{\prime
}-\mathcal{G}^{\prime2}+\frac{3}{2r}\mathcal{G}^{\prime}\\
&  +\frac{3}{2r}(-C+J-L)^{\prime}+\frac{1}{2}\frac{E_{2}M_{2}+M_{1}E_{1}%
}{\mathcal{D}}(J+L)^{\prime}(\mathcal{G}-C+J-L)^{\prime}]
\end{align*}

\[
+\frac{\mathbf{L}\cdot(\mbox{\boldmath $\sigma$}_{1}\times
\mbox{\boldmath $\sigma$}_{2})}{r}\frac{i}{2}\frac{M_{2}E_{1}-M_{1}E_{2}%
}{\mathcal{D}}(J+L)^{\prime}-\frac{i(J-L)^{\prime}}{2}%
((\mbox{\boldmath $\sigma$}_{1}\cdot\hat{\mathbf{r}}%
)({\mbox{\boldmath $\sigma$}}_{2}\cdot\mathbf{p}%
)+(\mbox{\boldmath $\sigma$}_{2}\cdot\hat{\mathbf{r}}%
)(\mbox{\boldmath $\sigma$}_{1}\cdot\mathbf{p}))\}|\phi_{+}\rangle
\]

\begin{equation}
=\mathrm{\exp}(-2\mathcal{G})\mathcal{B}^{2}|\phi_{+}\rangle.\label{main1}%
\end{equation}
where%
\begin{align}
\mathcal{D} &  \equiv E_{1}M_{2}+E_{2}M_{1}\nonumber\\
\mathcal{B}^{2} &  =E_{1}{}^{2}-M_{1}{}^{2}=E_{2}{}^{2}-M_{2}{}^{2}\nonumber\\
&  =b^{2}(w)+(\epsilon_{1}^{2}+\epsilon_{2}^{2})\mathrm{\sinh}^{2}%
(J)+2\epsilon_{1}\epsilon_{2}\sinh(J)\cosh(J)\nonumber\\
&  -(m_{1}^{2}+m_{2}^{2})\mathrm{\sinh}^{2}(L)-2m_{1}m_{2}\sinh(L)\cosh
(L).\label{spnin}%
\end{align}
$E_{i},M_{i},C,J,L,\mathcal{G}$ are all functions of the invariant $r$. We
point out that Eq.(\ref{main1}) differs from the forms presented in
\cite{long1}. \ Whereas the above equation involve four component spinor wave
functions, the ones given in \cite{long1} are obtained in terms of matrix wave
functions involving one component scalar and three component vector wave
functions. The form we choose in this paper is easier to compare with the
earlier existing nonrelativistic forms.

All of above equations when reduced to radial form have first derivative terms
(from the $\mathbf{\hat{r}}\cdot\mathbf{p}$ and
$(\mbox{\boldmath $\sigma$}_{1}\cdot\hat{\mathbf{r}}%
)(\mbox{\boldmath $\sigma$}_{2}\cdot\mathbf{p})+(\mbox{\boldmath $\sigma$}_{2}%
\cdot\hat{\mathbf{r}})(\mbox{\boldmath $\sigma$}_{1}\cdot\mathbf{p})$
terms)\textbf{.} \ These can be easily eliminated for the uncoupled equations
but are problematic for the coupled equations. The variable phase method
developed by Calogero \cite{cal} for computation of phases shifts starts with
coupled and uncoupled stationary state nonrelativistic Schr\"{o}dinger
equations which do not include the first derivative terms in their radial
forms. An advantage of the above for the relativistic case is that they are
Schr\"{o}dinger-like equations. Before we can apply the techniques for phase
shift calculations which have been already developed for the
Schr\"{o}dinger-like system in nonrelativistic quantum mechanics, we must get
rid of these first derivative terms. In terms of the above equations we seek a
matrix transformation that eliminates the terms first order in $\mathbf{p}$. \ 

The general form of the eigenvalue equation given in Eq.(\ref{main1}) is:
\begin{align}
&  [\mathbf{p}^{2}-ig^{\prime}\mathbf{\hat{r}}\cdot\mathbf{p}+\frac{g^{\prime
}}{2r}\vec{L}\cdot(\mbox{\boldmath $\sigma$}_{1}+\mbox{\boldmath $\sigma$}_{2}%
)-ih^{\prime}(\mbox{\boldmath $\sigma$}_{1}\cdot\mathbf{\hat{r}}%
\mbox{\boldmath $\sigma$}_{2}\cdot\mathbf{p}+\mbox{\boldmath $\sigma$}_{2}%
\cdot\mathbf{\hat{r}}\mbox{\boldmath $\sigma$}_{1}\cdot\mathbf{p})\nonumber\\
&  +k\mbox{\boldmath $\sigma$}_{1}\cdot\mbox{\boldmath $\sigma$}_{2}%
+n\mbox{\boldmath $\sigma$}_{1}\cdot\mathbf{\hat{r}}%
\mbox{\boldmath $\sigma$}_{2}\cdot\mathbf{\hat{r}}+l\vec{L}\cdot
(\mbox{\boldmath $\sigma$}_{1}-\mbox{\boldmath $\sigma$}_{2})+ij\vec{L}%
\cdot(\mbox{\boldmath $\sigma$}_{1}\times\mbox{\boldmath $\sigma$}_{2}%
)+m]|\phi_{+}\rangle\nonumber\\
&  =\mathcal{B}^{2}e^{-2\mathcal{G}}|\phi_{+}\rangle.
\end{align}
The $m$ term is the spin independent part involving derivatives of the
potentials. For the equal mass case, two terms drop out (see Eq.(\ref{main1}%
)), and the above equation becomes
\begin{align}
&  [\mathbf{p}^{2}-ig^{\prime}\mathbf{\hat{r}}\cdot\mathbf{p}+\frac{g^{\prime
}}{2r}\vec{L}\cdot(\mbox{\boldmath $\sigma$}_{1}+\mbox{\boldmath $\sigma$}_{2}%
)-ih^{\prime}(\mbox{\boldmath $\sigma$}_{1}\cdot\mathbf{\hat{r}}%
\mbox{\boldmath $\sigma$}_{2}\cdot\mathbf{p}+\mbox{\boldmath $\sigma$}_{2}%
\cdot\mathbf{\hat{r}}\mbox{\boldmath $\sigma$}_{1}\cdot\mathbf{p}%
)\nonumber\label{general}\\
&  +k\mbox{\boldmath $\sigma$}_{1}\cdot\mbox{\boldmath $\sigma$}_{2}%
+n\mbox{\boldmath $\sigma$}_{1}\cdot\mathbf{\hat{r}}%
\mbox{\boldmath $\sigma$}_{2}\cdot\mathbf{\hat{r}}+m]|\phi_{+}\rangle
=\mathcal{B}^{2}e^{-2\mathcal{G}}|\phi_{+}\rangle.
\end{align}
We introduce the spin-dependent scale change
\begin{equation}
|\phi_{+}\rangle\equiv\exp(F+K\mbox{\boldmath $\sigma$}_{1}\cdot
\mathbf{\hat{r}}\mbox{\boldmath $\sigma$}_{2}\cdot\mathbf{\hat{r}})|\psi
_{+}\rangle\equiv(A+B\mbox{\boldmath $\sigma$}_{1}\cdot\mathbf{\hat{r}%
}\mbox{\boldmath $\sigma$}_{2}\cdot\mathbf{\hat{r}})|\psi_{+}\rangle.
\label{exp_1}%
\end{equation}
with $F,K,A,B$ to be determined. \ We find that
\begin{align}
\mathbf{p}|\phi_{+}\rangle &  =(A+B\mbox{\boldmath $\sigma$}_{1}%
\cdot\mathbf{\hat{r}}\mbox{\boldmath $\sigma$}_{2}\cdot\mathbf{\hat{r}%
})\mathbf{p}|\psi_{+}\rangle-i(A^{\prime}+B^{\prime}%
\mbox{\boldmath $\sigma$}_{1}\cdot\mathbf{\hat{r}}%
\mbox{\boldmath $\sigma$}_{2}\cdot\mathbf{\hat{r}})\mathbf{\hat{r}}|\psi
_{+}\rangle\nonumber\\
&  -i\frac{B}{r}[(\mbox{\boldmath $\sigma$}_{1}-\mbox{\boldmath $\sigma$}_{1}%
\cdot\mathbf{\hat{r}\hat{r}})\mbox{\boldmath $\sigma$}_{2}\cdot\mathbf{\hat
{r}}+(\mbox{\boldmath $\sigma$}_{2}-\mbox{\boldmath $\sigma$}_{2}%
\cdot\mathbf{\hat{r}\hat{r}})\mbox{\boldmath $\sigma$}_{1}\cdot\mathbf{\hat
{r}}]|\psi_{+}\rangle,
\end{align}
and
\[
\frac{g^{\prime}}{2r}\mathbf{L}\cdot(\mbox{\boldmath $\sigma$}_{1}%
+\mbox{\boldmath $\sigma$}_{2})|\phi_{+}\rangle
=(A+B\mbox{\boldmath $\sigma$}_{1}\cdot\mathbf{\hat{r}}%
\mbox{\boldmath $\sigma$}_{2}\cdot\mathbf{\hat{r}})\frac{g^{\prime}}%
{2r}\mathbf{L}\cdot({\mbox{\boldmath $\sigma$}}_{1}
+\mbox{\boldmath $\sigma$}_{2})|\psi_{+}\rangle
\]%
\begin{equation}
+\frac{g^{\prime}}{2r}B[2\mbox{\boldmath $\sigma$}_{1}\cdot
\mbox{\boldmath $\sigma$}_{2}-4ir\mbox{\boldmath $\sigma$}_{1}\cdot
\mathbf{\hat{r}}\mbox{\boldmath $\sigma$}_{2}\cdot\mathbf{\hat{r}%
}\,\mathbf{\hat{r}}\cdot\mathbf{p}+2ir(\mbox{\boldmath $\sigma$}_{1}%
\cdot\mathbf{\hat{r}}\mbox{\boldmath $\sigma$}_{2}\cdot\mathbf{p}%
+\mbox{\boldmath $\sigma$}_{2}\cdot\mathbf{\hat{r}}%
\mbox{\boldmath $\sigma$}_{1}\cdot\mathbf{p})-6\mbox{\boldmath $\sigma$}_{1}%
\cdot\mathbf{\hat{r}}\mbox{\boldmath $\sigma$}_{2}\cdot\mathbf{\hat{r}}%
]|\psi_{+}\rangle.
\end{equation}
We thus find that
\begin{equation}
-ig^{\prime}\mathbf{\hat{r}}\cdot\mathbf{p}|\phi_{+}\rangle
=(A+B{\mbox{\boldmath $\sigma$}}_{1}\cdot\mathbf{\hat{r}}%
\mbox{\boldmath $\sigma$}_{2}\cdot\mathbf{\hat{r}})(-ig^{\prime}%
\mathbf{\hat{r}}\cdot\mathbf{p})|\psi_{+}\rangle+C|\psi_{+}\rangle
\end{equation}
and
\begin{align}
&  -ih^{\prime}(\mbox{\boldmath $\sigma$}_{1}\cdot\mathbf{\hat{r}%
}\mbox{\boldmath $\sigma$}_{2}\cdot\mathbf{p}+\mbox{\boldmath $\sigma$}_{2}%
\cdot\mathbf{\hat{r}}\mbox{\boldmath $\sigma$}_{1}\cdot\mathbf{p})|\phi
_{+}\rangle\\
&  =(A+B\mbox{\boldmath $\sigma$}_{1}\cdot\mathbf{\hat{r}}%
\mbox{\boldmath $\sigma$}_{2}\cdot\mathbf{\hat{r}})(-ih^{\prime}%
[\mbox{\boldmath $\sigma$}_{1}\cdot\mathbf{\hat{r}}%
\mbox{\boldmath $\sigma$}_{2}\cdot\mathbf{p}+\mbox{\boldmath $\sigma$}_{2}%
\cdot\mathbf{\hat{r}}\mbox{\boldmath $\sigma$}_{1}\cdot\mathbf{p}])|\psi
_{+}\rangle+D|\psi_{+}\rangle\nonumber
\end{align}
and finally
\begin{align}
\mathbf{p}^{2}|\phi_{+}\rangle &  =(A+B\mbox{\boldmath $\sigma$}_{1}%
\cdot\mathbf{\hat{r}}\mbox{\boldmath $\sigma$}_{2}\cdot\mathbf{\hat{r}%
})\mathbf{p}^{2}|\psi_{+}\rangle-2i(A^{\prime}+B^{\prime}%
\mbox{\boldmath $\sigma$}_{1}\cdot\mathbf{\hat{r}}%
\mbox{\boldmath $\sigma$}_{2}\cdot\mathbf{\hat{r}})\mathbf{\hat{r}}%
\cdot\mathbf{p}|\psi_{+}\rangle\\
&  +i\frac{2B}{r}[2\mbox{\boldmath $\sigma$}_{1}\cdot\mathbf{\hat{r}%
}\mbox{\boldmath $\sigma$}_{2}\cdot\mathbf{\hat{r}}\,\mathbf{\hat{r}}%
\cdot\mathbf{p}-(\mbox{\boldmath $\sigma$}_{1}\cdot\mathbf{\hat{r}%
}\mbox{\boldmath $\sigma$}_{2}\cdot\mathbf{p}+\mbox{\boldmath $\sigma$}_{2}%
\cdot\mathbf{\hat{r}}\mbox{\boldmath $\sigma$}_{1}\cdot\mathbf{p}]|\psi
_{+}\rangle+E|\psi_{+}\rangle\nonumber
\end{align}
where $C$ and $D$ and $E$ do not involve $\mathbf{p}$ and are given by
\begin{equation}
C=-g^{\prime}(A^{\prime}+B^{\prime}\mbox{\boldmath $\sigma$}_{1}%
\cdot\mathbf{\hat{r}}\mbox{\boldmath $\sigma$}_{2}\cdot\mathbf{\hat{r}}),
\end{equation}%
\begin{equation}
D=-2h^{\prime}(\mbox{\boldmath $\sigma$}_{1}\cdot\mathbf{\hat{r}%
}\mbox{\boldmath $\sigma$}_{2}\cdot\mathbf{\hat{r}}A^{\prime}+B^{\prime
})-2h^{\prime}\frac{B}{r}[\mathbf{L}\cdot(\mbox{\boldmath $\sigma$}_{1}%
+\mbox{\boldmath $\sigma$}_{2})+2-{\mbox{\boldmath $\sigma$}}_{1}%
\cdot\mathbf{\hat{r}}\mbox{\boldmath $\sigma$}_{2}\cdot\mathbf{\hat{r}%
}+\mbox{\boldmath $\sigma$}_{1}\cdot\mbox{\boldmath $\sigma$}_{2}],
\end{equation}
and
\begin{equation}
E=-(A^{\prime\prime}+B^{\prime\prime}\mbox{\boldmath $\sigma$}_{1}%
\cdot\mathbf{\hat{r}}\mbox{\boldmath $\sigma$}_{2}\cdot\mathbf{\hat{r}}%
)-\frac{2}{r}(A^{\prime}+B^{\prime}\mbox{\boldmath $\sigma$}_{1}%
\cdot\mathbf{\hat{r}}\mbox{\boldmath $\sigma$}_{2}\cdot\mathbf{\hat{r}%
})-2\frac{B}{r^{2}}(\mbox{\boldmath $\sigma$}_{1}\cdot
\mbox{\boldmath $\sigma$}_{2}-3\mbox{\boldmath $\sigma$}_{1}\cdot
\mathbf{\hat{r}}\mbox{\boldmath $\sigma$}_{2}\cdot\mathbf{\hat{r}}).
\end{equation}
The general form of the eigenvalue equation then becomes after some detail
\cite{liu}
\begin{align}
&  (A+B\mbox{\boldmath $\sigma$}_{1}\cdot\mathbf{\hat{r}}%
\mbox{\boldmath $\sigma$}_{2}\cdot\mathbf{\hat{r}})[\mathbf{p}^{2}-ig^{\prime
}\mathbf{\hat{r}}\cdot\mathbf{p}+\frac{g^{\prime}}{2r}\mathbf{L}%
\cdot(\mbox{\boldmath $\sigma$}_{1}+\mbox{\boldmath $\sigma$}_{2})-ih^{\prime
}(\mbox{\boldmath $\sigma$}_{1}\cdot\mathbf{\hat{r}}%
\mbox{\boldmath $\sigma$}_{2}\cdot\mathbf{p}+\mbox{\boldmath $\sigma$}_{2}%
\cdot\mathbf{\hat{r}}\mbox{\boldmath $\sigma$}_{1}\cdot\mathbf{p})]|\psi
_{+}\rangle\nonumber\\
&  +{\large (}\frac{g^{\prime}}{2r}B[2\mbox{\boldmath $\sigma$}_{1}%
\cdot{\mbox{\boldmath $\sigma$} }_{2}-4ir\mbox{\boldmath $\sigma$}_{1}%
\cdot\mathbf{\hat{r}}\mbox{\boldmath $\sigma$}_{2}\cdot\mathbf{\hat{r}%
}\,\mathbf{\hat{r}}\cdot\mathbf{p}+2ir(\mbox{\boldmath $\sigma$}_{1}%
\cdot\mathbf{\hat{r}}\mbox{\boldmath $\sigma$}_{2}\cdot\mathbf{p}%
+\mbox{\boldmath $\sigma$}_{2}\cdot\mathbf{\hat{r}}%
\mbox{\boldmath $\sigma$}_{1}\cdot\mathbf{p})-6\mbox{\boldmath $\sigma$}_{1}%
\cdot\mathbf{\hat{r}}\mbox{\boldmath $\sigma$}_{2}\cdot\mathbf{\hat{r}%
}]\nonumber\\
&  -2i(A^{\prime}+B^{\prime}\mbox{\boldmath $\sigma$}_{1}\cdot\mathbf{\hat{r}%
}\mbox{\boldmath $\sigma$}_{2}\cdot\mathbf{\hat{r}})\mathbf{\hat{r}}%
\cdot\mathbf{p}+i\frac{2B}{r}[2\mbox{\boldmath $\sigma$}_{1}\cdot
\mathbf{\hat{r}}\mbox{\boldmath $\sigma$}_{2}\cdot\mathbf{\hat{r}%
}\,\mathbf{\hat{r}}\cdot\mathbf{p}-(\mbox{\boldmath $\sigma$}_{1}%
\cdot\mathbf{\hat{r}}\mbox{\boldmath $\sigma$}_{2}\cdot\mathbf{p}%
+\mbox{\boldmath $\sigma$}_{2}\cdot\mathbf{\hat{r}}%
\mbox{\boldmath $\sigma$}_{1}\cdot\mathbf{p})]\nonumber\\
&  +(k\mbox{\boldmath $\sigma$}_{1}\cdot\mbox{\boldmath $\sigma$}_{2}%
+n\mbox{\boldmath $\sigma$}_{1}\cdot\mathbf{\hat{r}}%
\mbox{\boldmath $\sigma$}_{2}\cdot\mathbf{\hat{r}}%
)(A+B\mbox{\boldmath $\sigma$}_{1}\cdot\mathbf{\hat{r}}%
\mbox{\boldmath $\sigma$}_{2}\cdot\mathbf{\hat{r}})+R+m{\large )}|\psi
_{+}\rangle\nonumber\\
&  =\mathcal{B}^{2}\exp(-2\mathcal{G})(A+B\mbox{\boldmath $\sigma$}_{1}%
\cdot\mathbf{\hat{r}}\mbox{\boldmath $\sigma$}_{2}\cdot\mathbf{\hat{r}}%
)|\psi_{+}\rangle\label{e217}%
\end{align}
in which $R=C+D+E$.

\ Now, to bring this equation to the desired Schr\"{o}dinger-like form with no
linear\textbf{\ }$\mathbf{p}$ term we multiply both sides by
\begin{equation}
(A+B\mbox{\boldmath $\sigma$}_{1}\cdot\mathbf{\hat{r}}%
\mbox{\boldmath $\sigma$}_{2}\cdot\mathbf{\hat{r}})^{-1}=\frac
{(A-B\mbox{\boldmath $\sigma$}_{1}\cdot\mathbf{\hat{r}}%
\mbox{\boldmath $\sigma$}_{2}\cdot\mathbf{\hat{r}})}{A^{2}-B^{2}}%
\end{equation}
and find, using the exponential form above that appears in Eq.(\ref{exp_1}),
(and some detail \cite{liu})
\begin{align}
&  (A+B\mbox{\boldmath $\sigma$}_{1}\cdot\mathbf{\hat{r}}%
\mbox{\boldmath $\sigma$}_{2}\cdot\mathbf{\hat{r}})^{-1}[-2i(A^{\prime
}+B^{\prime}\mbox{\boldmath $\sigma$}_{1}\cdot\mathbf{\hat{r}}%
\mbox{\boldmath $\sigma$}_{2}\cdot\mathbf{\hat{r}})]\mathbf{\hat{r}}%
\cdot\mathbf{p}\nonumber\\
&  =-2i(F^{\prime}+K^{\prime}\mbox{\boldmath $\sigma$}_{1}\cdot\mathbf{\hat
{r}}\mbox{\boldmath $\sigma$}_{2}\cdot\mathbf{\hat{r}})\mathbf{\hat{r}}%
\cdot\mathbf{p},
\end{align}
and%
\begin{align}
&  (A+B\mbox{\boldmath $\sigma$}_{1}\cdot\mathbf{\hat{r}}%
\mbox{\boldmath $\sigma$}_{2}\cdot\mathbf{\hat{r}})^{-1}i\frac{2B}%
{r}[2\mbox{\boldmath $\sigma$}_{1}\cdot\mathbf{\hat{r}}%
\mbox{\boldmath $\sigma$}_{2}\cdot\mathbf{\hat{r}}\,\mathbf{\hat{r}}%
\cdot\mathbf{p}-(\mbox{\boldmath $\sigma$}_{1}\cdot\mathbf{\hat{r}%
}\mbox{\boldmath $\sigma$}_{2}\cdot\mathbf{p}+\mbox{\boldmath $\sigma$}_{2}%
\cdot\mathbf{\hat{r}}\mbox{\boldmath $\sigma$}_{1}\cdot\mathbf{p})]\nonumber\\
&  =\frac{2i\sinh(K)\cosh(K)}{r}[2\mbox{\boldmath $\sigma$}_{1}\cdot
\mathbf{\hat{r}}\mbox{\boldmath $\sigma$}_{2}\cdot\mathbf{\hat{r}%
}\,\mathbf{\hat{r}}\cdot\mathbf{p}-(\mbox{\boldmath $\sigma$}_{1}%
\cdot\mathbf{\hat{r}}\mbox{\boldmath $\sigma$}_{2}\cdot\mathbf{p}%
+\mbox{\boldmath $\sigma$}_{2}\cdot\mathbf{\hat{r}}%
\mbox{\boldmath $\sigma$}_{1}\cdot\mathbf{p})]\nonumber\\
&  +G
\end{align}
where(\cite{liu})%

\begin{equation}
G=-\frac{2\sinh^{2}(K)}{r^{2}}\mathbf{L}\cdot(\mbox{\boldmath $\sigma$}_{1}%
+\mbox{\boldmath $\sigma$}_{2}),
\end{equation}
and
\begin{align}
&  (A+B\mbox{\boldmath $\sigma$}_{1}\cdot\mathbf{\hat{r}}%
\mbox{\boldmath $\sigma$}_{2}\cdot\mathbf{\hat{r}})^{-1}\frac{g^{\prime}}%
{2r}B[2\mbox{\boldmath $\sigma$}_{1}\cdot\mbox{\boldmath $\sigma$}_{2}%
-4ir\mbox{\boldmath $\sigma$}_{1}\cdot\mathbf{\hat{r}}%
\mbox{\boldmath $\sigma$}_{2}\cdot\mathbf{\hat{r}}\,\mathbf{\hat{r}}%
\cdot\mathbf{p}\nonumber\\
&  +2ir(\mbox{\boldmath $\sigma$}_{1}\cdot\mathbf{\hat{r}}%
\mbox{\boldmath $\sigma$}_{2}\cdot\mathbf{p}+\mbox{\boldmath $\sigma$}_{2}%
\cdot\mathbf{\hat{r}}\mbox{\boldmath $\sigma$}_{1}\cdot\mathbf{p}%
)-6\mbox{\boldmath $\sigma$}_{1}\cdot\mathbf{\hat{r}}%
\mbox{\boldmath $\sigma$}_{2}\cdot\mathbf{\hat{r}}]\nonumber\\
&  =\frac{ig^{\prime}\sinh(K)\cosh(K)}{2r}[-4r\mbox{\boldmath $\sigma$}_{1}%
\cdot\mathbf{\hat{r}}\mbox{\boldmath $\sigma$}_{2}\cdot\mathbf{\hat{r}%
}\,\mathbf{\hat{r}}\cdot\mathbf{p}+2r(\mbox{\boldmath $\sigma$}_{1}%
\cdot\mathbf{\hat{r}}\mbox{\boldmath $\sigma$}_{2}\cdot\mathbf{p}%
+\mbox{\boldmath $\sigma$}_{2}\cdot\mathbf{\hat{r}}%
\mbox{\boldmath $\sigma$}_{1}\cdot\mathbf{p})\nonumber\\
&  -2i\mbox{\boldmath $\sigma$}_{1}\cdot\mbox{\boldmath $\sigma$}_{2}%
+6i\mbox{\boldmath $\sigma$}_{1}\cdot\mathbf{\hat{r}}%
\mbox{\boldmath $\sigma$}_{2}\cdot\mathbf{\hat{r}}]+H
\end{align}
where (\cite{liu})
\[
H=\frac{g^{\prime}\sinh^{2}(K)}{2r}[2\mathbf{L}\cdot
(\mbox{\boldmath $\sigma$}_{1}+\mbox{\boldmath $\sigma$}_{2}%
)-2\mbox{\boldmath $\sigma$}_{1}\cdot\mathbf{\hat{r}}%
\mbox{\boldmath $\sigma$}_{2}\cdot\mathbf{\hat{r}}%
+2\mbox{\boldmath $\sigma$}_{1}\cdot\mbox{\boldmath $\sigma$}_{2}+4].
\]
Note that $G$ and $H$ do not contain linear $\mathbf{p}$ type of \ terms. Now
collect the three different linear $\mathbf{p}$ type of \ terms in equation
(\ref{e217}):%

\begin{equation}
(-2iF^{\prime}-ig^{\prime})\mathbf{\hat{r}}\cdot\mathbf{p,}%
\end{equation}%
\begin{equation}
(-2i\frac{\sinh(K)\cosh(K)}{r}-ih^{\prime}+ig^{\prime}\sinh(K)\cosh
(K))(\mbox{\boldmath $\sigma$}_{1}\cdot\mathbf{\hat{r}}%
\mbox{\boldmath $\sigma$}_{2}\cdot\mathbf{p}+\mbox{\boldmath $\sigma$}_{2}%
\cdot\mathbf{\hat{r}}\mbox{\boldmath $\sigma$}_{1}\cdot\mathbf{p}),
\label{ex1}%
\end{equation}%
\begin{equation}
(4i\frac{\sinh(K)\cosh(K)}{r}-2i\sinh(K)\cosh(K)g^{\prime}-2iK^{\prime
})\mbox{\boldmath $\sigma$}_{1}\cdot\mathbf{\hat{r}}%
\mbox{\boldmath $\sigma$}_{2}\cdot\mathbf{\hat{r}}\,\mathbf{\hat{r}}%
\cdot\mathbf{p.} \label{ex2}%
\end{equation}
If we set the first of the above equations to $0$, we obtain the expected
result (for the uncoupled portion of the equation)
\begin{equation}
F^{\prime}=-g^{\prime}/2.
\end{equation}
If we set $h^{\prime}=-K^{\prime}$ and use $\mathbf{p}=\mathbf{\hat{r}%
}(\mathbf{\hat{r}}.\mathbf{p})-\frac{\mathbf{\hat{r}}\times L}{r}$ then the
two expressions(\ref{ex1}) and (\ref{ex2}) combine to
\begin{equation}
(2\frac{\sinh(K)\cosh(K)}{r}+h^{\prime}-g^{\prime}\sinh(K)\cosh(K))\frac
{\mbox{\boldmath $\sigma$}_{1}\cdot\mathbf{\hat{r}}%
\mbox{\boldmath $\sigma$}_{2}\cdot\mathbf{\hat{r}}\vec{L}\cdot
(\mbox{\boldmath $\sigma$}_{1}+\mbox{\boldmath $\sigma$}_{2})}{r}%
\end{equation}
which contains no $\,\mathbf{\hat{r}}\cdot\mathbf{p.}$ Thus the matrix scale
change
\begin{equation}
|\phi_{+}\rangle=\exp(-g/2)\exp(-h\mbox{\boldmath $\sigma$}_{1}\cdot
\mathbf{\hat{r}}\mbox{\boldmath $\sigma$}_{2}\cdot\mathbf{\hat{r}})|\psi
_{+}\rangle
\end{equation}
eliminates the linear $\mathbf{p}$ terms$.$

Further note \ that
\begin{align*}
&  (A+B\mbox{\boldmath $\sigma$}_{1}\cdot\mathbf{\hat{r}}%
\mbox{\boldmath $\sigma$}_{2}\cdot\mathbf{\hat{r}})^{-1}%
(k\mbox{\boldmath $\sigma$}_{1}\cdot\mbox{\boldmath $\sigma$}_{2}%
+n\mbox{\boldmath $\sigma$}_{1}\cdot\mathbf{\hat{r}}%
\mbox{\boldmath $\sigma$}_{2}\cdot\mathbf{\hat{r}}%
)(A+B\mbox{\boldmath $\sigma$}_{1}\cdot\mathbf{\hat{r}}%
\mbox{\boldmath $\sigma$}_{2}\cdot\mathbf{\hat{r}})\\
&  =(k\mbox{\boldmath $\sigma$}_{1}\cdot\mbox{\boldmath $\sigma$}_{2}%
+n\mbox{\boldmath $\sigma$}_{1}\cdot\mathbf{\hat{r}}%
\mbox{\boldmath $\sigma$}_{2}\cdot\mathbf{\hat{r}}),
\end{align*}%
\begin{equation}
(A+B\mbox{\boldmath $\sigma$}_{1}\cdot\mathbf{\hat{r}}%
\mbox{\boldmath $\sigma$}_{2}\cdot\mathbf{\hat{r}})^{-1}C|\psi_{+}%
\rangle=-g^{\prime}(F^{\prime}+K^{\prime}\mbox{\boldmath $\sigma$}_{1}%
\cdot\mathbf{\hat{r}}\mbox{\boldmath $\sigma$}_{2}\cdot\mathbf{\hat{r}}%
)|\psi_{+}\rangle,
\end{equation}
and (after some algebraic detail \cite{liu})%

\[
(A+B\mbox{\boldmath $\sigma$}_{1}\cdot\mathbf{\hat{r}}%
\mbox{\boldmath $\sigma$}_{2}\cdot\mathbf{\hat{r}})^{-1}D|\psi_{+}%
\rangle=-2h^{\prime}(K^{\prime}+F^{\prime}\mbox{\boldmath $\sigma$}_{1}%
\cdot\mathbf{\hat{r}}\mbox{\boldmath $\sigma$}_{2}\cdot\mathbf{\hat{r}}%
)|\psi_{+}\rangle
\]

\[
-2h^{\prime}\frac{\cosh(K)\sinh(K)}{r}[\mathbf{L}\cdot
(\mbox{\boldmath $\sigma$}_{1}+\mbox{\boldmath $\sigma$}_{2}%
)+2-\mbox{\boldmath $\sigma$}_{1}\cdot\mathbf{\hat{r}}%
\mbox{\boldmath $\sigma$}_{2}\cdot\mathbf{\hat{r}}%
+\mbox{\boldmath $\sigma$}_{1}\cdot\mbox{\boldmath $\sigma$}_{2}]|\psi
_{+}\rangle
\]

\begin{equation}
+2h^{\prime}\frac{\sinh^{2}(K)}{r}[\mbox{\boldmath $\sigma$}_{1}%
\cdot\mathbf{\hat{r}}\mbox{\boldmath $\sigma$}_{2}\cdot\mathbf{\hat{r}L}%
\cdot(\mbox{\boldmath $\sigma$}_{1}+{\mbox{\boldmath $\sigma$} }%
_{2})+3\mbox{\boldmath $\sigma$}_{1}\cdot\mathbf{\hat{r}}%
\mbox{\boldmath $\sigma$}_{2}\cdot\mathbf{\hat{r}}%
-\mbox{\boldmath $\sigma$}_{1}\cdot\mbox{\boldmath $\sigma$}_{2}]|\psi
_{+}\rangle.
\end{equation}
also
\[
(A+B\mbox{\boldmath $\sigma$}_{1}\cdot\mathbf{\hat{r}}%
\mbox{\boldmath $\sigma$}_{2}\cdot\mathbf{\hat{r}})^{-1}E|\psi_{+}%
\rangle=-[F^{\prime\prime}+F^{\prime2}+K^{\prime2}+(2F^{\prime}K^{\prime
}+K^{\prime\prime})\mbox{\boldmath $\sigma$}_{1}\cdot\mathbf{\hat{r}%
}\mbox{\boldmath $\sigma$}_{2}\cdot\mathbf{\hat{r}}]
\]

\[
-\frac{2}{r}[F^{\prime}+K^{\prime}\mbox{\boldmath $\sigma$}_{1}\cdot
\mathbf{\hat{r}}\mbox{\boldmath $\sigma$}_{2}\cdot\mathbf{\hat{r}}%
]-2\frac{\cosh(K)\sinh(K)}{r^{2}}(\mbox{\boldmath $\sigma$}_{1}\cdot
\mbox{\boldmath $\sigma$}_{2}-3\mbox{\boldmath $\sigma$}_{1}\cdot
\mathbf{\hat{r}}\mbox{\boldmath $\sigma$}_{2}\cdot\mathbf{\hat{r}})
\]

\begin{equation}
+2\frac{\sinh^{2}(K)}{r^{2}}(\mbox{\boldmath $\sigma$}_{1}\cdot\mathbf{\hat
{r}}\mbox{\boldmath $\sigma$}_{2}\cdot\mathbf{\hat{r}}%
-\mbox{\boldmath $\sigma$}_{1}\cdot\mbox{\boldmath $\sigma$}_{2}-2).
\end{equation}
So combining all terms and grouping by $\mathbf{p}^{2}$ term , spin
independent terms, spin-orbit angular momentum term $\mathbf{L}\cdot
(\mbox{\boldmath $\sigma$}_{1} +\mbox{\boldmath $\sigma$}_{2})$, spin-spin
term ${(\mbox{\boldmath $\sigma$}}_{1}{\cdot\mbox{\boldmath $\sigma$}}_{2})$,
tensor term $(\mbox{\boldmath $\sigma$}_{1}\mathbf{\cdot}\hat{\mathbf{r}%
})({\mbox{\boldmath $\sigma$}}_{2}\cdot\hat{\mathbf{r}})$, additional spin
independent term we have our Schr\"{o}dinger-like equation%

\begin{align}
&  \{\mathbf{p}^{2}+\frac{2g^{\prime}\sinh^{2}(K)}{r}-g^{\prime}F^{\prime
}-2h^{\prime}K^{\prime}-4h^{\prime}\frac{\cosh(K)\sinh(K)}{r}\nonumber\\
&  -F^{\prime\prime}-F^{\prime2}-K^{\prime2}-\frac{2}{r}F^{\prime}%
-4\frac{\sinh^{2}(K)}{r^{2}}\nonumber\\
&  +\mathbf{L}\cdot(\mbox{\boldmath $\sigma$}_{1}%
+\mbox{\boldmath $\sigma$}_{2})[\frac{g^{\prime}}{2r}+\frac{g^{\prime}%
\sinh^{2}(K)}{r}-\frac{2\sinh^{2}(K)}{r^{2}}-2h^{\prime}\frac{\cosh
(K)\sinh(K)}{r}]\nonumber\\
&  +\mbox{\boldmath $\sigma$}_{1}\cdot\mathbf{\hat{r}}%
\mbox{\boldmath $\sigma$}_{2}\cdot\mathbf{\hat{r}L}\cdot
(\mbox{\boldmath $\sigma$}_{1}+\mbox{\boldmath $\sigma$}_{2})(2h^{\prime}%
\frac{\sinh^{2}(K)}{r}+2\frac{\sinh(K)\cosh(K)}{r^{2}}+\frac{h^{\prime}}%
{r}-\frac{g^{\prime}\sinh(K)\cosh(K)}{r})\nonumber\\
&  +\mbox{\boldmath $\sigma$}_{1}\cdot\mbox{\boldmath $\sigma$}_{2}%
[k+\frac{g^{\prime}\cosh(K)\sinh(K)}{r}+\frac{g^{\prime}\sinh^{2}(K)}%
{r}-2h^{\prime}\frac{\cosh(K)\sinh(K)}{r}\nonumber\\
&  -2h^{\prime}\frac{\sinh^{2}(K)}{r}-2\frac{\cosh(K)\sinh(K)}{r^{2}}%
-2\frac{\sinh^{2}(K)}{r^{2}}]+\nonumber\\
&  \mbox{\boldmath $\sigma$}_{1}\cdot\mathbf{\hat{r}}%
\mbox{\boldmath $\sigma$}_{2}\cdot\mathbf{\hat{r}}[n-\frac{3g^{\prime}%
\cosh(K)\sinh(K)}{r}-\frac{g^{\prime}\sinh^{2}K}{r}-g^{\prime}K^{\prime
}-2h^{\prime}F^{\prime}+\frac{2h^{\prime}\cosh K\sinh K}{r}\nonumber\\
&  +6h^{\prime}\frac{\sinh^{2}(K)}{r}-(2F^{\prime}K^{\prime}+K^{\prime\prime
})-\frac{2}{r}K^{\prime}+6\frac{\cosh(K)\sinh(K)}{r^{2}}+2\frac{\sinh^{2}%
(K)}{r^{2}}]+m\}|\psi_{+}\rangle\nonumber\\
&  =\mathcal{B}^{2}e^{-2\mathcal{G}}|\psi_{+}\rangle\label{master_eq}%
\end{align}
Comparing Eq.(\ref{general}) with Eq.(\ref{main1}) we find
\begin{equation}
g^{\prime}=2\mathcal{G}^{\prime}-\frac{E_{2}M_{2}+M_{1}E_{1}}{\mathcal{D}%
}(J+L)^{\prime}=2\mathcal{G}^{\prime}-\log^{\prime}\mathcal{D=-}2F^{\prime},
\end{equation}

\begin{equation}
h^{\prime}=\frac{(J-L)^{\prime}}{2}=-K^{\prime},
\end{equation}

\begin{equation}
k=\frac{1}{2}\nabla^{2}\mathcal{G+}\frac{1}{2}\mathcal{G}^{\prime2}-\frac
{1}{2}\mathcal{G}^{\prime}\log^{\prime}\mathcal{D}-\frac{1}{2}\mathcal{G}%
^{\prime}C^{\prime}-\frac{1}{2}\frac{\mathcal{G}^{\prime}}{r}-\frac{1}{2}%
\frac{(-C+J-L)^{\prime}}{r}, \label{k}%
\end{equation}

\[
n=-\frac{1}{2}\nabla^{2}(-C+J-L)-\frac{1}{2}\nabla^{2}\mathcal{G-G}^{\prime
}(-C+J-L)^{\prime}-\mathcal{G}^{\prime2}+\frac{3}{2r}\mathcal{G}^{\prime}%
\]%
\begin{equation}
+\frac{3}{2r}(-C+J-L)^{\prime}+\frac{1}{2}\log^{\prime}\mathcal{D}%
(\mathcal{G}-C+J-L)^{\prime}, \label{j}%
\end{equation}

\begin{equation}
m=-\frac{1}{2}\nabla^{2}\mathcal{G-}\frac{1}{4}\mathcal{G}^{\prime2}-\frac
{1}{4}(C+J-L)^{\prime}(-C+J-L)^{\prime}+\frac{1}{2}\mathcal{G}^{\prime}%
\log^{\prime}\mathcal{D}\text{.}%
\end{equation}

Eq.(\ref{master_eq}) and it's derivation is an important part of this paper.
It will provide us with a way to derive phase shift equations using work by
other authors who developed methods for the nonrelativistic Schr\"{o}dinger
equation. First we need the radial form of the coordinate space form of this equation.

The following are the radial eigenvalue equations for singlet states
$^{1}S_{0}$, $^{1}P_{1}$, $^{1}D_{2}$ and triplet states $^{3}P_{0}$,
$^{3}P_{1}$, $^{3}S_{1}$, $^{3}D_{1}$ corresponding to Eq.(\ref{master_eq})
with the above substitutions. We emphasis that unlike the potentials used by
Reid, Hamada-Johnson and the Yale group \cite{6,4,7}, our potentials are fixed
by the structures of the relativistic two-body Dirac equations and we do not
have the freedom of choosing different potentials for different angular
momentum states.

$^{1}S_{0}$, $^{1}P_{1}$, $^{1}D_{2}$ ( a general singlet $^{1}J_{j}$): \ For
these states $\mathbf{L}\cdot(\mbox{\boldmath $\sigma$}_{1}%
+\mbox{\boldmath $\sigma$}_{2})=0$, $\mbox{\boldmath $\sigma$}_{1}%
\cdot\mbox{\boldmath $\sigma$}_{2}=-3$, $\mbox{\boldmath $\sigma$}_{1}%
\cdot\mathbf{\hat{r}}\mbox{\boldmath $\sigma$}_{2}\cdot\mathbf{\hat{r}=-}1$.
There is no off diagonal term. We find (adding and subtracting the $b^{2}$
term)
\[
\{-\frac{d^{2}}{dr^{2}}+\frac{j(j+1)}{r^{2}}{+}\Phi(r)\}v(r)=b^{2}v(r)
\]
where our effective potential for above equation is
\[
\Phi(r)=\frac{(2\mathcal{G}-\log(\mathcal{D})-J+L)^{^{\prime}2}}{4}%
+\frac{(2\mathcal{G}-\log(\mathcal{D})-J+L)^{\prime\prime}}{2}+\frac
{(2\mathcal{G}-\log(\mathcal{D})-J+L)^{^{\prime}}}{r}\newline%
\]%
\begin{equation}
+\frac{1}{2}\nabla^{2}(-C+J\mathcal{-}L-3\mathcal{G})-\frac{1}{4}%
(C+J-L-\mathcal{G}+2\ln D)^{^{\prime}}(-C+J-L-3\mathcal{G})^{\prime
}-\mathcal{B}^{2}e^{-2\mathcal{G}}+b^{2}(w).
\end{equation}
Our radial eigenvalue equations for singlet states $^{1}S_{0}$, $^{1}P_{1}$,
$^{1}D_{2}$ have the same potential forms except for the $\frac{j(j+1)}{r^{2}%
}$ angular momentum barrier term. Later, we shall show that their potentials
actually are different due to the inclusion of isospin $\mathbf{\tau}%
_{1}\mathbf{\cdot\tau}_{2}$ terms.

$^{3}P_{1}$( a general triplet $^{3}J_{j}$): For these, $\mathbf{L}%
\cdot(\mbox{\boldmath $\sigma$}_{1}+\mbox{\boldmath $\sigma$}_{2})=-2$,
$\mbox{\boldmath $\sigma$}_{1}\cdot\mbox{\boldmath $\sigma$}_{2}=1$,
$\mbox{\boldmath $\sigma$}_{1}\cdot\mathbf{\hat{r}}{\mbox{\boldmath $\sigma$}}%
_{2}\cdot\mathbf{\hat{r}=}1$

For the $^{3}P_{1}$ state the radial eigenvalue equation is%

\[
\{-\frac{d^{2}}{dr^{2}}+\frac{j(j+1)}{r^{2}}{+}\Phi(r)\}v(r)=b^{2}v(r)
\]
with
\[
\Phi(r)=\frac{(2\mathcal{G}-\log(\mathcal{D})+J-L)^{\prime2}}{4}%
+\frac{(2\mathcal{G}-\log(\mathcal{D})+J-L)^{\prime\prime}}{2}+\frac
{(\mathcal{G}+J-L-C)^{\prime}}{r}\newline%
\]%
\begin{equation}
-{\frac{1}{2}}\nabla^{2}(-C+J-L+\mathcal{G})+\frac{1}{4}(2\log(\mathcal{D}%
)-(C+J-L+3\mathcal{G))}^{\prime}(J-L-C+\mathcal{G})^{\prime}-\mathcal{B}%
^{2}\exp(-2\mathcal{G})+b^{2}(w),
\end{equation}

The $^{3}S_{1}$ and $^{3}D_{1}$ are coupled states described by $u_{-}(r)$ and
$u_{+}(r)$ and their radial eigenvalue equations are (using $\mathbf{L}%
\cdot(\mbox{\boldmath $\sigma$}_{1}+\mbox{\boldmath $\sigma$}_{2})=2(j-1)$,
$\mbox{\boldmath $\sigma$}_{1}\cdot\mbox{\boldmath $\sigma$}_{2}=1$,
$\mbox{\boldmath $\sigma$}_{1}\cdot\mathbf{\hat{r}}%
\mbox{\boldmath $\sigma$}_{2}\cdot\mathbf{\hat{r}=}\frac{1}{2j+1}$ (diagonal
term), and $\mbox{\boldmath $\sigma$}_{1}\cdot\mathbf{\hat{r}}%
\mbox{\boldmath $\sigma$}_{2}\cdot\mathbf{\hat{r}=}\frac{2\sqrt{j(j+1)}}%
{2j+1}$(off diagonal term)) in the form%

\begin{equation}
\{-\frac{d^{2}}{dr^{2}}+\Phi_{11}(r)\}u_{-}+\Phi_{12}(r)u_{+}=b^{2}u_{-}
\label{B21112}%
\end{equation}

\begin{equation}
\{-\frac{d^{2}}{dr^{2}}{+}\frac{6}{r^{2}}+\Phi_{22}(r)\}u_{+}+\Phi
_{21}(r)u_{-}=b^{2}u_{+} \label{B22221}%
\end{equation}
where
\begin{align}
\Phi_{11}(r)  &  =\{\frac{8}{3}\frac{(2\mathcal{G}^{\prime}-\log^{\prime
}\mathcal{D})\sinh^{2}(h)}{r}+\frac{8}{3}\frac{(J-L)^{\prime}\cosh(h)\sinh
(h)}{r}-\frac{16}{3}\frac{\sinh^{2}(h)}{r^{2}}+\frac{(2\mathcal{G}^{\prime
}-\log^{\prime}(\mathcal{D}))^{2}}{4}\nonumber\\
&  +\frac{(J-L)^{\prime2}}{4}+\frac{(2\mathcal{G}^{\prime}-\log^{\prime
}(\mathcal{D}))(J-L)^{\prime}}{6}+\frac{(2\mathcal{G}^{\prime\prime}%
-\log^{\prime\prime}(\mathcal{D}))}{2}\nonumber\\
&  +\frac{(J-L)^{\prime\prime}}{6}+\frac{(2\mathcal{G}^{\prime}-\log^{\prime
}(\mathcal{D}))}{r}+\frac{(J-L)^{\prime}}{3r}\nonumber\\
&  \frac{1}{3}[-{\frac{1}{2}}\nabla^{2}(-C+J-L+\mathcal{G})-\mathcal{G}%
^{\prime}(J-L-C+\mathcal{G})^{\prime}+{\frac{1}{2}}\log^{\prime}%
(\mathcal{D})(\mathcal{G}+J-L-C)^{\prime})]\nonumber\\
&  \frac{1}{4}\mathcal{G}^{\prime2}-{\frac{1}{2}}\mathcal{G}^{\prime}%
C^{\prime}-\frac{1}{4}(C+J-L)^{\prime}(-C+J-L)^{\prime}-\mathcal{B}^{2}%
\exp(-2\mathcal{G})+b^{2}(w)\}\\
& \nonumber\\
\Phi_{12}(r)  &  =\frac{2\sqrt{2}}{3}\{(2\mathcal{G}^{\prime}-\log^{\prime
}\mathcal{D})(\frac{3\cosh(h)\sinh(h)}{r}-\frac{\sinh^{2}(h)}{r}%
)+(J-L)^{\prime}(\frac{3\sinh^{2}(h)}{r}-\frac{\cosh(h)\sinh(h)}%
{r})\nonumber\\
&  -\frac{6\cosh(h)\sinh(h)}{r^{2}}+\frac{2\sinh^{2}(h)}{r^{2}}-{\frac{1}{2}%
}\nabla^{2}(-C+J-L+\mathcal{G})\nonumber\\
&  -\mathcal{G}^{\prime}(J-L-C+\mathcal{G})^{\prime}+\frac{3(\mathcal{G}%
+J-L-C)^{\prime}}{2r}+{\frac{1}{2}}\log^{\prime}(\mathcal{D})(\mathcal{G}%
+J-L-C)^{\prime}\nonumber\\
&  +\frac{(2\mathcal{G}^{\prime}-\log^{\prime}(\mathcal{D}))(J-L)^{\prime}}%
{2}+\frac{(J-L)^{\prime\prime}}{2}+\frac{(J-L)^{\prime}}{r}\} \label{phi1}%
\end{align}

\begin{align}
\Phi_{22}(r)  &  =\{-\frac{8}{3}\frac{(2\mathcal{G}^{\prime}-\log^{\prime
}\mathcal{D})\sinh^{2}(h)}{r}-\frac{8}{3}\frac{(J-L)^{\prime}\cosh(h)\sinh
(h)}{r}+\frac{16}{3}\frac{\sinh^{2}(h)}{r^{2}}+\frac{(2\mathcal{G}^{\prime
}-\log^{\prime}(\mathcal{D}))^{2}}{4}\nonumber\\
&  +\frac{(J-L)^{\prime2}}{4}-\frac{(2\mathcal{G}^{\prime}-\log^{\prime
}(\mathcal{D}))(J-L)^{\prime}}{6}+\frac{(2\mathcal{G}^{\prime\prime}%
-\log^{\prime\prime}(\mathcal{D}))}{2}-\frac{(J-L)^{\prime\prime}}%
{6}\nonumber\\
&  -\frac{2(2\mathcal{G}^{\prime}-\log^{\prime}(\mathcal{D}))}{r}%
+\frac{2(J-L)^{\prime}}{3r}-\frac{(\mathcal{G}+J-L-C)^{\prime}}{r}\nonumber\\
&  -\frac{1}{3}[-{\frac{1}{2}}\nabla^{2}(-C+J-L+\mathcal{G})-\mathcal{G}%
^{\prime}(J-L-C+\mathcal{G})^{\prime}+{\frac{1}{2}}\log^{\prime}%
(\mathcal{D})(\mathcal{G}+J-L-C)^{\prime})]\nonumber\\
&  \frac{1}{4}\mathcal{G}^{\prime2}-{\frac{1}{2}}\mathcal{G}^{\prime}%
C^{\prime}-\frac{1}{4}(C+J-L)^{\prime}(-C+J-L)^{\prime}-\mathcal{B}^{2}%
\exp(-2\mathcal{G})+b^{2}(w)\}\\
& \nonumber\\
\Phi_{21}(r)  &  =\Phi_{12}\label{phi2}\\
&  -4\sqrt{2}[(J-L)^{\prime}(\frac{\sinh^{2}(h)}{r}+\frac{1}{2r})-\frac
{2\cosh(h)\sinh(h)}{r^{2}}+\frac{(2\mathcal{G}^{\prime}-\log^{\prime
}\mathcal{D})\cosh(h)\sinh(h)}{r}]\}\nonumber
\end{align}
(Note that \ because of the spin-orbit-tensor term, the potential is not
symmetric.) In Appendix B we give the coupled equations for triplet
$^{3}j_{j-1}$ and $^{3}j_{j+1}$ for general $j.~$The remaining special case is
for the $^{3}P_{0}$ state and has the form%

\[
\{-\frac{d^{2}}{dr^{2}}+\frac{2}{r^{2}}{+}\Phi(r)\}v=b^{2}(w)v
\]
where%

\[
\Phi(r)=\frac{(2\mathcal{G}-\log(\mathcal{D})-J+L)^{\prime2}}{4}%
+\frac{(2\mathcal{G}-\log(\mathcal{D})-J+L)^{\prime\prime}}{2}+\frac
{(\log(\mathcal{D})-(4\mathcal{G}+J-L-2C))^{\prime}}{r}\newline%
\]%
\[
+{\frac{1}{2}}\nabla^{2}(-C+J-L+\mathcal{G})-{\frac{1}{2}}\mathcal{G}^{\prime
}C^{\prime}+{\frac{1}{4}(C}^{\prime2}-(J-L)^{\prime2})+\mathcal{G}^{\prime
}(\frac{5}{4}\mathcal{G}+J-L-C)^{\prime}\newline%
\]%
\begin{equation}
-\frac{1}{2}\log^{\prime}(\mathcal{D})(J-L-C+\mathcal{G})^{\prime}%
-\mathcal{B}^{2}\exp(-2\mathcal{G})+b^{2}(w),
\end{equation}

Now we can apply the techniques already developed for the radial
Schr\"{o}dinger equation
\begin{equation}
(-\frac{d^{2}}{dr^{2}}+\frac{l(l+1)}{r^{2}}+2mV_{lsj}(r))v=2mEv
\end{equation}
in nonrelativistic quantum mechanics to the above radial equations by the
substitutions
\begin{equation}
2mV_{lsj}(r)\rightarrow\Phi_{lsj}(r),\qquad2mE\longrightarrow b^{2}(w).
\end{equation}
By comparing $\Phi$ and $2m$ one could determine whether our $\Phi$ is similar
to standard type of phenomenological potentials such as Reid's potentials. But
first, in the next section, we discuss the models we used in our calculation.
This includes how we choose the $\mathcal{G}$, $L$ and $C$ invariant potential
functions, the mesons we used in our calculation, and the way they enter into
the two-body Dirac equations.

\section{The Invariant Interaction Functions}

\subsection{The $\mathcal{G}$ and $L$ Interaction Functions}

Our dynamics depends on how we parametrize the invariant interaction functions
$\mathcal{G},L$ and $C.$ We first consider how to model $\mathcal{G}$ and $L,$
corresponding to vector and scalar interactions. As we have seen, in order
that Eq.(\ref{D-a}) and Eq.(\ref{D-b}) satisfy Eq.(\ref{S1S2}), it is
necessary that the invariant functions $G$, $E_{1},E_{2},$ $M_{1}$ and $M_{2}$
depend on the relative separation, $x=x_{1}-x_{2}$, only through the
space-like coordinate four vector $x_{\bot}^{\mu}=x^{\mu}+\hat{P}^{\mu}%
(\hat{P}\cdot x),$ perpendicular to the total four-momentum $P.$ For QCD and
QED applications, $G$, $E_{1},E_{2}$ are functions \cite{c2}\cite{c4} of an
invariant $\mathcal{A}$. The explicit forms for functions $E_{1},E_{2},$ $G$
are
\begin{align}
E_{1}  &  =G(\epsilon_{1}-\mathcal{A})\label{E1-b}\\
E_{2}  &  =G(\epsilon_{2}-\mathcal{A})\nonumber
\end{align}
and
\begin{equation}
G^{2}=\frac{1}{(1-\frac{2\mathcal{A}}{w})}. \label{G2-c}%
\end{equation}
The function $\mathcal{A}(r)$ is responsible for the covariant
electromagnetic-like $A_{i}^{\mu}$. Even though the dependencies of
$E_{1},E_{2},G$ on $\mathcal{A}$ is not unique, they are constrained by the
requirement that they yield an effective Hamiltonian with the correct
nonrelativistic and semi-relativistic limits (classical and quantum mechanical
\cite{c14,saz3}). For QCD and QED application , $M_{1}$ and $M_{2}$ are
functions of two invariant functions \cite{c1}\cite{c4}, $\mathcal{A}(r)$ and
$S(r)$
\[
M_{1}^{2}(\mathcal{A},S)=m_{1}^{2}+G^{2}(2m_{w}S+S^{2})
\]%
\begin{equation}
M_{2}^{2}(\mathcal{A},S)=m_{2}^{2}+G^{2}(2m_{w}S+S^{2}). \label{M22}%
\end{equation}
The invariant function $S(r)$ is responsible for the scalar potential since
$S_{i}=0,$ if $S(r)=0,$ while $\mathcal{A}(r)$ contributes to the $S_{i}~$(if
$S(r)\neq0$ ) as well as to the vector potential $A_{i}^{\mu}.$ So, finally,
the five invariant functions $G$, $E_{1},E_{2},$ $M_{1}$ and $M_{2}$ (or
$\mathcal{G}=-J,L)$ depend on two independent invariant potential functions
$S$ and $\mathcal{A}$. (Compare also the spin independent portions to
Eqs.(\ref{scl},\ref{faiaa}) through calculation of $E_{i}^{2}-M_{i}^{2}%
-b^{2}.$)

Expressing $G$, $E_{1},E_{2},$ $M_{1}$ and $M_{2}$ in terms of $S$ and
$\mathcal{A}$ is important for semi-phenomenological and other applications
that emphasize the relationship of the interactions to effective external
potentials of the two associated one-body problems. However, the five
invariants $G$, $E_{1},E_{2},$ $M_{1}$ and $M_{2}$ can also be expressed in
the hyperbolic representation \cite{c12} in terms of the three invariants $L,$
$J$ and $\mathcal{G}$ (see Eqs.(\ref{m}), (\ref{e}) and (\ref{Ge}))$.~L,$ $J$
and $\mathcal{G}$ generate scalar, time-like vector and space-like vector
interactions respectively and enter into our Dirac equations via the sum
$\Delta_{L}+\Delta_{J}+\Delta_{\mathcal{G}}$ where Eqs(\ref{interL}%
,\ref{interJ},\ref{interG}) define $\Delta_{L},\Delta_{J},\Delta_{\mathcal{G}%
}$.

We may use Eqs.(\ref{d4}) to relate the matrix potentials $\Delta$ to a given
field theoretical or semi-phenomenological Feynman amplitude. As mentioned
earlier, a matrix amplitude proportional to $\gamma_{1}^{\mu}\cdot\gamma
_{2\mu}$ corresponding to an electromagnetic-like interaction would require
\cite{c6} $J=-\mathcal{G}$. Matrix amplitude proportional to either
$I_{1}I_{2}$ or $\gamma_{1}\cdot\hat{P}\gamma_{2}\cdot\hat{P}$ would
correspond to semi-phenomenological scalar or time-like vector interactions.
The two-body Dirac equations in the hyperbolic form of Eq.(\ref{d4}) give a
simple version \cite{c12} for the norm of the sixteen component Dirac spinor.
The two-body Dirac equations in \textquotedblleft\ external potential
\textquotedblright\ form, Eq.(\ref{D-a}) and Eq.(\ref{D-b}), (or more
generally (\ref{f1}) and (\ref{f2})) are simpler to reduce to the
Schr\"{o}dinger-like form and are useful for numerical calculations (see
Sazdjian \cite{Saz2} for a related reduction). We describe the parametrization
of the pseudoscalar interaction $C$ below in Eq.(\ref{C-t1t1}).

\subsection{Mesons Used in the Phase Shift Calculations}

We obtain our semi-phenomenological potentials for two nucleon interactions by
incorporating the meson exchange model and the two-body Dirac equations.
Because the pion is the lightest meson, its exchange is associated with the
longest range nuclear force. The shortest range behaviors of our
semi-phenomenological potentials are modified by the form factors, which are
treated purely phenomenologically. We exclude heavy mesons that mediate the
ranges shorter than that modified by the form factors. The intermediate range
part of our semi-phenomenological potentials comes from exchange of mesons
which are heavier than the pion. We use a total of 9 mesons in our fits. These
include scalar mesons $\sigma,$ $a_{0}$ and $f_{0},$ vector mesons $\rho,$
$\omega$ and $\phi,$ and pseudoscalar mesons $\pi,$ $\eta$ and $\eta^{\prime
}.$ In this paper, we are ignoring tensor and pseudovector interactions,
limiting ourselves to vector, scalar and pseudoscalar interactions, all with
masses less than about 1000 $\mathrm{MeV}$. See the Table below for the
detailed features of the mesons we used \cite{cn} .%

\begin{table}[H] \centering
\caption{ Data On Mesons(T=isospin, G=G-parity, J=spin, $\protect\pi $
=parity)\label{tab:pions}}
\begin{tabular}
[c]{lllll}\hline\hline
Particles & Mass(MeV) & $T^{G}$ & $J^{\pi}$ & Width(MeV)\\\hline
$\pi^{\pm}$ & 139.57018$\pm$0.00035 & $1^{-}$ & $0^{-}$ & ---\\
$\pi^{0}$ & 134.9766 $\pm$ 0.0006 & $1^{-}$ & $0^{-}$ & ---\\
$\eta$ & 547.3 $\pm$ 0.12 & $0^{+}$ & $0^{-}$ & $(1.18\pm0.11)\times10^{-3}$\\
$\rho$ & 769.3 $\pm$ 0.8 & $1^{+}$ & $1^{-}$ & 150.2 $\pm$ 0.8\\
$\omega$ & 782.57 $\pm$ 0.12 & $0^{-}$ & $1^{-}$ & 8.44 $\pm$ 0.09\\
$\eta^{^{\prime}}$ & 957.78 $\pm$ 0.14 & $0^{+}$ & $0^{-}$ & 0.202 $\pm$
0.016\\
$\phi$ & 1019.417$\pm$ 0.014 & $0^{-}$ & $1^{-}$ & 4.458 $\pm$ 0.032\\
$f_{0}$ & 980 $\pm$ 10 & $0^{+}$ & $0^{+}$ & 40 to 100\\
$a_{0}$ & 984.8 $\pm$ 1.4 $\hspace{0in}$ & $1^{-}$ & $0^{+}$ & 50 to 100\\
$\sigma$ & 500--700 & $0^{+}$ & $0^{+}$ & 600 to 1000\\\hline\hline
\end{tabular}
\end{table}%

\subsection{ \bigskip Modeling the Invariant Interaction Functions}

We initially assume the following introduction of scalar interactions into
two-body Dirac equations (see Eqs.(\ref{S1-1}, \ref{S1-2}),\ref{M22})):%

\begin{equation}
S=-g_{\sigma}^{2}\frac{e^{-m_{\sigma}r}}{r}-(\mathbf{\tau}_{1}\mathbf{\cdot
\tau}_{2})g_{a_{0}}^{2}\frac{e^{-m_{a_{0}}r}}{r}-g_{f_{0}}^{2}\frac
{e^{-m_{f_{0}}r}}{r} \label{S-g2}%
\end{equation}
where $g_{\sigma}^{2},$ $g_{a_{0}}^{2},$ $g_{f_{0}}^{2}$ are coupling
constants for the $\sigma,$ $a_{0}$ and $f_{0}$ mesons and $m_{\sigma},$
$m_{a_{0}}$ and $m_{f_{0}}$ the corresponding masses. $(\mathbf{\tau}%
_{1}\mathbf{\cdot\tau}_{2})$ is $1$ or $-3$ for isospin triplet or singlet states.

Pseudoscalar interactions are assumed to enter into two-body Dirac equations
in the form (see Eq.(\ref{main1}))%

\begin{equation}
C=(\mathbf{\tau}_{1}\mathbf{\cdot\tau}_{2})\frac{g_{\pi}^{2}}{w}%
\frac{e^{-m_{\pi}r}}{r}+\frac{g_{\eta}^{2}}{w}\frac{e^{-m_{\eta}r}}{r}%
+\frac{g_{\eta^{\prime}}^{2}}{w}\frac{e^{-m_{\eta^{\prime}}r}}{r},
\label{C-t1t1}%
\end{equation}
where $w=\epsilon_{1}+\epsilon_{2}$ is total energy of two nucleon system.
$g_{\pi}^{2},$ $g_{\eta}^{2},$ $g_{\eta^{\prime}}^{2}$ are coupling constants
for mesons $\pi,$ $\eta$ and $\eta^{\prime}$ respectively and $m_{\pi},$
$m_{\eta}$ and $m_{\eta^{\prime}}$ the corresponding masses. This form for $C$
yields the correct limit at low energy.

We also initially assume that our vector interactions enter into two-body
Dirac equations in the form (see Eqs. (\ref{E1-b}) and (\ref{G2-c}))%

\begin{equation}
A=(\mathbf{\tau}_{1}\mathbf{\cdot\tau}_{2})g_{\rho}^{2}\frac{e^{-m_{\rho}r}%
}{r}+g_{w}^{2}\frac{e^{-m_{w}r}}{r}+g_{\phi}^{2}\frac{e^{-m_{\phi}r}}{r}
\label{A-t1t2}%
\end{equation}
where $g_{\rho}^{2},$ $g_{\omega}^{2},$ $g_{\phi}^{2}$ are coupling constants
for mesons $\rho,$ $\omega$ and $\phi$ and $m_{\rho},$ $m_{\omega}$ and
$m_{\phi}$ are the corresponding masses.

We use form factors to modify the small $r$ behaviors in $S,$ $C$
and$\mathcal{~}A$, that is, the shortest range part of nucleon-nucleon
interaction. We choose our form factors by replacing $r$ in $S,$ $C$ and $A$ with%

\begin{equation}
r\longrightarrow\sqrt{r^{2}+r_{0}^{2}}. \label{rr0}%
\end{equation}
In our first model, we just use two different $r_{0}^{\prime}s$ to fit the
experimental data, one $r_{0}$ for the pion, one for all the other 8 mesons
whose masses are heavier than pion's mass. We set these two $r_{0}^{\prime}s$
as two free parameters in our fit. These form factors are different from the
conventional choices, usually given in momentum space, but the effects are similar.

In the constraint equations, $A$ and $S$ are relativistic invariant functions
of the invariant separation $r=\sqrt{x_{\perp}^{2}}$ (see below for the
distinction between $\mathcal{A}$ and $A$)$.$ Since it is possible that $A$
and $S$ as identified from the nonrelativistic limit can take on large
positive and negative values, it is necessary to modify $G$, $E_{1},E_{2},$
$M_{1}$ and $M_{2}$ so that the interaction functions remain real when $A$
become large and repulsive \cite{c15}. These modifications are not unique but
must maintain correct limits.

We have tested several models, two of which can give us fair to good fit to
the experimental data.

\paragraph{Model 1}

For $E_{i}=G(\epsilon_{i}-\mathcal{A})$ to be real, we need only require that
$G$ be real or $\mathcal{A}$ $<w/2$. This restriction on $\mathcal{A}$ is
enough to ensure that $M_{i}=G\sqrt{m_{i}^{2}(1-2\mathcal{A}/w)+2m_{w}S+S^{2}%
}$ be real as well (as long as $S\geq$ $0$ ). In order that $\mathcal{A}$ be
so restricted we choose to redefine it as
\begin{equation}
\mathcal{A}=A,\hspace{0.5in}A\leq0 \label{ALT0}%
\end{equation}%
\begin{equation}
\mathcal{A}=\frac{A}{\sqrt{4A^{2}+w^{2}}},\hspace{0.5in}A\geq0. \label{tenor}%
\end{equation}
This parametrization gives an $\mathcal{A}$ that is continuous through its
second derivative.

We next consider the problems that may arise in the limit when one of the
masses becomes very large \cite{c15}. Even though both our masses used in this
paper are equal, we demand that our equations display correct limits. We must
modify $M_{1}$ and $M_{2}$ so that it has the correct static limit (say
$m_{2}$ $\rightarrow\infty$). It does appear that $M_{1}\rightarrow m_{1}+S$
when $m_{2}$ $\rightarrow\infty$. However this is only true if $m_{1}+S\geq0.$
In other words, in the limit $m_{2}$ $\rightarrow\infty,$ the two-body Dirac
equations would reduce to
\begin{equation}
(\gamma\cdot p_{1}+|m_{1}+S|)|\psi\rangle=0.
\end{equation}
This would deviate from the standard one-body Dirac equation in the region of
strong attractive scalar potential ($S<$ $-m_{1}$). In order to correct this
problem, we take advantage of the hyperbolic parametrization. We desire a form
for $M_{i}$ that has the expected behavior ($M_{i}\rightarrow m_{i}+S$ in the
limit when $S$ becomes large and negative and one of the masses is large). So
we modify our $L$ in the following way \cite{c15}%

\begin{equation}
\sinh L=\frac{SG^{2}}{w}(1+\frac{G^{2}(\epsilon_{w}-A)S}{m_{w}\sqrt
{w^{2}+S^{2}}}),\hspace{0.5in}S<0 \label{shl0}%
\end{equation}
and for
\[
S>0
\]%
\[
M_{1}^{2}=m_{1}^{2}+G^{2}(2m_{w}S+S^{2})
\]%
\begin{equation}
M_{2}^{2}=m_{2}^{2}+G^{2}(2m_{w}S+S^{2}) \label{M23}%
\end{equation}
with Eqs.(\ref{m}).

A crucial feature of this $\sinh L$ extrapolation is that for fixed $S,$ the
static limit($m_{2}\gg m_{1}$) form is $\sinh L$ $\rightarrow S/w$ which leads
to $M_{1}\rightarrow m_{1}+S.$ The above modifications are not unique give the
correct semirelativistic limits \cite{c15}.

\paragraph{Model 2}

This model comes from the work of H. Sazdjian \cite{saz3}. Using a special
techniques of amplitude summation, he is able to sum an infinite number of
Feynman diagrams (of the ladder and cross ladder variety). For the vector
interactions, he obtained results that correspond to Eq.(\ref{faiaa}) to
Eq.(\ref{faiaa1}) and Eq.(\ref{E1-b}) to Eq.(\ref{G2-c})(modified here in
Eq.(\ref{tenor}) for $A\geq0$). For scalar interactions $(L(S,A))$ he obtained
two results. One again agrees with Eq.(\ref{scl}) and Eqs.(\ref{M22}). As we
have seen above this must be modified (see Eq.(\ref{shl0})) for $S\leq0$. His
second result is the one we use here for our second model for $(L(S,A)).$ That
replaces Eq.(\ref{shl0})and Eq.(\ref{M23}) with the model:
\[
S+A>0
\]
then
\begin{equation}
S\longrightarrow-\mathcal{A+}\frac{(S+A)w}{\sqrt{4(S+A)^{2}+w^{2}}}%
\end{equation}
while if
\[
S+A<0
\]
we let
\begin{equation}
S\longrightarrow-\mathcal{A}+S+A.
\end{equation}
In both case we let
\begin{equation}
\sinh L=\sinh(-\frac{1}{2}\ln(1-\frac{2(S+\mathcal{A})}{w})-\mathcal{G}).
\label{shl2}%
\end{equation}

\subsection{Non-minimal Coupling Of Vector Mesons}

The coupling of the vector mesons in Eq.(\ref{A-t1t2}) corresponds in quantum
field theory to the minimal coupling $g_{\rho}V_{\mu}\overline{\psi}%
\gamma^{\mu}$ $\psi$ analogous to $eA_{\mu}\overline{\psi}\gamma^{\mu}$ $\psi$
in QED. In our model, we are not concerned about renormalization, since the
quantum field theory is not fundamental, so that we cannot rule out the
non-minimal coupling of the $\rho,$ $\omega,$ $\phi$ analogous to%

\begin{equation}
i\frac{e}{2M}\overline{\psi}[\gamma^{\mu},\gamma^{\nu}]\psi F_{\mu\nu}.
\label{tensor}%
\end{equation}

We can convert the above expressions to something simpler by integration by
parts and using the free Dirac equation for the spinor field. This
nonrenormalizable interaction becomes
\begin{equation}
i\frac{e}{2M}\overline{\psi}[\gamma^{\mu},\gamma^{\nu}]\psi F_{\mu\nu
}\rightarrow-i\frac{4em_{N}}{M}\overline{\psi}\gamma^{\mu}\psi A_{\mu}%
-i\frac{2e}{M}(\overline{\psi}\partial^{\mu}\psi-(\partial^{\mu}\overline
{\psi})\psi)A_{\mu}. \label{q-sqrt0}%
\end{equation}
The first term can be absorbed into the standard minimal coupling while the
second term gives rise to an amplitude written below. Changing from photon to
vector mesons ($\rho$) and using on shell features we find
\begin{equation}
\frac{4f_{\rho}^{2}(\eta_{\mu\nu}+\frac{q_{\mu}q_{\nu}}{m_{\rho}^{2}%
})(p+p^{\prime})^{\mu}(p+p^{\prime})^{\nu}}{M^{2}(q^{2}+m_{\rho}%
^{2}-i\varepsilon)}=\frac{4f_{\rho}^{2}(p+p^{\prime})^{2}}{M^{2}(q^{2}%
+m_{\rho}^{2}-i\varepsilon)}=\frac{-4f_{\rho}^{2}(4m_{N}^{2}+q^{2})}%
{M^{2}(q^{2}+m_{\rho}^{2}-i\varepsilon)} \label{q-sqrt}%
\end{equation}
where $q=p-p^{\prime}$. The mass $M$ is a mass scale for the interaction,
$m_{N}$ is the fermion (nucleon) mass and $m_{\rho}$ is the $\rho$ meson mass.

How does this interaction modify our Dirac equations? Which of the 8 or so
invariants are affected (see Eq.(\ref{interL}) to Eq.(\ref{interA}))? In terms
of its matrix structure, the above would appear to contribute to what we
called $\Delta_{L}$ (see Eq.(\ref{interL})). It is as if we include an
additional scalar interaction with an exchanged mass of a $\rho$ and subtract
from it the Laplacian (the $q^{2}$ terms in Eq.(\ref{q-sqrt})). That is
\begin{equation}
S\rightarrow S+S^{\prime}-\nabla^{2}S^{\prime}/4m_{N}^{2}%
\end{equation}
where
\begin{equation}
S^{\prime}=-\frac{16m_{N}^{2}}{M^{2}}\frac{f_{\rho}^{2}\exp(-m_{\rho}r)}{r}%
\end{equation}
so that the modification is rather simple. \ It has the opposite sign as the
vector interaction. \ That is, it would produce an attractive interaction for
$pp$ scattering. But to lowest order, its attractive effects are canceled by
the contribution of the first term on the right hand side of Eq.(\ref{q-sqrt0}%
).\ In our application, this means that Eq.(\ref{S-g2}) and Eq.(\ref{A-t1t2})
are replaced (including the $r_{0}$ by Eq.(\ref{rr0})) by
\begin{equation}
S=-g_{\sigma}^{2}\frac{e^{-m_{\sigma}\overline{r}}}{\overline{r}}-g_{a_{0}%
}^{2}\frac{e^{-m_{a_{0}}\overline{r}}}{\overline{r}}-g_{f_{0}}^{2}%
\frac{e^{-m_{f_{0}}\overline{r}}}{\overline{r}}-S^{\prime}\label{sg1}%
\end{equation}
where
\begin{equation}
S^{\prime}=(\tau_{1}\cdot\tau_{2})g_{\rho}^{\prime2}(1-\frac{\nabla^{2}%
}{4m_{N}^{2}})\frac{e^{-m_{\rho}\overline{r}}}{\overline{r}}+g_{w}^{\prime
2}(1-\frac{\nabla^{2}}{4m_{N}^{2}})\frac{e^{-m_{w}\overline{r}}}{\overline{r}%
}+g_{\phi}^{\prime2}(1-\frac{\nabla^{2}}{4m_{N}^{2}})\frac{e^{-m_{\phi
}\overline{r}}}{\overline{r}}%
\end{equation}
and
\begin{equation}
A=(\tau_{1}\cdot\tau_{2})(g_{\rho}^{2}+g_{\rho}^{\prime2})\frac{e^{-m_{\rho
}\overline{r}}}{\overline{r}}+(g_{w}^{2}+g_{w}^{\prime2})\frac{e^{-m_{w}%
\overline{r}}}{\overline{r}}+(g_{\phi}^{2}+g_{\phi}^{\prime2})\frac
{e^{-m_{\phi}\overline{r}}}{\overline{r}}\label{ag1}%
\end{equation}
where $g_{\rho}^{\prime2},$ $g_{\omega}^{\prime2},$ $g_{\phi}^{\prime2}$ are
also coupling constants we will fit.

\section{Variable Phase Approach for Calculating Phase Shifts}

\bigskip

In this section, we discuss and review the phase shift methods which we used
in our numerical calculations, which include phase shift equations for
uncoupled and coupled states and the phase shift equations with Coulomb
potentials. The variable phase approach developed by Calogero has several
advantages over the traditional approach. In the traditional approach, one
integrates the radial Schr\"{o}dinger equation from the origin to the
asymptotic region where the potential is negligible, and then compares the
phase of the radial wave function with that of a free wave and thus obtain the
phase shift. In the variable phase approach we need only integrate a first
order non-linear differential equation from the origin to the asymptotic
region, thereby obtaining directly the value of the scattering phase shift.

This method is very convenient for us since we can reduce our two-body Dirac
equations to a Schr\"{o}dinger-like form for which the variable phase approach
was developed. Thus, we can conveniently use this variable phase method to
compute the phase shift for our relativistic two-body equations.

\subsection{ Phase Shift Equation For Uncoupled Schr\"{o}dinger Equation}

\label{sec:pse}

Ref. \cite{cal} gives a derivation of a nonlinear equation for the phase shift
for the scattering on a spherically symmetrical potential with the boundary condition%

\begin{equation}
u_{l}(0)=0
\end{equation}
of the radial uncoupled Schr\"{o}dinger equation%

\begin{equation}
u_{l}^{\prime\prime}(r)+[k^{2}-\frac{l(l+1)}{r^{2}}-V(r)]u_{l}(r)=0.
\label{un}%
\end{equation}
The radial wave function is real, and it defines the \textquotedblleft
scattering phase shift\textquotedblright\ $\delta_{l}$ through the comparison
of its asymptotic behavior with that of the sine function:%

\begin{equation}
u_{l}(r)\overset{r\rightarrow\infty}{\longrightarrow}const\cdot\sin
(kr-\frac{l\pi}{2}+\delta_{l}) \label{infin}%
\end{equation}
The equation that Calogero derives is%

\begin{equation}
t_{l}^{\prime}(r)=-\frac{1}{k}V(r)[\hat{\jmath}_{l}(kr)-t_{l}(r)\hat{n}%
_{l}(kr)]^{2} \label{diff}%
\end{equation}
where $t_{l}(r)$ has the limiting value $\tan\delta_{l}$ with the boundary
condition $t_{l}(0)=0.$ This is a first-order nonlinear differential equation
and can be rewritten \cite{cal} in terms of another function $\delta_{l}(r)$
defined by
\begin{equation}
t_{l}(r)=\tan\delta_{l}(r) \label{tantl}%
\end{equation}
with the boundary condition%

\begin{equation}
\delta_{l}(r)\overset{r\rightarrow0}{\longrightarrow}0
\end{equation}
and limiting value%

\begin{equation}
\lim_{r\rightarrow\infty}\delta_{l}(r)\equiv\delta_{l}(\infty)=\delta_{l}%
\end{equation}
The differential equation for $\delta_{l}(r)$ is \cite{cal}%

\begin{equation}
\delta_{l}^{\prime}(r)=-k^{-1}V(r)\left[  \cos\delta_{l}(r)\hat{\jmath}%
_{l}(kr)-\sin\delta_{l}(r)\hat{n}_{l}(kr)\right]  ^{2} \label{ph-eq}%
\end{equation}
The solution of this first order nonlinear differential equation yields
asymptotically the value of the scattering phase shift. The function
$\delta_{l}(r)$ is named the \textquotedblleft phase function
\textquotedblright\ and Eq.(\ref{ph-eq}) is called the \textquotedblleft phase
equation \textquotedblright. It is our main tool for studying the properties
of scattering phase shifts. Eq.(\ref{ph-eq}) becomes particular simple in the
case of $S$ waves%

\begin{equation}
\delta_{0}^{\prime}(r)=-k^{-1}V(r)\sin^{2}[kr+\delta_{0}(r)]. \label{s-ph-eq}%
\end{equation}

Now, since our Schr\"{o}dinger-like equations in CM system has the form%

\begin{equation}
\lbrack\mathbf{\nabla}^{2}-b^{2}-\Phi]|\psi\rangle=0
\end{equation}
we can directly follow the above steps to obtain the phase shift by swapping
$k\rightarrow b$ , and $V\rightarrow\Phi.$ There is no change in the phase
shift equation, even though our quasipotential $\Phi$ depends on the CM system
energy $w.$

We have found in convenient to put all the angular momentum barrier terms in
the potentials, and change all the phase shift equations to the form of $S$
state-like phase shift equations \cite{cal}. \ This puts our phase shift
equations in a much simpler form. For spin singlet states, our phase shift
equations become just%

\begin{equation}
\delta_{l}^{\prime}(r)=-b^{-1}\Phi_{l}(r)\sin^{2}[br+\delta_{l}(r)].
\label{mes}%
\end{equation}
This equation is similar to the $^{1}S_{0}$ state phase equation(see
Eq.(\ref{s-ph-eq})), but it works well for all the singlet states when the
angular momentum barrier term ($\frac{l(l+1)}{r^{2}}$) is included in
$\Phi_{l}(r),$
\begin{equation}
\Phi_{l}(r)=\Phi(r)+\frac{l(l+1)}{r^{2}}.
\end{equation}

Because the nucleon-nucleon interactions are short range, we integrate our
phase shift equations (for both the singlet and triplet states) to a distance
(for example 6 fermis) where the nucleon-nucleon potential becomes very weak.
Then the angular momentum barrier terms $\frac{l(l+1)}{r^{2}}$ dominate the
potential $\Phi_{l}(r)$ and we let our potential $\Phi_{l}(r)=\frac
{l(l+1)}{r^{2}}$ and integrate our phase shift equations from 6 fm to infinity
to get our phase shift. (This \ can be done analytically in the case of the
uncoupled equations \cite{cal}.)

Because of the modification of our phase shift equations, we also need to
modify our boundary conditions for phase shift equations. For the uncoupled
singlet states $^{1}P_{1}$, $^{1}D_{2}$ and triplet states $^{3}P_{0}$,
$^{3}P_{1}$, the modified boundary conditions are \cite{cal}%

\begin{equation}
\delta_{l}^{\prime}(0)=-\frac{l}{l+1}b.
\end{equation}
This is implemented numerically by an additional boundary conditions at $r=h,
$ so our boundary conditions for uncoupled singlet states $^{1}P_{1}$,
$^{1}D_{2}$ and triplet states $^{3}P_{0}$, $^{3}P_{1}$ are%

\begin{equation}
\delta_{l}(h)=-\frac{l}{l+1}bh \label{bh}%
\end{equation}
where $h$ is stepsize in our calculation, $b=\sqrt{b^{2}}$ and of course
$\delta_{l}(0)=0.$ So for $P$ and $D$ states, the new boundary conditions are
$\delta_{1}(h)=-\frac{1}{2}bh,$ $\delta_{2}(h)=-\frac{2}{3}bh$ respectively.

\subsection{ Phase Shift Equation For Coupled Schr\"{o}dinger Equations}

\label{sec:psecs}

For coupled Schr\"{o}dinger-like equations, the phase shift equation involves
coupled phase shift functions. \ We discuss an approach here different \ from
that originally presented in \cite{dega}. The key idea for this new derivation
is taken from one\ presented in a well known quantum text \cite{quant}. We
present an appropriate adaptation of this idea here in the uncoupled case to
demonstrate the general idea and then extend it to the coupled case. Consider
a radial equation of the form%
\[
(-\frac{d^{2}}{dr^{2}}+\Phi_{l}(r))u=b^{2}u
\]
Following \cite{quant} we assume
\begin{align}
u(r)  &  =A(r)\sin(br+\delta_{l}(r))\label{u1}\\
u^{\prime}(r)  &  =bA(r)\cos(br+\delta_{l}(r)). \label{ub}%
\end{align}
Taking the derivative of the first equation we find that
\begin{equation}
A^{\prime}=-A\delta_{l}^{\prime}\cot(br+\delta) \label{phs}%
\end{equation}
and then using this and Eq.(\ref{ub}) the above radial Schr\"{o}dinger
equation reduces to Eq.(\ref{mes}).

The coupled radial Schr\"{o}dinger equation has the form%
\begin{equation}
U^{\prime\prime}=-b^{2}U+\frac{1}{2}(\Phi_{L}U+U\Phi_{L}) \label{sch}%
\end{equation}
where both $U$ and $\Phi_{L}$ \ are two by two matrices. \ The effective
quasipotential matrix is of the form%
\begin{equation}
\Phi_{L}=\left(
\begin{array}
[c]{cc}%
\Phi_{11}+\frac{l_{1}(l_{1}+1)}{r^{2}} & \Phi_{12}\\
\Phi_{21} & \Phi_{22}+\frac{l_{2}(l_{2}+1)}{r^{2}}%
\end{array}
\right)  \label{quasi}%
\end{equation}
while the matrix wave function is assumed to be of the form
\begin{align}
U  &  =\frac{1}{2}[\mathcal{A}\sin(br+\mathcal{D}\mathbb{)+}\sin
(br+\mathcal{D}\mathbb{)}\mathcal{A}]\label{cua}\\
U^{\prime}  &  =\frac{b}{2}[\mathcal{A}\cos(br+\mathcal{D}\mathbb{)}%
+\cos(br+\mathcal{D}\mathbb{)}\mathcal{A}] \label{cub}%
\end{align}
with (using Pauli matrices to designate the matrix structure)
\begin{align}
\mathcal{D}  &  \mathbb{=}\delta+\mathbf{D}\cdot
\mbox{\boldmath $\sigma$}\nonumber\\
\mathcal{A}  &  \mathbb{=}a+\mathbf{A}\cdot\mbox{\boldmath $\sigma$}.
\end{align}
(The functions $\mathcal{D}$, $\mathcal{A}$ are not related to earlier
functions which use the same symbols). We further assume (for real and
symmetric potentials) that both the phase and amplituded functions are
diagonalized by the same orthogonal matrix%
\begin{equation}
\tilde{U}\mathbf{=}RUR^{-1}=(a+A\sigma_{3})\sin(br+\delta+D\sigma_{3}).
\end{equation}
Combining Eqs.(\ref{cua},\ref{cub}) together with Eq.(\ref{sch}) so as to
produce the analogue of the phase shift Eq.(\ref{phs}) requires we use the
following properties of the orthogonal matrix $R$
\begin{align}
R  &  =\left(
\begin{array}
[c]{cc}%
\cos\varepsilon(r) & \sin\varepsilon(r)\\
-\sin\varepsilon(r) & \cos\varepsilon(r)
\end{array}
\right) \nonumber\\
R^{\prime}R^{-1}  &  =\varepsilon^{\prime}\left(
\begin{array}
[c]{cc}%
0 & 1\\
-1 & 0
\end{array}
\right)  =\varepsilon^{\prime}i\sigma_{2}%
\end{align}
In appendix C we derive the coupled phase shift equations ($\delta=(\delta
_{1}+\delta_{2})/2,~D=(\delta_{1}-\delta_{2})/2$) below:
\begin{align}
\delta_{1}^{\prime}(r)  &  =-\frac{1}{b}[(\Phi_{11}+\frac{l_{1}(l_{1}%
+1)}{r^{2}})\cos^{2}\varepsilon(r)\label{psa}\\
&  +(\Phi_{22}+\frac{l_{2}(l_{2}+1)}{r^{2}})\sin^{2}\varepsilon(r)+\Phi
_{12}\sin2\varepsilon(r)]\sin^{2}(br+\delta_{1}(r)\mathbb{)}\nonumber\\
\delta_{2}^{\prime}(r)  &  =-\frac{1}{b}[(\Phi_{22}+\frac{l_{2}(l_{2}%
+1)}{r^{2}})\cos^{2}\varepsilon(r)\label{psb}\\
&  +(\Phi_{11}+\frac{l_{1}(l_{1}+1)}{r^{2}})\sin^{2}\varepsilon(r)-\Phi
_{12}\sin2\varepsilon(r)]\sin^{2}(br+\delta_{2}(r)\mathbb{)}\nonumber\\
\varepsilon^{\prime}(r)  &  =\frac{1}{b\sin(\delta_{1}(r)-\delta_{2}%
(r))}\{\frac{1}{2}[\Phi_{11}+\frac{l_{1}(l_{1}+1)}{r^{2}}-\Phi_{22}\frac
{l_{2}(l_{2}+1)}{r^{2}}]\sin2\varepsilon(r)\nonumber\\
&  -\Phi_{12}\cos2\varepsilon(r)\}\sin(br+\delta_{1}(r)\mathbb{)}%
\sin(br+\delta_{2}(r)\mathbb{)} \label{psc}%
\end{align}
Similar coupled equations are derived by \cite{dega} for coupled $S-$ wave
equations. Since our potentials include the angular momentum barrier terms we
use simple trigonometric functions in place of spherical Bessel and Hankel
functions. \ This requires a modification of the boundary conditions just as
in the uncoupled case. For this end we find it most convenient to rewrite the
above three equations in the matrix form
\begin{equation}
T_{L}^{\prime}=-\frac{1}{b}[\sin^{2}(br)\Phi_{L}+\sin(br)\cos(br)(\Phi
_{L}T_{L}+T_{L}\Phi_{L})+\cos^{2}(br)T_{L}\Phi_{L}T_{L}] \label{tp}%
\end{equation}
in which the matrix $T_{L}$ has eigenvalues of $\tan\delta_{1}$ and
$\tan\delta_{2}$. \ The actual phase shifts are
\begin{align}
\delta_{1}  &  =\delta_{1}(r\rightarrow\infty)\nonumber\\
\delta_{2}  &  =\delta_{2}(r\rightarrow\infty)\nonumber\\
\varepsilon &  =\varepsilon(r\rightarrow\infty),
\end{align}
The first boundary conditions on the above equations is%

\begin{equation}
T_{L}(0)=0. \label{Tis0}%
\end{equation}
The further numerical boundary conditions which we need \ for $\varepsilon
(h),$ $\delta_{1}(h)$ and $\delta_{2}(h)$ are from (for small $h$)
\begin{equation}
T_{L}(h)=hT_{L}^{\prime}(0). \label{Thh}%
\end{equation}
At small $r,$ we can approximate our $\Phi_{L}$ for coupled $S$ and $D$ states
in terms of their small $r$ behavior. \ \ We find \cite{smlr}
\begin{equation}
\Phi_{L}=\frac{1}{r^{2}}\left(
\begin{array}
[c]{ll}%
\eta_{-} & \eta_{0}\\
\eta_{0} & 6+\eta_{+}%
\end{array}
\right)  . \label{FIP}%
\end{equation}
Substitute Eq.(\ref{Thh}) and Eq.(\ref{FIP}) into Eq.(\ref{tp}) and we find%

\begin{equation}
T_{L}(h)=hT_{L}^{\prime}(0)=bh\left(
\begin{array}
[c]{ll}%
\alpha & \beta\\
\beta & -\frac{2}{3}+\gamma
\end{array}
\right)
\end{equation}
where
\[
\alpha=-\eta_{-},
\]%
\[
\beta=-\frac{1}{3}\eta_{0}%
\]%
\begin{equation}
\gamma=-\frac{\eta_{+}}{45}.
\end{equation}
Then we can find $\varepsilon(h),$ $\tan\delta_{-}(h)$ and $\tan\delta_{+}(h)$
by diagonalizing the matrix $T_{L}(h)$. The matrix diagonalizing $T_{L}(h)$ is%

\[
\left(
\begin{array}
[c]{ll}%
\cos\varepsilon & -\sin\varepsilon\\
\sin\varepsilon & \cos\varepsilon
\end{array}
\right)  .
\]
This leads to the initial conditions%

\[
\tan(2\varepsilon)=\frac{\frac{2}{3}\eta_{0}}{\eta_{-}-(\frac{2}{3}+\frac
{\eta_{+}}{45})}%
\]

\[
\tan\delta_{-}(h)=T_{11}=bh(-\eta_{-}\cos^{2}\varepsilon-\frac{2}{3}\eta
_{0}\cos\varepsilon\sin\varepsilon-(\frac{2}{3}+\frac{\eta_{+}}{45})\sin
^{2}\varepsilon)
\]

\begin{equation}
\tan\delta_{+}(h)=T_{22}=bh(-\eta_{-}\sin^{2}\varepsilon+\frac{2}{3}\eta
_{0}\cos\varepsilon\sin\varepsilon-(\frac{2}{3}+\frac{\eta_{+}}{45})\cos
^{2}\varepsilon)
\end{equation}
and from these initial conditions we can then integrate our equations
(\ref{psa}-\ref{psc}) for the coupled system. (Note how these reduce to the
uncoupled initial condition Eq.(\ref{bh}) with no coupling.)

\bigskip

\section{Phase Shift Calculations}

It is our aim to determine if an adequate description of the nucleon-nucleon
phase shifts can be obtained by the use of the CTBDE to incorporate the meson
exchange model. In contrast to the relativistic equations used in other
approaches \cite{mac}\cite{koch}\cite{arndt}\cite{berger}\cite{gross1,gross2},
the CTBDE can be exactly reduced to a local Schr\"{o}dinger-like form. This
allows us to gain additional physical insight into the nucleon-nucleon
interactions. We test our two models to find which one give us the best fit to
the experimental phase shift data in nucleon-nucleon scattering. These two
models are among many which have been tested.

The data set{\ \cite{stoks}} which we used in our test consist of $pp$ and
$np$ nucleon-nucleon scattering phase shift data up to $T_{Lab}=350$ MeV
published in physics journals between 1955 and 1992. In our fits, we use
experimental phase shift data for $NN$ scattering in the singlet states
$^{1}S_{0}$, $^{1}P_{1}$, $^{1}D_{2}$ and triplet states $^{3}P_{0}$,
$^{3}P_{1}$, $^{3}S_{1}$, $^{3}D_{1}$. We use our parameter fit results from
$np$ scattering to predicate the result in $pp$ scattering. (The variable
phase method for potentials including the Coulomb potential is reviewed in
Appendix D.) Thus we did not put the $pp$ scattering data of singlet states
$^{1}S_{0}$, $^{1}D_{2}$ and triplet states $^{3}P_{0}$, $^{3}P_{1}$ into our
fits. (There is no $pp$ scattering in $^{1}P_{1}$, $^{3}S_{1}$ and $^{3}D_{1}$
states because of the consideration of the Pauli principle).

We use 7 angular momentum states in our fit. There are $11$ data points for
every angular momentum state, in the energy range from $1$ to $350$ MeV, so
the total number of data points in our fits is 77. To determine the free
coupling constant (and the sigma mass $m_{\sigma}$) in our potentials, we have
to perform a best fit to the experimentally measured phase shift data. The
coupling constant are generally searched by minimizing the quantity $\chi^{2}$
. The definition of our $\chi^{2}$ is%

\begin{equation}
\chi^{2}=\sum_{i}\{\frac{\delta_{i}^{th}-\delta_{i}^{\exp}}{\Delta\delta_{i}%
}\}^{2} \label{Kai-sqare}%
\end{equation}
where the $\delta_{i}^{th}$ is theoretical phase shifts, the $\delta_{i}%
^{\exp}$ is experimental phase shifts and we let $\Delta\delta_{i}=1$ degree.
(Our model at this stage is too simplified to perform a fit that involves the
actual experimental errors).

We have tried several methods to minimize our $\chi^{2}$: the gradient method,
grid method and Monte Carlo simulations. Our $\chi^{2}$ drops very quickly at
the beginning if we search by the gradient method, then it always hits some
local minima and cannot jump out. Obviously, the grid method should lead us to
the global minimum. The problem is that if we want to find the best fit
parameters we must let the mesh size be small. But then the calculation time
becomes unbearably long. On the other hand if we choose a larger mesh, we will
miss the parameters which we are looking for.

We found that the Monte Carlo method can solve above dilemma. We set a
reasonable range for all the parameters which we want to fit and generate all
our fitting parameters randomly. Initially, the calculation time is also very
long for this method, but it can leads us to a rough area where our fitting
parameters are located in. Then we shrink the range for all our fitting
parameters and do our calculation again (or use the gradient method in
tandem), our calculation time then being greatly reduced. By repeating several
time in the same way, we can finally find the parameters.

To expedite our calculations further, we put restrictions on $^{1}S_{0}$ and
$^{3}S_{1}$ states. After every set of parameters is generated randomly, we
first test it on the $^{1}S_{0}$ state at $1$ MeV. For $^{1}S_{0}$ state, if
\begin{equation}
\mid{\delta_{i}^{th}-\delta_{i}^{\exp}}\mid>0.2\mid\delta_{i}^{\exp}\mid
\end{equation}
we let the computer jump out of this loop and generate another set of
parameters and test it again until a set of parameters passes this
restriction. Then we test it on the $^{3}S_{1}$ states at $1$ MeV with the
same restriction. We only calculate $\delta_{i}^{th}$ at higher energy if a
set of parameters pass these two restrictions. Our code can run at least $50 $
times faster by this two restrictions. After we shrink our parameter ranges 2
or 3 times, all of our parameters are confined in a small region. At this
time, we may change our restriction to
\begin{equation}
\mid{\delta_{i}^{th}-\delta_{i}^{\exp}}\mid>0.15\mid\delta_{i}^{\exp}\mid
\end{equation}
and put restriction on $^{1}P_{1}$ states or any other states to let our code
run more efficiently.

Using this method we tried several different models to fit the phase shift
experimental data of seven different angular momentum states including the
singlet states $^{1}S_{0}$, $^{1}P_{1}$, $^{1}D_{2}$ and triplet states
$^{3}P_{0}$, $^{3}P_{1}$, $^{3}S_{1}$, $^{3}D_{1}$. Two models which we
discussed above can give us a fairly good fit to the experimental data. The
parameters which we obtained for model 1 are listed in table
\ref{tab:parameter}, and for model 2 are listed in table \ref{tab:parameter2}.
For the features of mesons in table \ref{tab:parameter} and
\ref{tab:parameter2}, please refer to table \ref{tab:pions} and Eq.(\ref{S-g2}%
) to Eq.(\ref{A-t1t2}). The sigma mass is in MeV while the structure parameter
$r_{0}$ is in inverse MeV.%

\begin{table}[H] \centering
\caption{Parameters From Fitting
Experimental Data(Model 1).\label{tab:parameter}}
\begin{tabular}
[c]{||c|c|c|c|c|c|c|c||}\hline\hline
& $\eta$ & $\eta^{\prime}$ & $\sigma$ & $\rho$ & $\omega$ & $\pi$ & $a_{0}%
$\\\hline
$g^{2}$ & 2.25 & 4.80 & 47.9 & 11.6 & 16.5 & 13.3 & 0.13\\\hline
$r_{0}(\times10^{-3})$ & 2.843 & 2.843 & 2.843 & 2.843 & 2.843 & 0.645 &
2.843\\\hline
& $\phi$ & $f_{0}$ & $\rho^{\prime}$ & $\omega^{\prime}$ & $\phi^{\prime}$ &
$m_{\sigma}$ & \\\hline
$g^{2}$ & 5.64 & 19.9 & 0.34 & 20.6 & 3.10 & 724.1 & \\\hline
$r_{0}(\times10^{-3})$ & 2.843 & 2.843 & 2.843 & 2.843 & 2.843 & ------ &
\\\hline\hline
\end{tabular}
\end{table}%

.%

\begin{table}[H] \centering
\caption{Parameters From Fitting
Experimental Data(Model 2).\label{tab:parameter2}}
\begin{tabular}
[c]{||c|c|c|c|c|c|c|c||}\hline\hline
& $\eta$ & $\eta^{\prime}$ & $\sigma$ & $\rho$ & $\omega$ & $\pi$ & $a_{0}%
$\\\hline
$g^{2}$ & 0.88 & 1.70 & 54.7 & 2.58 & 18.3 & 13.6 & 10.5\\\hline
$r_{0}(\times10^{-3})$ & 1.336 & 1.264 & 3.180 & 6.640 & 2.627 & 1.717 &
9.282\\\hline
& $\phi$ & $f_{0}$ & $\rho^{\prime}$ & $\omega^{\prime}$ & $\phi^{\prime}$ &
$m_{\sigma}$ & \\\hline
$g^{2}$ & 9.12 & 33.5 & 5.11 & 28.6 & 12.1 & 694.3 & \\\hline
$r_{0}(\times10^{-3})$ & 11.45 & 4.447 & 6.640 & 2.627 & 11.45 & ------ &
\\\hline\hline
\end{tabular}
\end{table}%

\subsection{Model 1}

The theoretical phase shifts which we calculated by using the parameters for
model 1 and the experimental phase shifts for all the seven states are listed
in table \ref{tab:np-phase}. We use parameters given above to predict the
phase shift of $pp$ scattering. Our prediction for the four $pp$ scattering
states which include singlet states $^{1}S_{0}$, $^{1}D_{2}$ and triplet
states $^{3}P_{0}$, $^{3}P_{1}$ are listed in table \ref{tab:pp-phase}.%

\begin{table}[H] \centering
\caption{$np $ Scattering Phase Shift
Of $^{1}S_{0}$, $^{1}P_{1}$, $^{1}D_{2}$, $^{3}P_{0}$, $^{3}P_{1}$, $^{3}S_{1}$ And $^{3}D_{1}$ States(Model 1).\label{tab:np-phase}}
\begin{tabular}
[t]{||c|c|c|c|c||}\hline\hline
Energy & $^{1}S_{0}$ & $^{1}P_{1}$ & $^{1}D_{2}$ & $^{3}P_{0}$\\\hline
(MeV) &
\begin{tabular}
[c]{ll}%
Exp. & The.
\end{tabular}
&
\begin{tabular}
[c]{ll}%
Exp. & The.
\end{tabular}
&
\begin{tabular}
[c]{ll}%
Exp. & The.
\end{tabular}
&
\begin{tabular}
[c]{ll}%
Exp. & The.
\end{tabular}
\\\hline
1 &
\begin{tabular}
[c]{ll}%
62.07 & 59.96
\end{tabular}
&
\begin{tabular}
[c]{ll}%
-0.187 & -0.359
\end{tabular}
&
\begin{tabular}
[c]{ll}%
0.00 & 0.00
\end{tabular}
&
\begin{tabular}
[c]{ll}%
0.18 & 0.00
\end{tabular}
\\\hline
5 &
\begin{tabular}
[c]{ll}%
63.63 & 63.48
\end{tabular}
&
\begin{tabular}
[c]{ll}%
-1.487 & -1.169
\end{tabular}
&
\begin{tabular}
[c]{ll}%
0.04 & 0.00
\end{tabular}
&
\begin{tabular}
[c]{ll}%
1.63 & 1.55
\end{tabular}
\\\hline
10 &
\begin{tabular}
[c]{ll}%
59.96 & 60.40
\end{tabular}
&
\begin{tabular}
[c]{ll}%
-3.039 & -2.870
\end{tabular}
&
\begin{tabular}
[c]{ll}%
0.16 & 0.05
\end{tabular}
&
\begin{tabular}
[c]{ll}%
3.65 & 3.57
\end{tabular}
\\\hline
25 &
\begin{tabular}
[c]{ll}%
50.90 & 51.95
\end{tabular}
&
\begin{tabular}
[c]{ll}%
-6.311 & -6.641
\end{tabular}
&
\begin{tabular}
[c]{ll}%
0.68 & 0.52
\end{tabular}
&
\begin{tabular}
[c]{ll}%
8.13 & 8.72
\end{tabular}
\\\hline
50 &
\begin{tabular}
[c]{ll}%
40.54 & 41.65
\end{tabular}
&
\begin{tabular}
[c]{ll}%
-9.670 & -10.23
\end{tabular}
&
\begin{tabular}
[c]{ll}%
1.73 & 1.13
\end{tabular}
&
\begin{tabular}
[c]{ll}%
10.70 & 11.62
\end{tabular}
\\\hline
100 &
\begin{tabular}
[c]{ll}%
26.78 & 26.64
\end{tabular}
&
\begin{tabular}
[c]{ll}%
-14.52 & -13.49
\end{tabular}
&
\begin{tabular}
[c]{ll}%
3.90 & 2.00
\end{tabular}
&
\begin{tabular}
[c]{ll}%
8.460 & 10.17
\end{tabular}
\\\hline
150 &
\begin{tabular}
[c]{ll}%
16.94 & 15.18
\end{tabular}
&
\begin{tabular}
[c]{ll}%
-18.65 & -15.26
\end{tabular}
&
\begin{tabular}
[c]{ll}%
5.79 & 2.51
\end{tabular}
&
\begin{tabular}
[c]{ll}%
3.690 & 5.688
\end{tabular}
\\\hline
200 &
\begin{tabular}
[c]{ll}%
8.940 & 5.615
\end{tabular}
&
\begin{tabular}
[c]{ll}%
-22.18 & -16.49
\end{tabular}
&
\begin{tabular}
[c]{ll}%
7.29 & 2.91
\end{tabular}
&
\begin{tabular}
[c]{ll}%
-1.44 & 0.66
\end{tabular}
\\\hline
250 &
\begin{tabular}
[c]{ll}%
1.960 & -2.719
\end{tabular}
&
\begin{tabular}
[c]{ll}%
-25.13 & -17.60
\end{tabular}
&
\begin{tabular}
[c]{ll}%
8.53 & 3.11
\end{tabular}
&
\begin{tabular}
[c]{ll}%
-6.51 & -4.38
\end{tabular}
\\\hline
300 &
\begin{tabular}
[c]{ll}%
-4.460 & -10.16
\end{tabular}
&
\begin{tabular}
[c]{ll}%
-27.58 & -18.63
\end{tabular}
&
\begin{tabular}
[c]{ll}%
9.69 & 3.55
\end{tabular}
&
\begin{tabular}
[c]{ll}%
-11.47 & -9.206
\end{tabular}
\\\hline
350 &
\begin{tabular}
[c]{ll}%
-10.59 & -16.94
\end{tabular}
&
\begin{tabular}
[c]{ll}%
-29.66 & -19.68
\end{tabular}
&
\begin{tabular}
[c]{ll}%
10.96 & 3.311
\end{tabular}
&
\begin{tabular}
[c]{ll}%
-16.39 & -13.81
\end{tabular}
\\\hline
Energy & $^{3}P_{1}$ & $^{3}S_{1}$ & $^{3}D_{1}$ & $\varepsilon$\\\hline
(MeV) &
\begin{tabular}
[c]{ll}%
Exp. & The.
\end{tabular}
&
\begin{tabular}
[c]{ll}%
Exp. & The.
\end{tabular}
&
\begin{tabular}
[c]{ll}%
Exp. & The.
\end{tabular}
&
\begin{tabular}
[c]{ll}%
Exp. & The.
\end{tabular}
\\\hline
1 &
\begin{tabular}
[c]{ll}%
-0.11 & -0.33
\end{tabular}
&
\begin{tabular}
[c]{ll}%
147.747 & 142.692
\end{tabular}
&
\begin{tabular}
[c]{ll}%
-0.005 & 0.719
\end{tabular}
&
\begin{tabular}
[c]{ll}%
0.105 & 0.287
\end{tabular}
\\\hline
5 &
\begin{tabular}
[c]{ll}%
-0.94 & -0.88
\end{tabular}
&
\begin{tabular}
[c]{ll}%
118.178 & 112.670
\end{tabular}
&
\begin{tabular}
[c]{ll}%
-0.183 & -0.176
\end{tabular}
&
\begin{tabular}
[c]{ll}%
0.672 & 1.224
\end{tabular}
\\\hline
10 &
\begin{tabular}
[c]{ll}%
-2.06 & -2.26
\end{tabular}
&
\begin{tabular}
[c]{ll}%
102.611 & 98.215
\end{tabular}
&
\begin{tabular}
[c]{ll}%
-0.677 & -0.256
\end{tabular}
&
\begin{tabular}
[c]{ll}%
1.159 & 1.951
\end{tabular}
\\\hline
25 &
\begin{tabular}
[c]{ll}%
-4.88 & -5.70
\end{tabular}
&
\begin{tabular}
[c]{ll}%
80.63 & 78.38
\end{tabular}
&
\begin{tabular}
[c]{ll}%
-2.799 & -2.910
\end{tabular}
&
\begin{tabular}
[c]{ll}%
1.793 & 2.587
\end{tabular}
\\\hline
50 &
\begin{tabular}
[c]{ll}%
-8.25 & -10.18
\end{tabular}
&
\begin{tabular}
[c]{ll}%
62.77 & 62.00
\end{tabular}
&
\begin{tabular}
[c]{ll}%
-6.433 & -6.947
\end{tabular}
&
\begin{tabular}
[c]{ll}%
2.109 & 2.495
\end{tabular}
\\\hline
100 &
\begin{tabular}
[c]{ll}%
-13.24 & -16.66
\end{tabular}
&
\begin{tabular}
[c]{ll}%
43.23 & 43.18
\end{tabular}
&
\begin{tabular}
[c]{ll}%
-12.23 & -13.94
\end{tabular}
&
\begin{tabular}
[c]{ll}%
2.420 & 3.013
\end{tabular}
\\\hline
150 &
\begin{tabular}
[c]{ll}%
-17.46 & -22.12
\end{tabular}
&
\begin{tabular}
[c]{ll}%
30.72 & 30.64
\end{tabular}
&
\begin{tabular}
[c]{ll}%
-16.48 & -19.35
\end{tabular}
&
\begin{tabular}
[c]{ll}%
2.750 & 3.562
\end{tabular}
\\\hline
200 &
\begin{tabular}
[c]{ll}%
-21.30 & -26.98
\end{tabular}
&
\begin{tabular}
[c]{ll}%
21.22 & 20.95
\end{tabular}
&
\begin{tabular}
[c]{ll}%
-19.71 & -23.78
\end{tabular}
&
\begin{tabular}
[c]{ll}%
3.130 & 4.489
\end{tabular}
\\\hline
250 &
\begin{tabular}
[c]{ll}%
-24.84 & -31.46
\end{tabular}
&
\begin{tabular}
[c]{ll}%
13.39 & 12.95
\end{tabular}
&
\begin{tabular}
[c]{ll}%
-22.21 & -27.62
\end{tabular}
&
\begin{tabular}
[c]{ll}%
3.560 & 5.682
\end{tabular}
\\\hline
300 &
\begin{tabular}
[c]{ll}%
-28.07 & -35.67
\end{tabular}
&
\begin{tabular}
[c]{ll}%
6.600 & 6.127
\end{tabular}
&
\begin{tabular}
[c]{ll}%
-24.14 & -31.01
\end{tabular}
&
\begin{tabular}
[c]{ll}%
4.030 & 6.982
\end{tabular}
\\\hline
350 &
\begin{tabular}
[c]{ll}%
-30.97 & -39.58
\end{tabular}
&
\begin{tabular}
[c]{ll}%
0.502 & 0.171
\end{tabular}
&
\begin{tabular}
[c]{ll}%
-25.57 & -34.15
\end{tabular}
&
\begin{tabular}
[c]{ll}%
4.570 & 8.536
\end{tabular}
\\\hline\hline
\end{tabular}
\end{table}%
%

\begin{table}[H] \centering
\caption{$pp $ Scattering Phase Shift Of $^{1}S_{0} $, $^{1}D_{2}$,
$^{3}P_{0}$ And $^{3}P_{1}$ States(Model 1).\label{tab:pp-phase}}
\begin{tabular}
[t]{||c|c|c|c|c||}\hline\hline
Energy & $^{1}S_{0}$ & $^{1}D_{2}$ & $^{3}P_{0}$ & $^{3}P_{1}$\\\hline
MeV &
\begin{tabular}
[c]{ll}%
Exp. & The.
\end{tabular}
&
\begin{tabular}
[c]{ll}%
Exp. & The.
\end{tabular}
&
\begin{tabular}
[c]{ll}%
Exp. & The.
\end{tabular}
&
\begin{tabular}
[c]{ll}%
Exp. & The.
\end{tabular}
\\\hline
1 &
\begin{tabular}
[c]{ll}%
32.68 & 51.95
\end{tabular}
&
\begin{tabular}
[c]{ll}%
0.001 & -0.091
\end{tabular}
&
\begin{tabular}
[c]{ll}%
0.134 & 0.381
\end{tabular}
&
\begin{tabular}
[c]{ll}%
-0.081 & -1.215
\end{tabular}
\\\hline
5 &
\begin{tabular}
[c]{ll}%
54.83 & 55.47
\end{tabular}
&
\begin{tabular}
[c]{ll}%
0.043 & -0.183
\end{tabular}
&
\begin{tabular}
[c]{ll}%
1.582 & 0.954
\end{tabular}
&
\begin{tabular}
[c]{ll}%
-0.902 & -2.536
\end{tabular}
\\\hline
10 &
\begin{tabular}
[c]{ll}%
55.22 & 54.45
\end{tabular}
&
\begin{tabular}
[c]{ll}%
0.165 & -0.270
\end{tabular}
&
\begin{tabular}
[c]{ll}%
3.729 & 1.773
\end{tabular}
&
\begin{tabular}
[c]{ll}%
-2.060 & -3.864
\end{tabular}
\\\hline
25 &
\begin{tabular}
[c]{ll}%
48.67 & 47.64
\end{tabular}
&
\begin{tabular}
[c]{ll}%
0.696 & -0.441
\end{tabular}
&
\begin{tabular}
[c]{ll}%
8.575 & 5.422
\end{tabular}
&
\begin{tabular}
[c]{ll}%
-4.932 & -7.932
\end{tabular}
\\\hline
50 &
\begin{tabular}
[c]{ll}%
38.90 & 37.77
\end{tabular}
&
\begin{tabular}
[c]{ll}%
1.711 & -0.504
\end{tabular}
&
\begin{tabular}
[c]{ll}%
11.47 & 9.766
\end{tabular}
&
\begin{tabular}
[c]{ll}%
-8.317 & -13.15
\end{tabular}
\\\hline
100 &
\begin{tabular}
[c]{ll}%
24.97 & 23.63
\end{tabular}
&
\begin{tabular}
[c]{ll}%
3.790 & 0.511
\end{tabular}
&
\begin{tabular}
[c]{ll}%
9.450 & 7.862
\end{tabular}
&
\begin{tabular}
[c]{ll}%
-13.26 & -18.45
\end{tabular}
\\\hline
150 &
\begin{tabular}
[c]{ll}%
14.75 & 12.37
\end{tabular}
&
\begin{tabular}
[c]{ll}%
5.606 & 1.141
\end{tabular}
&
\begin{tabular}
[c]{ll}%
4.740 & 3.812
\end{tabular}
&
\begin{tabular}
[c]{ll}%
-17.43 & -24.42
\end{tabular}
\\\hline
200 &
\begin{tabular}
[c]{ll}%
6.550 & 3.024
\end{tabular}
&
\begin{tabular}
[c]{ll}%
7.058 & 2.407
\end{tabular}
&
\begin{tabular}
[c]{ll}%
--0.370 & -1.178
\end{tabular}
&
\begin{tabular}
[c]{ll}%
-21.25 & -28.50
\end{tabular}
\\\hline
250 &
\begin{tabular}
[c]{ll}%
-0.31 & -5.15
\end{tabular}
&
\begin{tabular}
[c]{ll}%
8.270 & 2.994
\end{tabular}
&
\begin{tabular}
[c]{ll}%
-5.430 & -6.193
\end{tabular}
&
\begin{tabular}
[c]{ll}%
-24.77 & -33.26
\end{tabular}
\\\hline
300 &
\begin{tabular}
[c]{ll}%
-6.15 & -12.55
\end{tabular}
&
\begin{tabular}
[c]{ll}%
9.420 & 3.136
\end{tabular}
&
\begin{tabular}
[c]{ll}%
-10.39 & -10.98
\end{tabular}
&
\begin{tabular}
[c]{ll}%
-27.99 & -37.63
\end{tabular}
\\\hline
350 &
\begin{tabular}
[c]{ll}%
-11.13 & -19.27
\end{tabular}
&
\begin{tabular}
[c]{ll}%
10.69 & 2.902
\end{tabular}
&
\begin{tabular}
[c]{ll}%
-15.30 & -15.42
\end{tabular}
&
\begin{tabular}
[c]{ll}%
-30.89 & -41.13
\end{tabular}
\\\hline\hline
\end{tabular}
\end{table}%

\begin{figure}
[ptb]
\begin{center}
\includegraphics[
height=4.7305in,
width=5.3195in
]%
{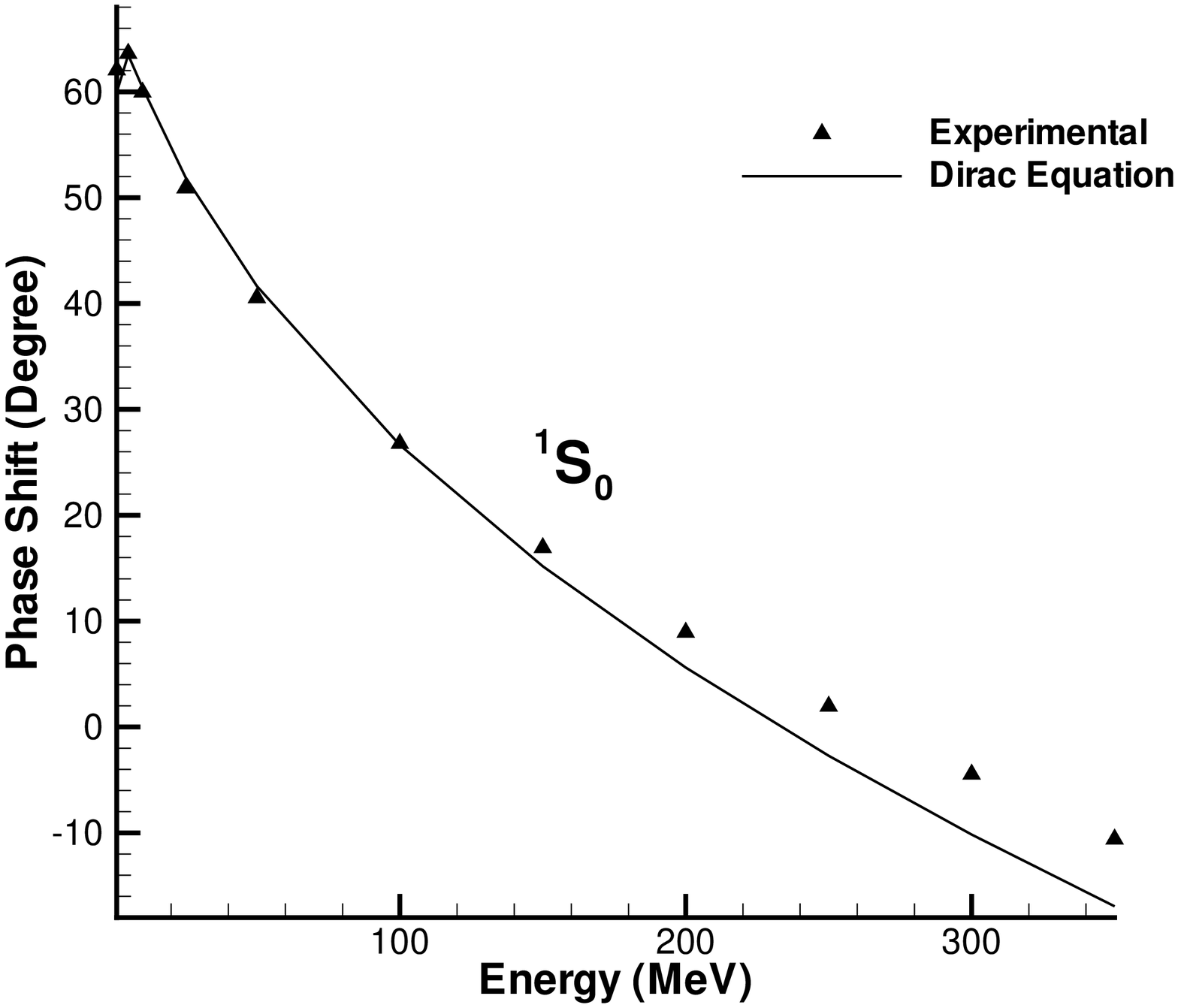}%
\caption{$np$ Scattering Phase Shift for $^{1}S_{0}$ State (Model 1)}%
\label{fig:s0}%
\end{center}
\end{figure}

\begin{figure}
[ptbptbptbptbptbptbptb]
\begin{center}
\includegraphics[
height=4.7305in,
width=5.3195in
]%
{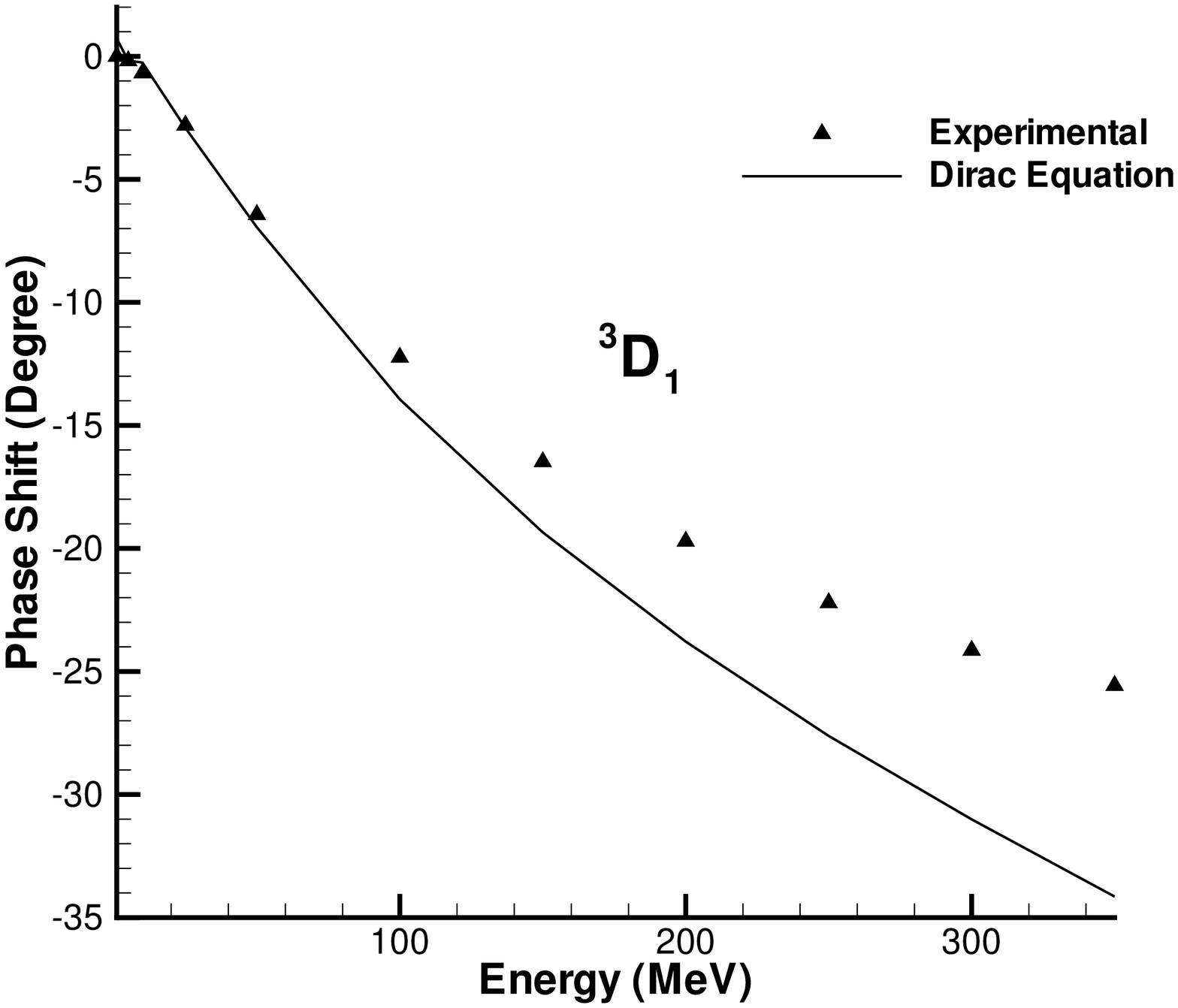}%
\caption{$np$ Scattering Phase Shift for $^{3}D_{1}$ State (Model 1)}%
\label{fig:d31}%
\end{center}
\end{figure}

\begin{figure}
[ptbptbptbptbptbptbptbptb]
\begin{center}
\includegraphics[
height=4.7305in,
width=5.3195in
]%
{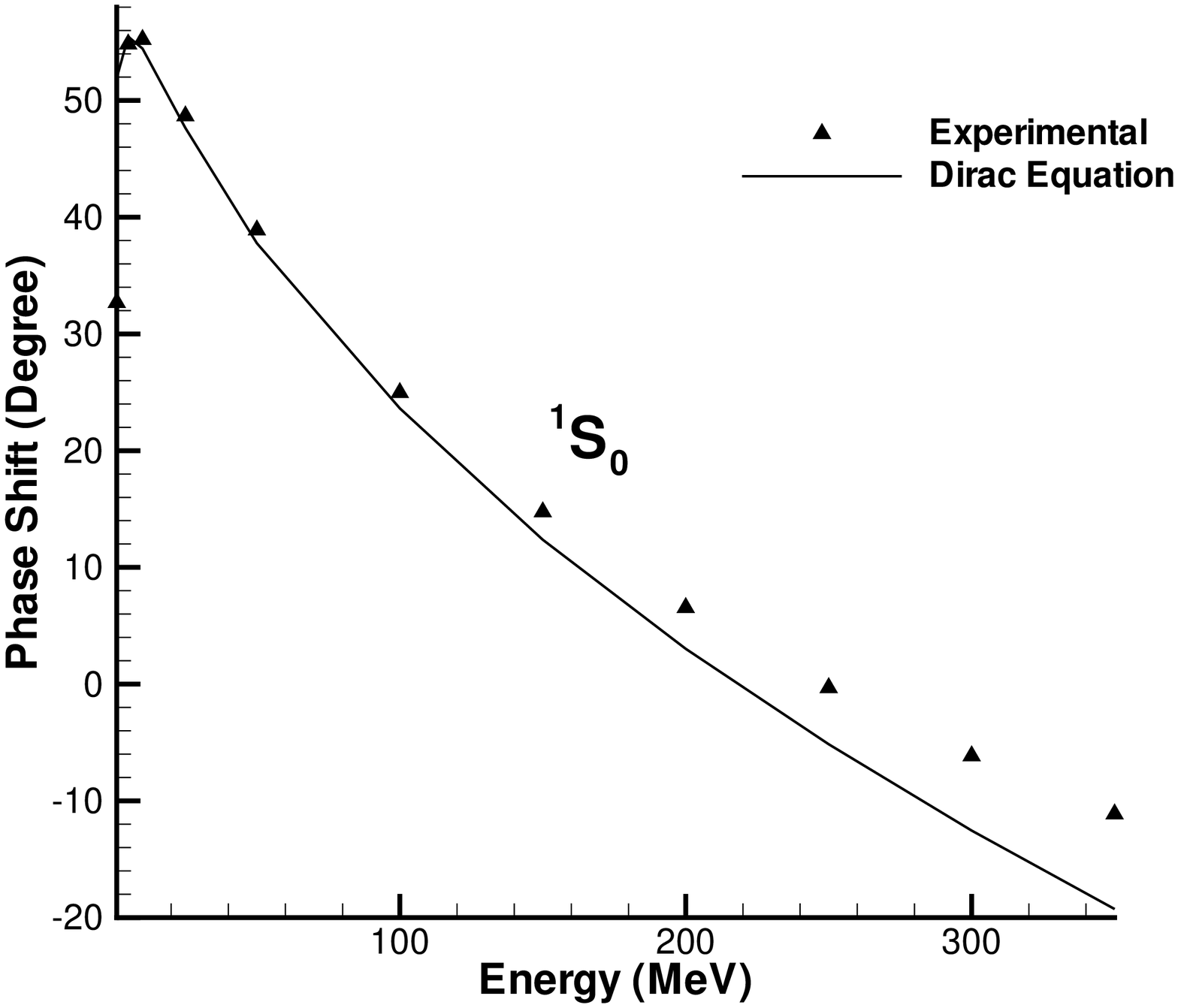}%
\caption{$pp$ Scattering Phase Shift for $^{1}S_{0}$ State (Model 1)}%
\label{fig:s0pp}%
\end{center}
\end{figure}

\begin{figure}
[ptbptbptbptbptbptbptbptbptbptbptb]
\begin{center}
\includegraphics[
height=4.7305in,
width=5.3195in
]%
{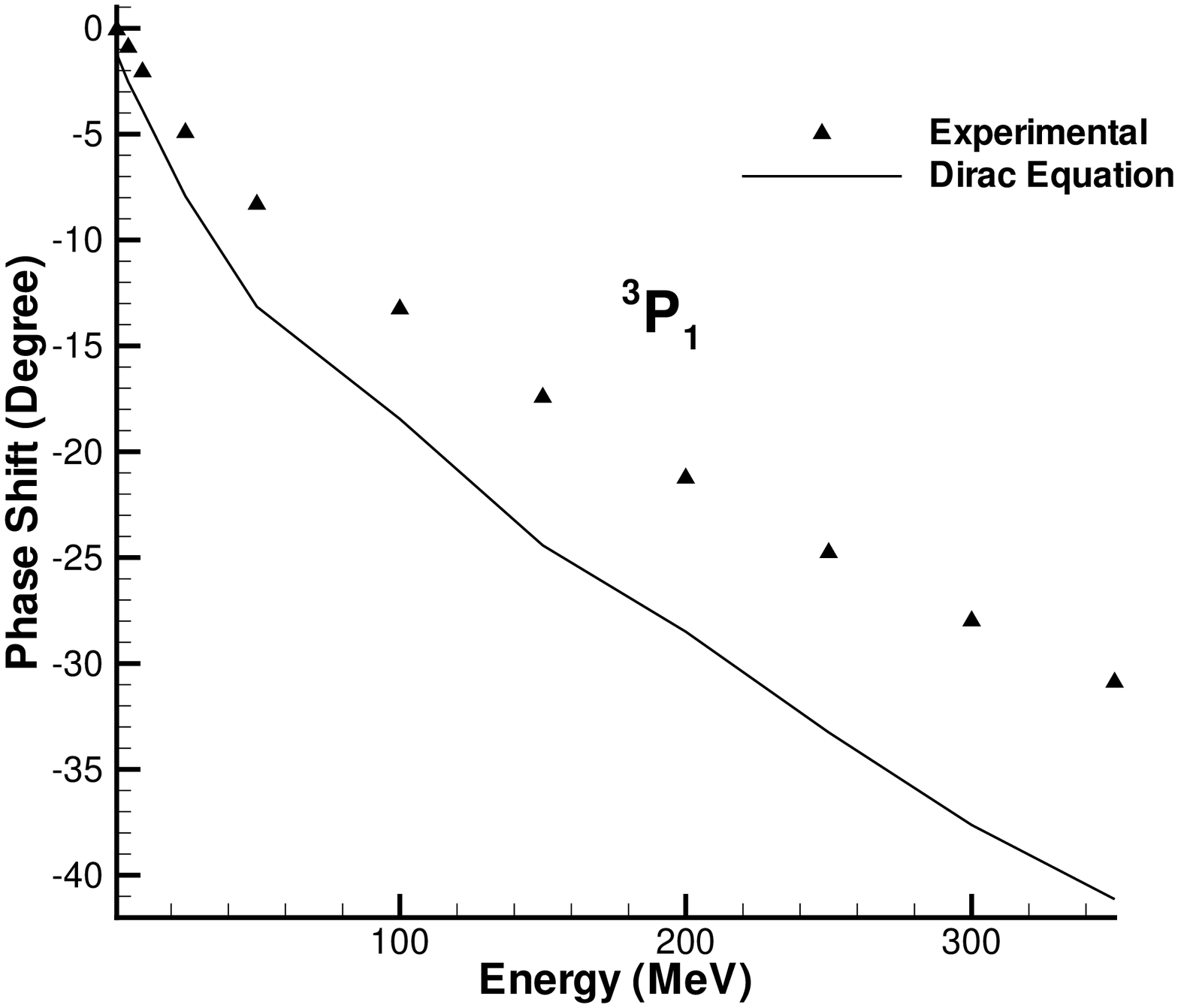}%
\caption{$pp$ Scattering Phase Shift for $^{3}P_{1}$ State (Model 1)}%
\label{fig:p31pp}%
\end{center}
\end{figure}

The results for $np$ scattering are also presented from figure \ref{fig:s0} to
figure \ref{fig:d31} and for $pp$ scattering from figure \ref{fig:s0pp} to
figure \ref{fig:p31pp}%

\begin{figure}
[ptbptb]
\begin{center}
\includegraphics[
height=4.7305in,
width=5.3195in
]%
{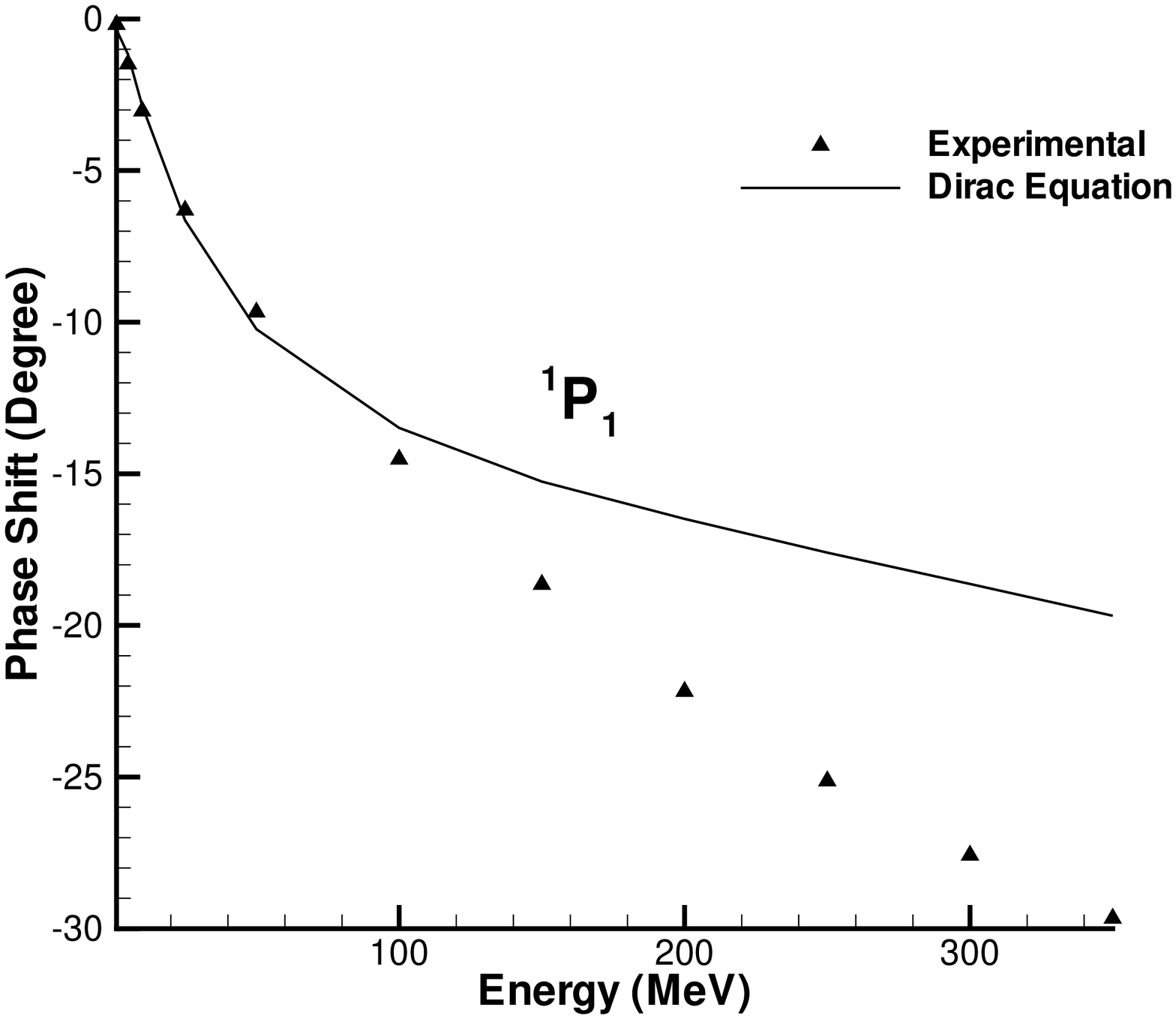}%
\caption{$np$ Scattering Phase Shift for $^{1}P_{1}$ State (Model 1)}%
\label{fig:p1}%
\end{center}
\end{figure}
\begin{figure}
[ptbptbptb]
\begin{center}
\includegraphics[
height=4.7305in,
width=5.3195in
]%
{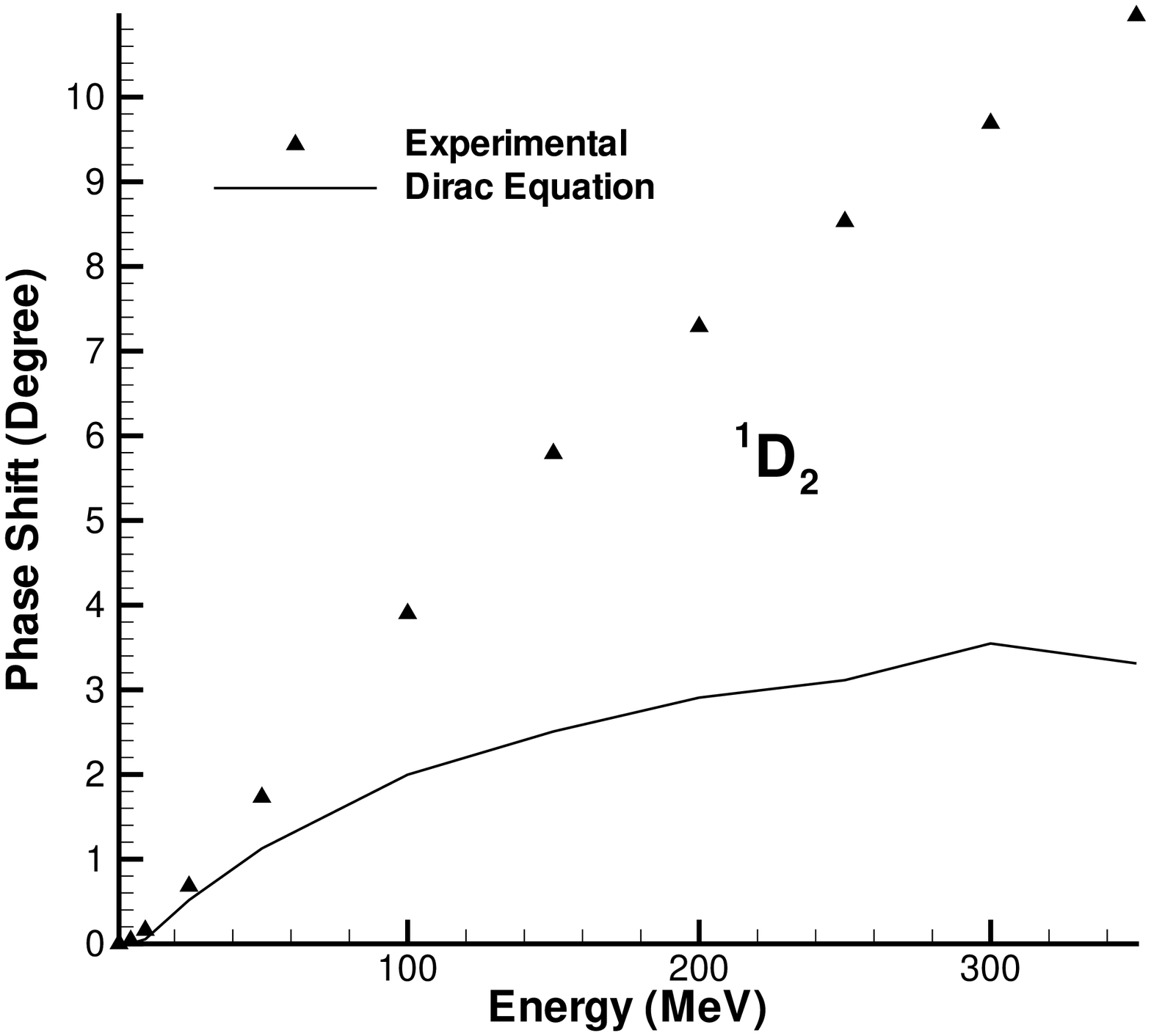}%
\caption{$np$ Scattering Phase Shift for $^{1}D_{2}$ State (Model 1)}%
\label{fig:d2}%
\end{center}
\end{figure}
\begin{figure}
[ptbptbptbptb]
\begin{center}
\includegraphics[
height=4.7305in,
width=5.3195in
]%
{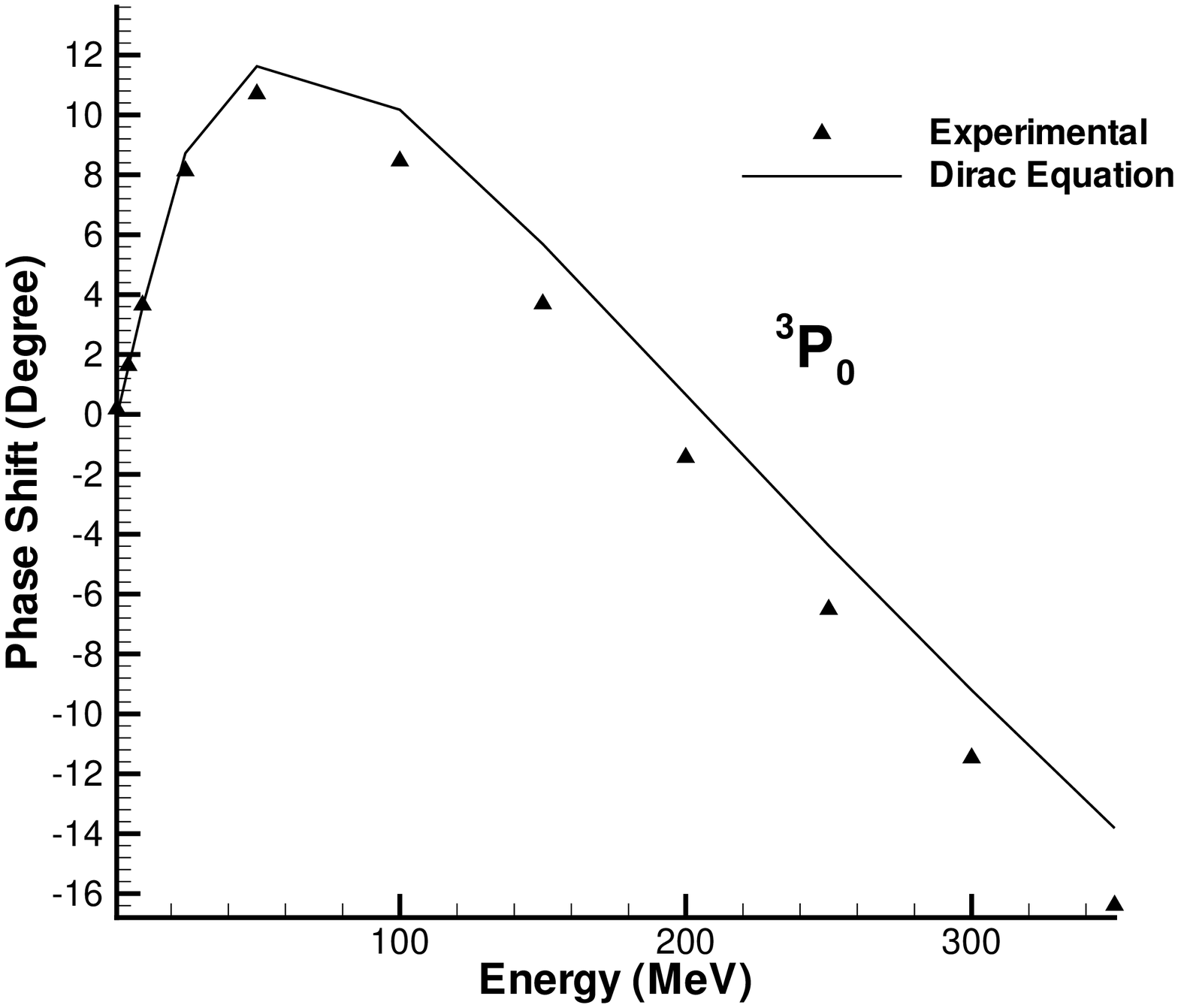}%
\caption{$np$ Scattering Phase Shift for $^{3}P_{0}$ State (Model 1)}%
\label{fig:p30}%
\end{center}
\end{figure}
\begin{figure}
[ptbptbptbptbptb]
\begin{center}
\includegraphics[
height=4.7305in,
width=5.3195in
]%
{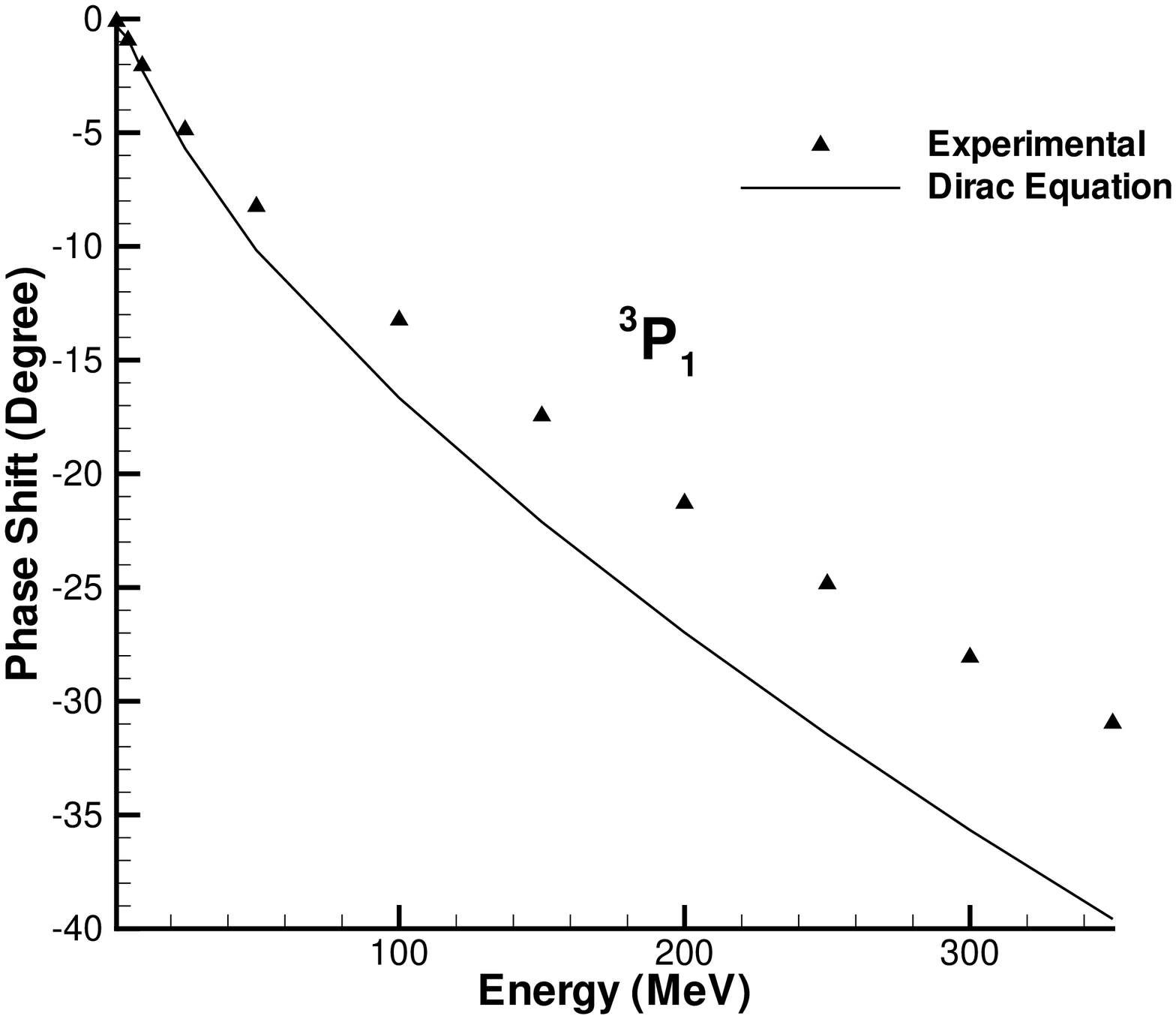}%
\caption{$np$ Scattering Phase Shift for $^{3}P_{1}$ State (Model 1)}%
\label{fig:p31}%
\end{center}
\end{figure}
\begin{figure}
[ptbptbptbptbptbptb]
\begin{center}
\includegraphics[
height=4.7305in,
width=5.3195in
]%
{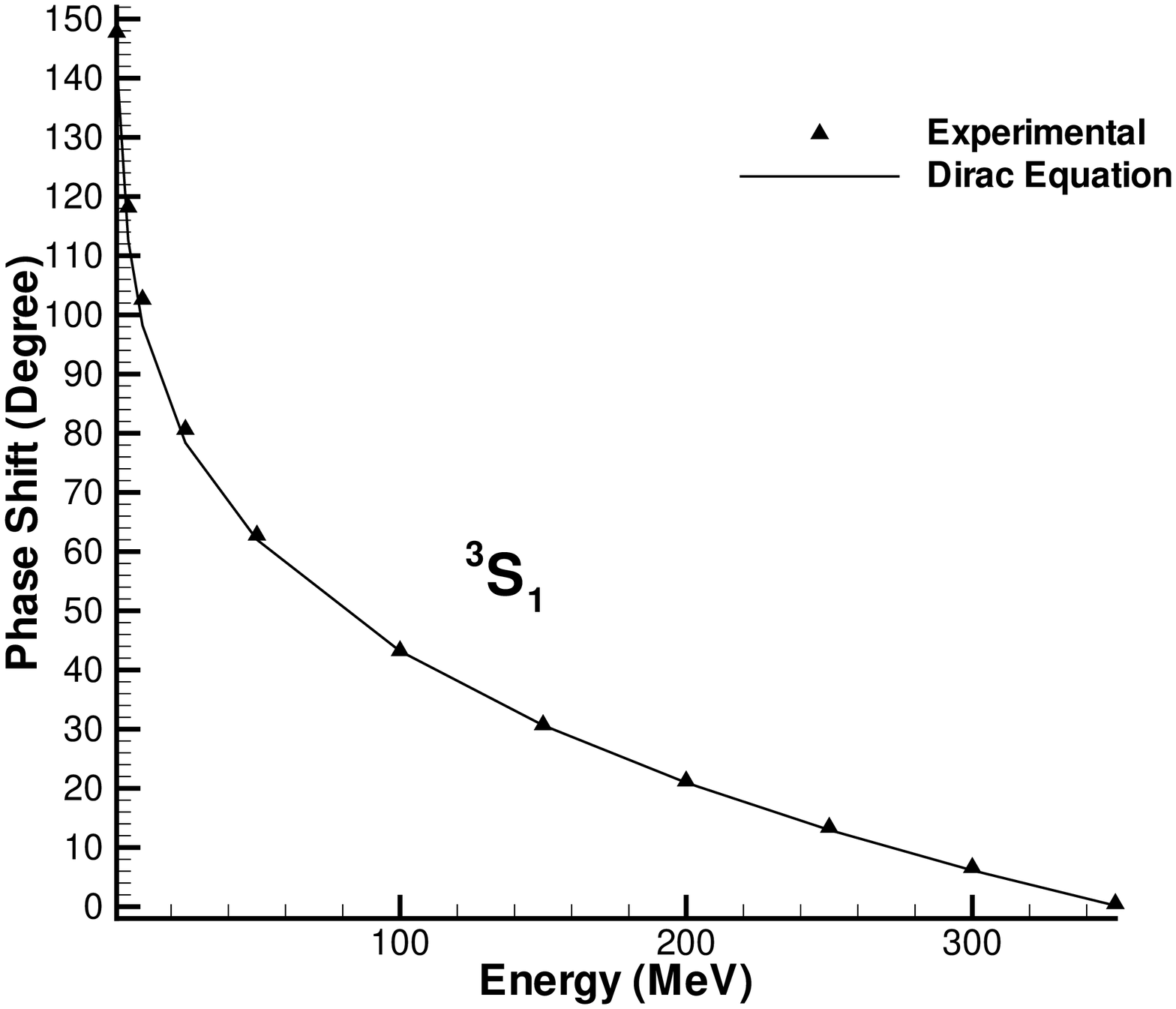}%
\caption{$np$ Scattering Phase Shift for $^{3}S_{1}$ State (Model 1)}%
\label{fig:s31}%
\end{center}
\end{figure}
\begin{figure}
[ptbptbptbptbptbptbptbptbptb]
\begin{center}
\includegraphics[
height=4.7305in,
width=5.3195in
]%
{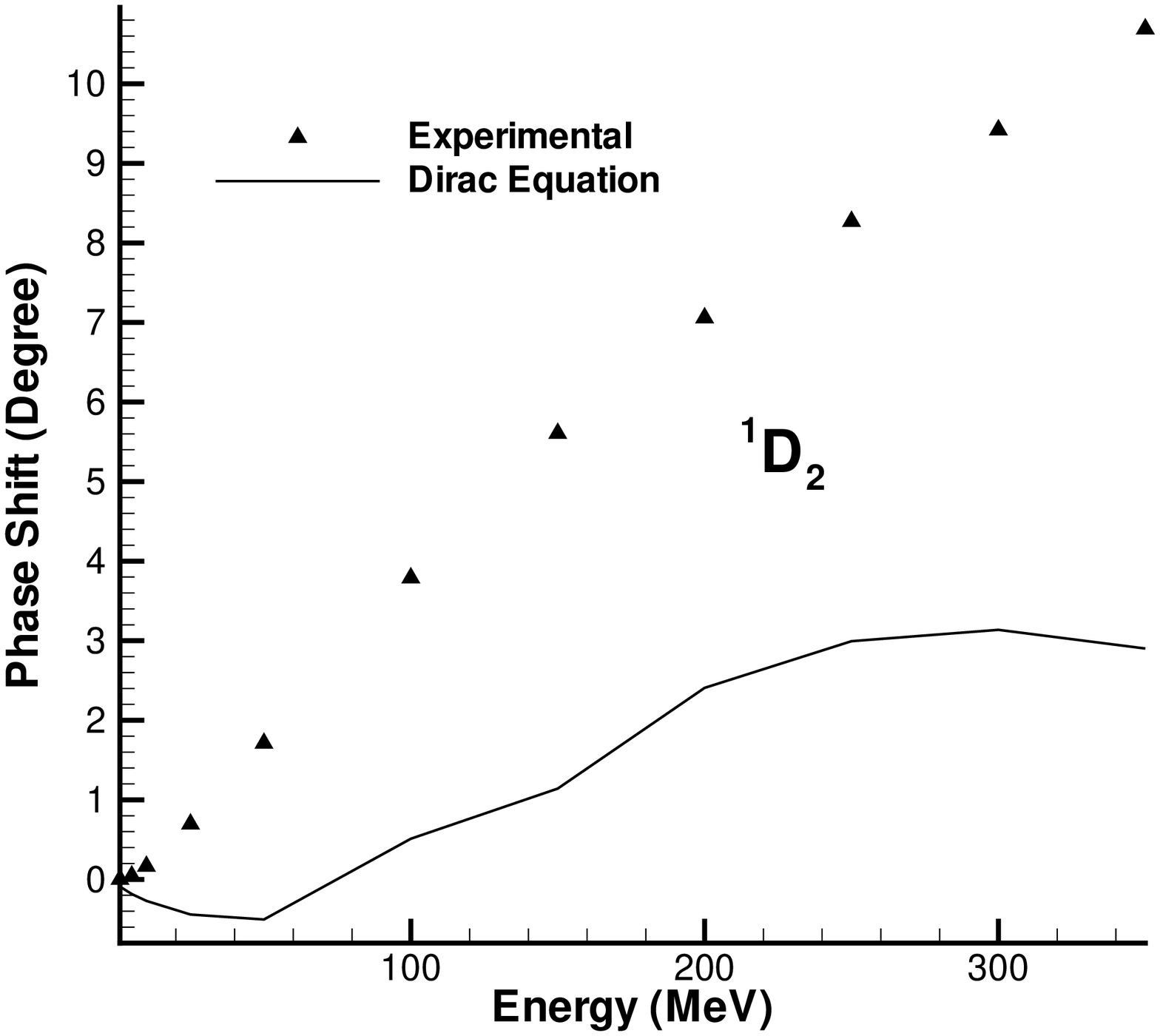}%
\caption{$np$ Scattering Phase Shift for $^{1}D_{2}$ State (Model 1)}%
\label{fig:d2pp}%
\end{center}
\end{figure}
\begin{figure}
[ptbptbptbptbptbptbptbptbptbptb]
\begin{center}
\includegraphics[
height=4.7305in,
width=5.3195in
]%
{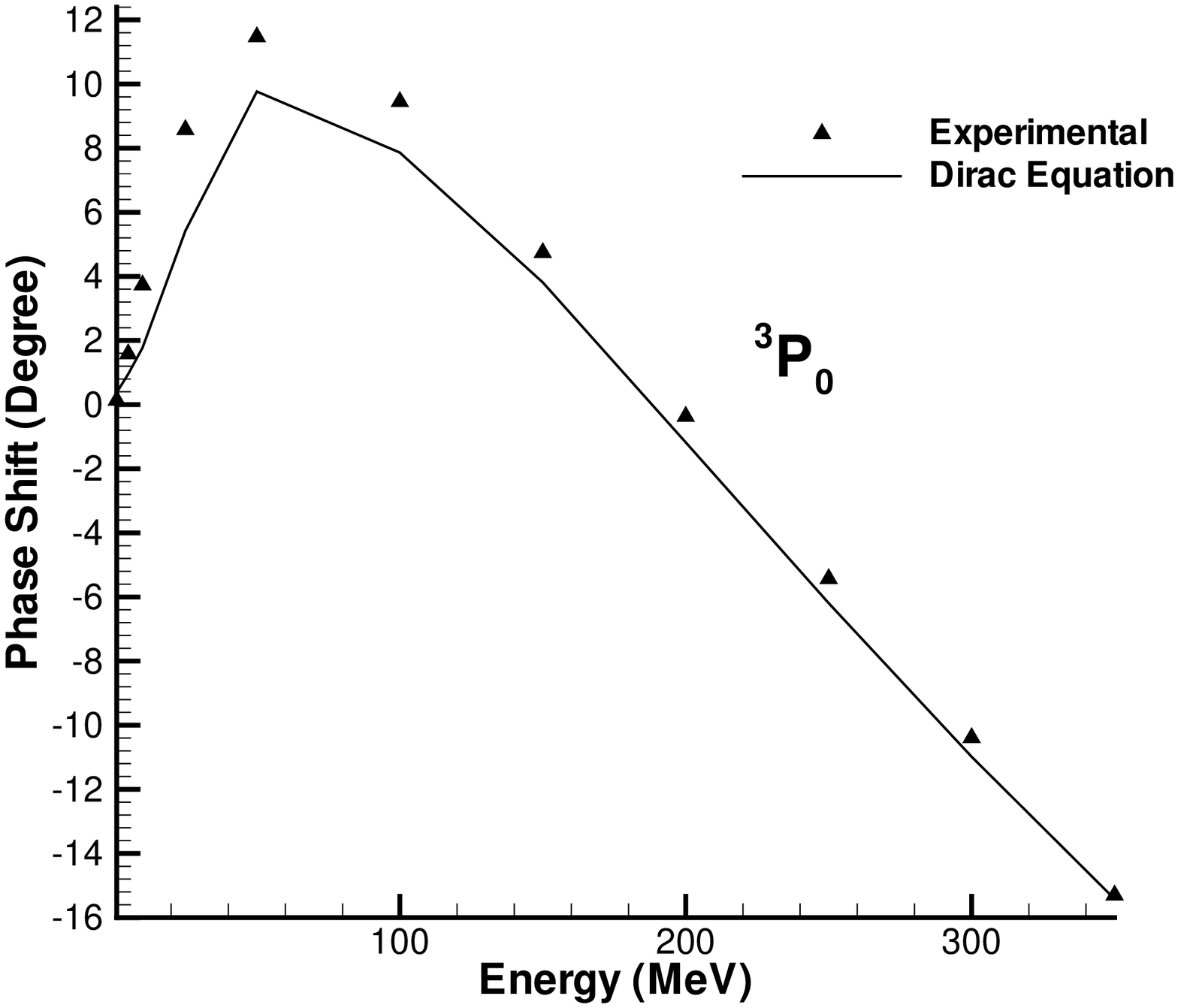}%
\caption{$pp$ Scattering Phase Shift for $^{3}P_{0}$ State (Model 1)}%
\label{fig:p30pp}%
\end{center}
\end{figure}

\clearpage

\subsection{Model 2}

The theoretical phase shifts which we calculated by using the parameters for
model 2 and the experimental phase shifts for all the seven states are listed
in table \ref{tab:np-phase2}. We also use the parameters for model 2 to
predict the phase shift of $pp$ scattering.

%

\begin{table}[H] \centering
\caption{$np $ Scattering Phase Shift
Of $^{1}S_{0}$, $^{1}P_{1}$, $^{1}D_{2}$, $^{3}P_{0}$, $^{3}P_{1}$, $^{3}S_{1}$ And $^{3}D_{1}$ States(Model 2).\label{tab:np-phase2}}
\begin{tabular}
[t]{||c|c|c|c|c||}\hline\hline
Energy & $^{1}S_{0}$ & $^{1}P_{1}$ & $^{1}D_{2}$ & $^{3}P_{0}$\\\hline
(MeV) &
\begin{tabular}
[c]{ll}%
Exp. & The.
\end{tabular}
&
\begin{tabular}
[c]{ll}%
Exp. & The.
\end{tabular}
&
\begin{tabular}
[c]{ll}%
Exp. & The.
\end{tabular}
&
\begin{tabular}
[c]{ll}%
Exp. & The.
\end{tabular}
\\\hline
1 &
\begin{tabular}
[c]{ll}%
62.07 & 60.60
\end{tabular}
&
\begin{tabular}
[c]{ll}%
-0.187 & -0.358
\end{tabular}
&
\begin{tabular}
[c]{ll}%
0.00 & 0.02
\end{tabular}
&
\begin{tabular}
[c]{ll}%
0.18 & 0.00
\end{tabular}
\\\hline
5 &
\begin{tabular}
[c]{ll}%
63.63 & 63.50
\end{tabular}
&
\begin{tabular}
[c]{ll}%
-1.487 & -1.163
\end{tabular}
&
\begin{tabular}
[c]{ll}%
0.04 & 0.15
\end{tabular}
&
\begin{tabular}
[c]{ll}%
1.63 & 1.61
\end{tabular}
\\\hline
10 &
\begin{tabular}
[c]{ll}%
59.96 & 60.20
\end{tabular}
&
\begin{tabular}
[c]{ll}%
-3.039 & -2.857
\end{tabular}
&
\begin{tabular}
[c]{ll}%
0.16 & 0.39
\end{tabular}
&
\begin{tabular}
[c]{ll}%
3.65 & 3.74
\end{tabular}
\\\hline
25 &
\begin{tabular}
[c]{ll}%
50.90 & 51.44
\end{tabular}
&
\begin{tabular}
[c]{ll}%
-6.311 & -6.629
\end{tabular}
&
\begin{tabular}
[c]{ll}%
0.68 & 0.40
\end{tabular}
&
\begin{tabular}
[c]{ll}%
8.13 & 9.28
\end{tabular}
\\\hline
50 &
\begin{tabular}
[c]{ll}%
40.54 & 40.91
\end{tabular}
&
\begin{tabular}
[c]{ll}%
-9.670 & -10.36
\end{tabular}
&
\begin{tabular}
[c]{ll}%
1.73 & 1.37
\end{tabular}
&
\begin{tabular}
[c]{ll}%
10.70 & 12.69
\end{tabular}
\\\hline
100 &
\begin{tabular}
[c]{ll}%
26.78 & 25.86
\end{tabular}
&
\begin{tabular}
[c]{ll}%
-14.52 & -14.44
\end{tabular}
&
\begin{tabular}
[c]{ll}%
3.90 & 2.42
\end{tabular}
&
\begin{tabular}
[c]{ll}%
8.460 & 11.74
\end{tabular}
\\\hline
150 &
\begin{tabular}
[c]{ll}%
16.94 & 14.62
\end{tabular}
&
\begin{tabular}
[c]{ll}%
-18.65 & -17.55
\end{tabular}
&
\begin{tabular}
[c]{ll}%
5.79 & 3.62
\end{tabular}
&
\begin{tabular}
[c]{ll}%
3.690 & 7.399
\end{tabular}
\\\hline
200 &
\begin{tabular}
[c]{ll}%
8.940 & 5.435
\end{tabular}
&
\begin{tabular}
[c]{ll}%
-22.18 & -20.37
\end{tabular}
&
\begin{tabular}
[c]{ll}%
7.29 & 4.55
\end{tabular}
&
\begin{tabular}
[c]{ll}%
-1.44 & 2.36
\end{tabular}
\\\hline
250 &
\begin{tabular}
[c]{ll}%
1.960 & -2.428
\end{tabular}
&
\begin{tabular}
[c]{ll}%
-25.13 & -23.15
\end{tabular}
&
\begin{tabular}
[c]{ll}%
8.53 & 5.24
\end{tabular}
&
\begin{tabular}
[c]{ll}%
-6.51 & -2.78
\end{tabular}
\\\hline
300 &
\begin{tabular}
[c]{ll}%
-4.460 & -9.330
\end{tabular}
&
\begin{tabular}
[c]{ll}%
-27.58 & -25.87
\end{tabular}
&
\begin{tabular}
[c]{ll}%
9.69 & 5.34
\end{tabular}
&
\begin{tabular}
[c]{ll}%
-11.47 & -7.746
\end{tabular}
\\\hline
350 &
\begin{tabular}
[c]{ll}%
-10.59 & -15.52
\end{tabular}
&
\begin{tabular}
[c]{ll}%
-29.66 & -28.54
\end{tabular}
&
\begin{tabular}
[c]{ll}%
10.96 & 5.30
\end{tabular}
&
\begin{tabular}
[c]{ll}%
-16.39 & -12.52
\end{tabular}
\\\hline
Energy & $^{3}P_{1}$ & $^{3}S_{1}$ & $^{3}D_{1}$ & $\varepsilon$\\\hline
(MeV) &
\begin{tabular}
[c]{ll}%
Exp. & The.
\end{tabular}
&
\begin{tabular}
[c]{ll}%
Exp. & The.
\end{tabular}
&
\begin{tabular}
[c]{ll}%
Exp. & The.
\end{tabular}
&
\begin{tabular}
[c]{ll}%
Exp. & The.
\end{tabular}
\\\hline
1 &
\begin{tabular}
[c]{ll}%
-0.11 & -0.32
\end{tabular}
&
\begin{tabular}
[c]{ll}%
147.747 & 144.797
\end{tabular}
&
\begin{tabular}
[c]{ll}%
-0.005 & 0.719
\end{tabular}
&
\begin{tabular}
[c]{ll}%
0.105 & 0.264
\end{tabular}
\\\hline
5 &
\begin{tabular}
[c]{ll}%
-0.94 & -0.81
\end{tabular}
&
\begin{tabular}
[c]{ll}%
118.178 & 115.232
\end{tabular}
&
\begin{tabular}
[c]{ll}%
-0.183 & -0.172
\end{tabular}
&
\begin{tabular}
[c]{ll}%
0.672 & 1.106
\end{tabular}
\\\hline
10 &
\begin{tabular}
[c]{ll}%
-2.06 & -2.08
\end{tabular}
&
\begin{tabular}
[c]{ll}%
102.611 & 100.668
\end{tabular}
&
\begin{tabular}
[c]{ll}%
-0.677 & -0.239
\end{tabular}
&
\begin{tabular}
[c]{ll}%
1.159 & 1.723
\end{tabular}
\\\hline
25 &
\begin{tabular}
[c]{ll}%
-4.88 & -5.07
\end{tabular}
&
\begin{tabular}
[c]{ll}%
80.63 & 80.66
\end{tabular}
&
\begin{tabular}
[c]{ll}%
-2.799 & -2.834
\end{tabular}
&
\begin{tabular}
[c]{ll}%
1.793 & 2.099
\end{tabular}
\\\hline
50 &
\begin{tabular}
[c]{ll}%
-8.25 & -8.68
\end{tabular}
&
\begin{tabular}
[c]{ll}%
62.77 & 64.30
\end{tabular}
&
\begin{tabular}
[c]{ll}%
-6.433 & -6.798
\end{tabular}
&
\begin{tabular}
[c]{ll}%
2.109 & 1.708
\end{tabular}
\\\hline
100 &
\begin{tabular}
[c]{ll}%
-13.24 & -13.55
\end{tabular}
&
\begin{tabular}
[c]{ll}%
43.23 & 45.68
\end{tabular}
&
\begin{tabular}
[c]{ll}%
-12.23 & -13.77
\end{tabular}
&
\begin{tabular}
[c]{ll}%
2.420 & 1.663
\end{tabular}
\\\hline
150 &
\begin{tabular}
[c]{ll}%
-17.46 & -17.74
\end{tabular}
&
\begin{tabular}
[c]{ll}%
30.72 & 33.35
\end{tabular}
&
\begin{tabular}
[c]{ll}%
-16.48 & -19.34
\end{tabular}
&
\begin{tabular}
[c]{ll}%
2.750 & 1.541
\end{tabular}
\\\hline
200 &
\begin{tabular}
[c]{ll}%
-21.30 & -21.67
\end{tabular}
&
\begin{tabular}
[c]{ll}%
21.22 & 23.80
\end{tabular}
&
\begin{tabular}
[c]{ll}%
-19.71 & -24.11
\end{tabular}
&
\begin{tabular}
[c]{ll}%
3.130 & 1.648
\end{tabular}
\\\hline
250 &
\begin{tabular}
[c]{ll}%
-24.84 & -25.47
\end{tabular}
&
\begin{tabular}
[c]{ll}%
13.39 & 15.90
\end{tabular}
&
\begin{tabular}
[c]{ll}%
-22.21 & -28.38
\end{tabular}
&
\begin{tabular}
[c]{ll}%
3.560 & 1.834
\end{tabular}
\\\hline
300 &
\begin{tabular}
[c]{ll}%
-28.07 & -29.14
\end{tabular}
&
\begin{tabular}
[c]{ll}%
6.600 & 9.099
\end{tabular}
&
\begin{tabular}
[c]{ll}%
-24.14 & -32.29
\end{tabular}
&
\begin{tabular}
[c]{ll}%
4.030 & 1.965
\end{tabular}
\\\hline
350 &
\begin{tabular}
[c]{ll}%
-30.97 & -32.67
\end{tabular}
&
\begin{tabular}
[c]{ll}%
0.502 & 3.095
\end{tabular}
&
\begin{tabular}
[c]{ll}%
-25.57 & -36.01
\end{tabular}
&
\begin{tabular}
[c]{ll}%
4.570 & 2.147
\end{tabular}
\\\hline\hline
\end{tabular}
\end{table}%
%

\begin{table}[H] \centering
\caption{$pp $ Scattering Phase Shift Of $^{1}S_{0} $, $^{1}D_{2}$,
$^{3}P_{0}$ And $^{3}P_{1}$ States(Model 2).\label{tab:pp-phase2}}
\begin{tabular}
[t]{||c|c|c|c|c||}\hline\hline
Energy & $^{1}S_{0}$ & $^{1}D_{2}$ & $^{3}P_{0}$ & $^{3}P_{1}$\\\hline
MeV &
\begin{tabular}
[c]{ll}%
Exp. & The.
\end{tabular}
&
\begin{tabular}
[c]{ll}%
Exp. & The.
\end{tabular}
&
\begin{tabular}
[c]{ll}%
Exp. & The.
\end{tabular}
&
\begin{tabular}
[c]{ll}%
Exp. & The.
\end{tabular}
\\\hline
1 &
\begin{tabular}
[c]{ll}%
32.68 & 52.40
\end{tabular}
&
\begin{tabular}
[c]{ll}%
0.001 & -0.116
\end{tabular}
&
\begin{tabular}
[c]{ll}%
0.134 & 0.417
\end{tabular}
&
\begin{tabular}
[c]{ll}%
-0.081 & -1.172
\end{tabular}
\\\hline
5 &
\begin{tabular}
[c]{ll}%
54.83 & 55.48
\end{tabular}
&
\begin{tabular}
[c]{ll}%
0.043 & -0.232
\end{tabular}
&
\begin{tabular}
[c]{ll}%
1.582 & 1.042
\end{tabular}
&
\begin{tabular}
[c]{ll}%
-0.902 & -2.434
\end{tabular}
\\\hline
10 &
\begin{tabular}
[c]{ll}%
55.22 & 54.24
\end{tabular}
&
\begin{tabular}
[c]{ll}%
0.165 & -0.327
\end{tabular}
&
\begin{tabular}
[c]{ll}%
3.729 & 1.934
\end{tabular}
&
\begin{tabular}
[c]{ll}%
-2.060 & -3.682
\end{tabular}
\\\hline
25 &
\begin{tabular}
[c]{ll}%
48.67 & 47.13
\end{tabular}
&
\begin{tabular}
[c]{ll}%
0.696 & -0.524
\end{tabular}
&
\begin{tabular}
[c]{ll}%
8.575 & 5.943
\end{tabular}
&
\begin{tabular}
[c]{ll}%
-4.932 & -7.355
\end{tabular}
\\\hline
50 &
\begin{tabular}
[c]{ll}%
38.90 & 37.04
\end{tabular}
&
\begin{tabular}
[c]{ll}%
1.711 & -0.505
\end{tabular}
&
\begin{tabular}
[c]{ll}%
11.47 & 10.88
\end{tabular}
&
\begin{tabular}
[c]{ll}%
-8.317 & -11.57
\end{tabular}
\\\hline
100 &
\begin{tabular}
[c]{ll}%
24.97 & 22.85
\end{tabular}
&
\begin{tabular}
[c]{ll}%
3.790 & 0.994
\end{tabular}
&
\begin{tabular}
[c]{ll}%
9.450 & 9.417
\end{tabular}
&
\begin{tabular}
[c]{ll}%
-13.26 & -15.41
\end{tabular}
\\\hline
150 &
\begin{tabular}
[c]{ll}%
14.75 & 11.82
\end{tabular}
&
\begin{tabular}
[c]{ll}%
5.606 & 2.036
\end{tabular}
&
\begin{tabular}
[c]{ll}%
4.740 & 5.543
\end{tabular}
&
\begin{tabular}
[c]{ll}%
-17.43 & -19.97
\end{tabular}
\\\hline
200 &
\begin{tabular}
[c]{ll}%
6.550 & 2.845
\end{tabular}
&
\begin{tabular}
[c]{ll}%
7.058 & 3.211
\end{tabular}
&
\begin{tabular}
[c]{ll}%
--0.370 & 0.495
\end{tabular}
&
\begin{tabular}
[c]{ll}%
-21.25 & -23.23
\end{tabular}
\\\hline
250 &
\begin{tabular}
[c]{ll}%
-0.31 & -4.86
\end{tabular}
&
\begin{tabular}
[c]{ll}%
8.270 & 3.648
\end{tabular}
&
\begin{tabular}
[c]{ll}%
-5.430 & -4.589
\end{tabular}
&
\begin{tabular}
[c]{ll}%
-24.77 & -27.28
\end{tabular}
\\\hline
300 &
\begin{tabular}
[c]{ll}%
-6.15 & -11.72
\end{tabular}
&
\begin{tabular}
[c]{ll}%
9.420 & 3.956
\end{tabular}
&
\begin{tabular}
[c]{ll}%
-10.39 & -9.516
\end{tabular}
&
\begin{tabular}
[c]{ll}%
-27.99 & -31.05
\end{tabular}
\\\hline
350 &
\begin{tabular}
[c]{ll}%
-11.13 & -17.85
\end{tabular}
&
\begin{tabular}
[c]{ll}%
10.69 & 4.014
\end{tabular}
&
\begin{tabular}
[c]{ll}%
-15.30 & -14.13
\end{tabular}
&
\begin{tabular}
[c]{ll}%
-30.89 & -34.22
\end{tabular}
\\\hline\hline
\end{tabular}
\end{table}%

\begin{figure}
[ptb]
\begin{center}
\includegraphics[
height=4.7202in,
width=5.3082in
]%
{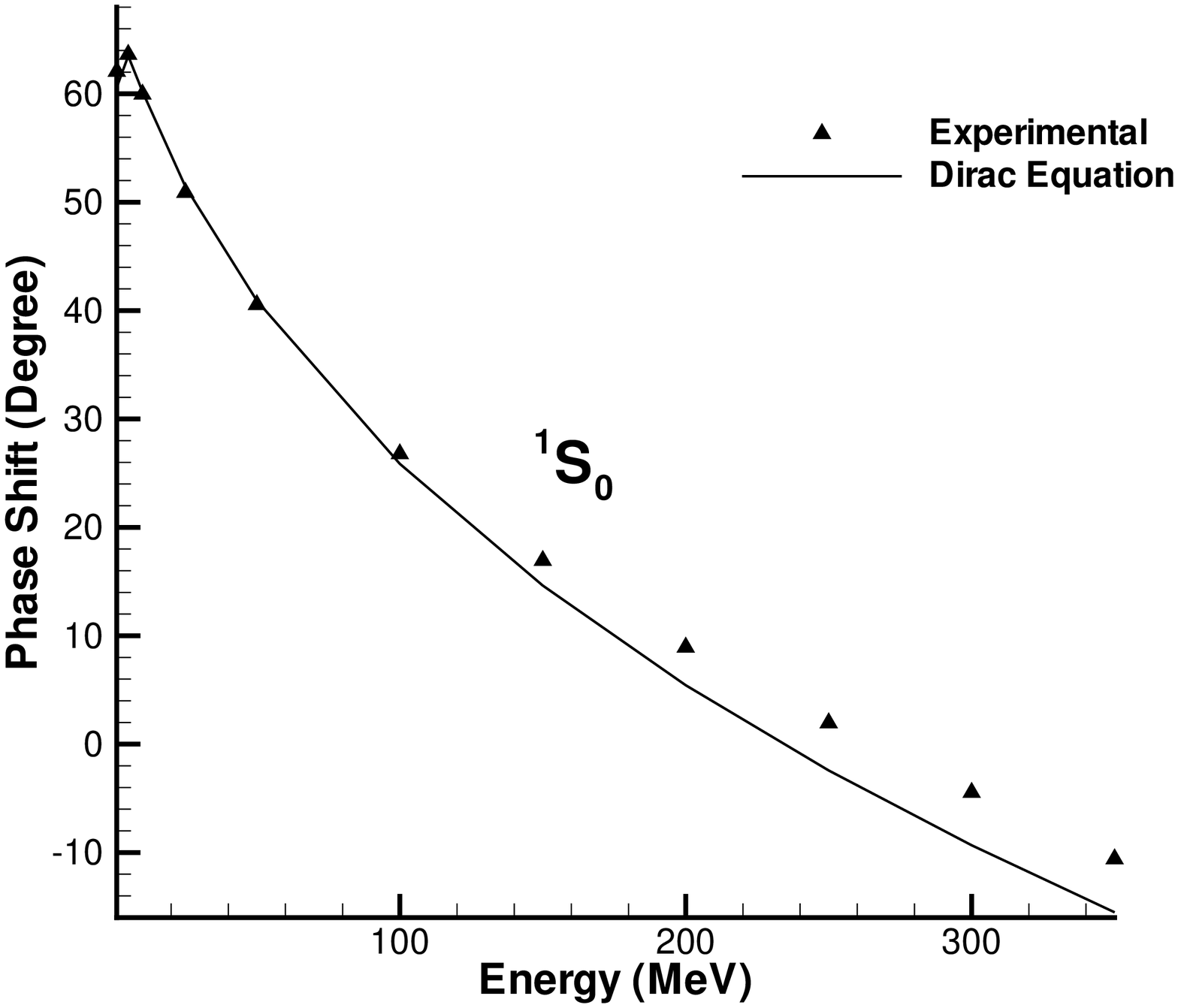}%
\caption{$np$ Scattering Phase Shift of $^{1}S_{0}$ State (Model 2)}%
\label{fig:s02}%
\end{center}
\end{figure}

\begin{figure}
[ptbptbptbptbptbptbptb]
\begin{center}
\includegraphics[
height=4.7202in,
width=5.3082in
]%
{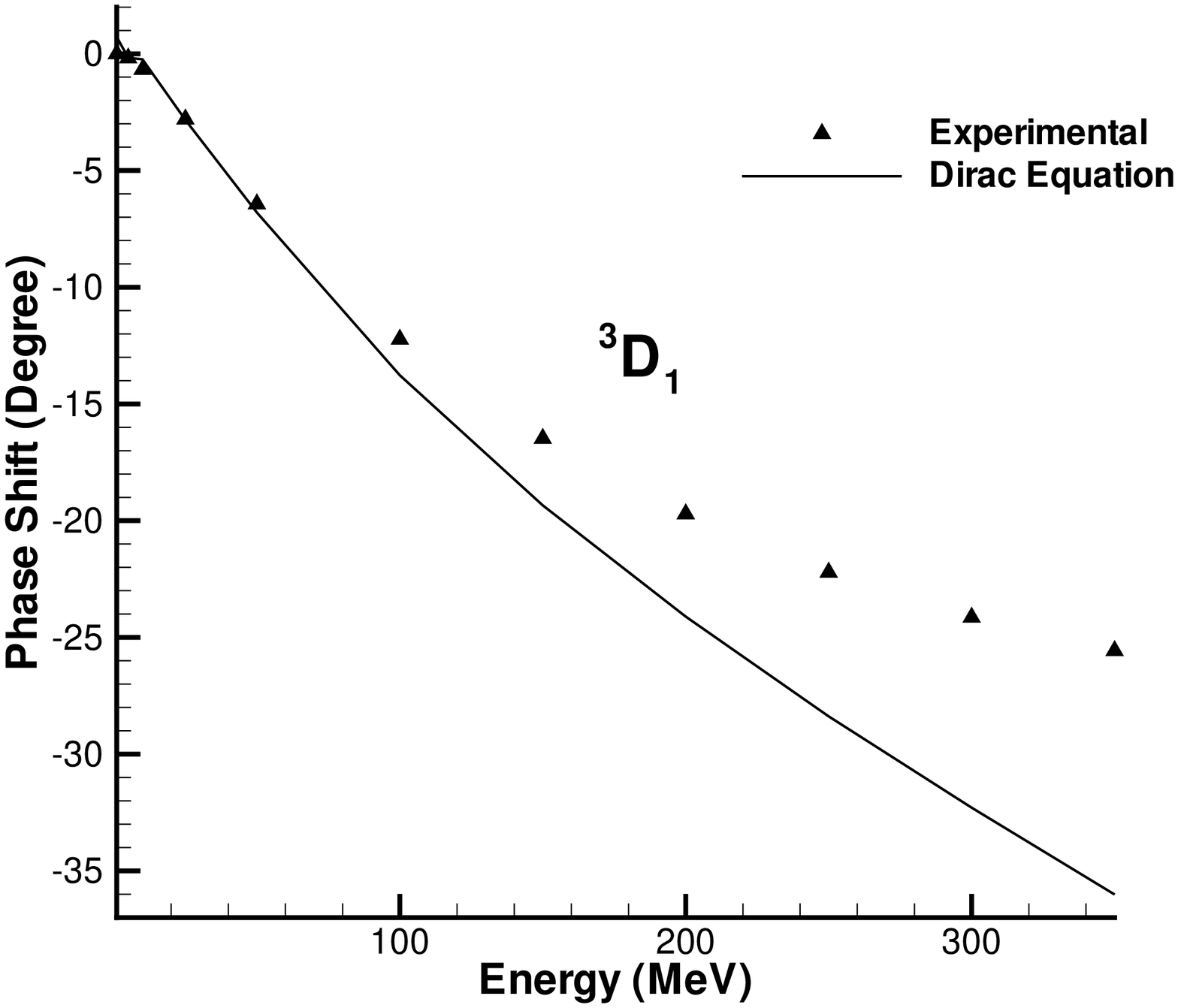}%
\caption{$np$ Scattering Phase Shift of $^{3}D_{1}$ State (Model 2)}%
\label{fig:d312}%
\end{center}
\end{figure}

\begin{figure}
[ptbptbptbptbptbptbptbptb]
\begin{center}
\includegraphics[
height=4.7202in,
width=5.3082in
]%
{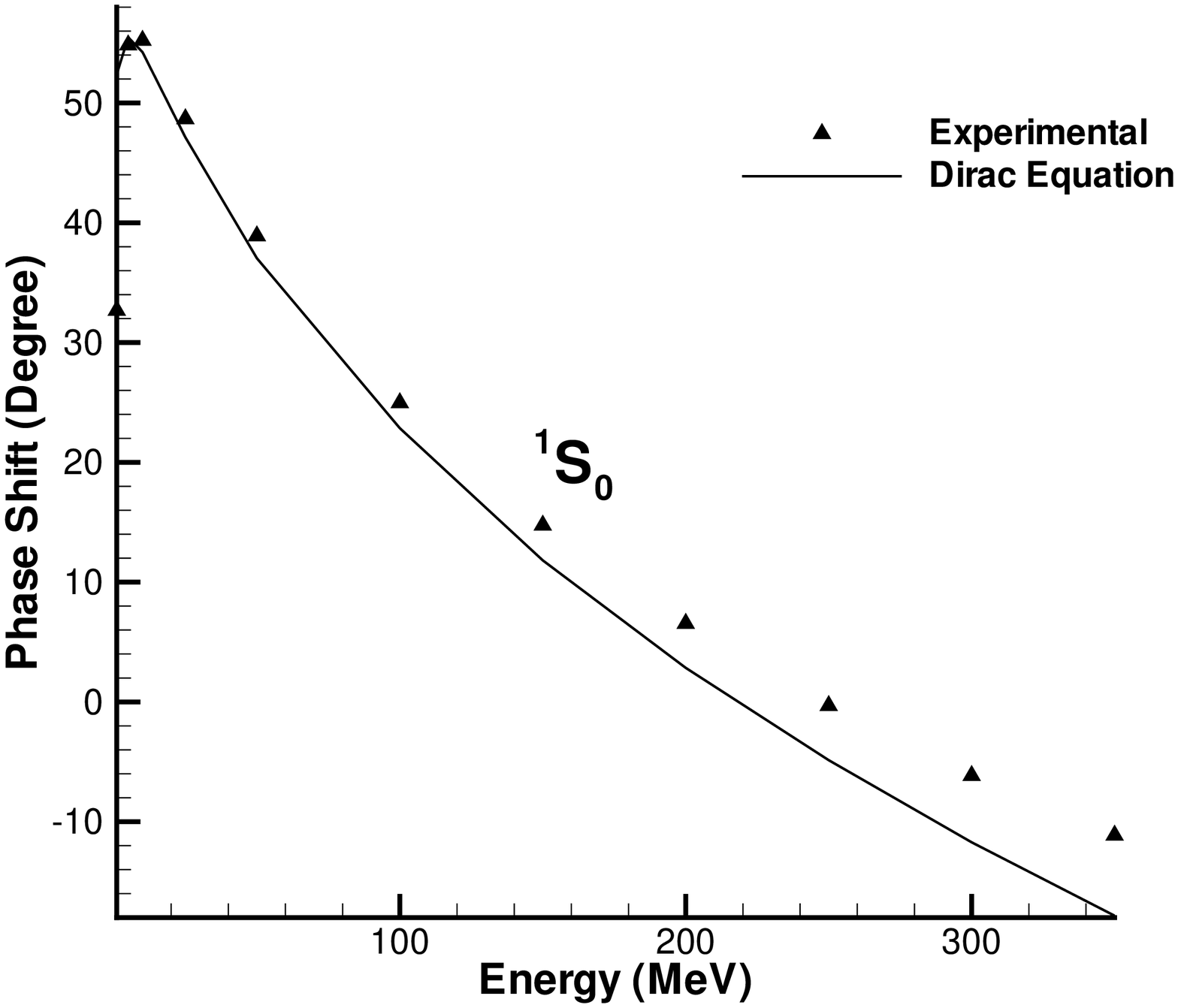}%
\caption{$pp$ Scattering Phase Shift of $^{1}S_{0}$ State (Model 2)}%
\label{fig:s0pp2}%
\end{center}
\end{figure}

\begin{figure}
[ptbptbptbptbptbptbptbptbptbptbptb]
\begin{center}
\includegraphics[
height=4.2203in,
width=4.7461in
]%
{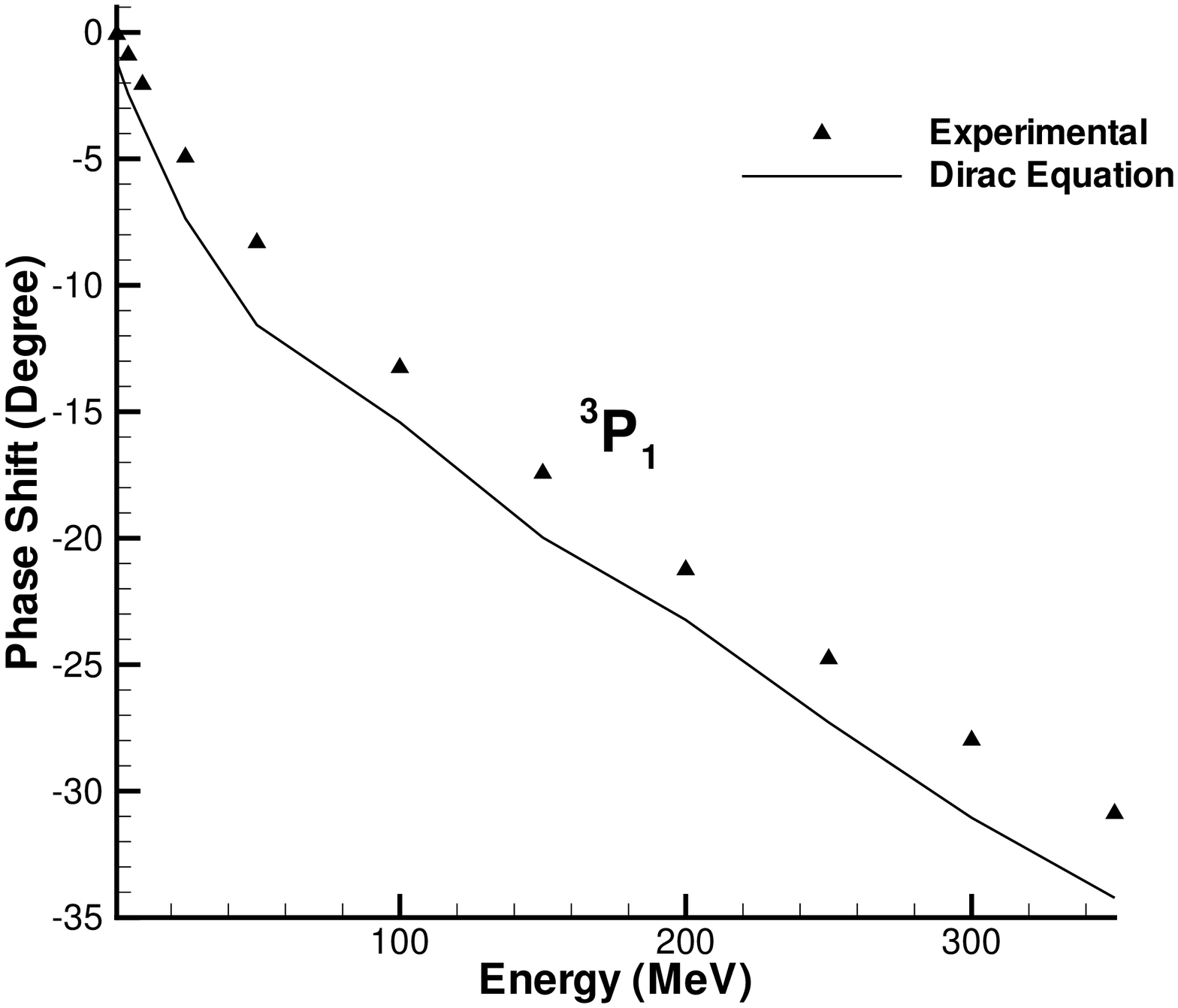}%
\caption{$pp$ Scattering Phase Shift of $^{3}P_{1}$ State (Model 2)}%
\label{fig:p31pp2}%
\end{center}
\end{figure}

The prediction for the four $pp$ scattering states are listed in table
\ref{tab:pp-phase2}. The results of model 2 for $np$ scattering are given in
figure \ref{fig:s02} to figure \ref{fig:d312} and for $pp$ scattering are from
figure \ref{fig:s0pp2} to figure \ref{fig:p31pp2}. Our results show for this
model an improvement over those of model 1 especially for the the singlet $P$
and $D$ states. \ However there is still much to be desired in the fit. One
possible cause of this problem is that we did not include tensor and
pseudovector interactions in our covariant potentials, limiting ourselves to
scalar, vector and pseudoscalar. Another may be the ignoring of the
pseudovector coupling of the pseudoscalar mesons to \ the nucleon. Our results
in $pp$ scattering show that if we obtain a good fit in $np$ scattering our
predicted results in $pp$ scattering will also be good. This means that it is
unnecessary to include $pp$ scattering in the our fit, we may use the
parameters obtained in $np$ scattering to predict the results in $pp$
scattering. Overall our results are promising and indicate that the two-body
Dirac equations of constraint dynamics together with the meson exchange model
are suitable to construct semi-phenomenological potential models for
nucleon-nucleon scattering.%

\begin{figure}
[ptbptb]
\begin{center}
\includegraphics[
height=4.7202in,
width=5.3082in
]%
{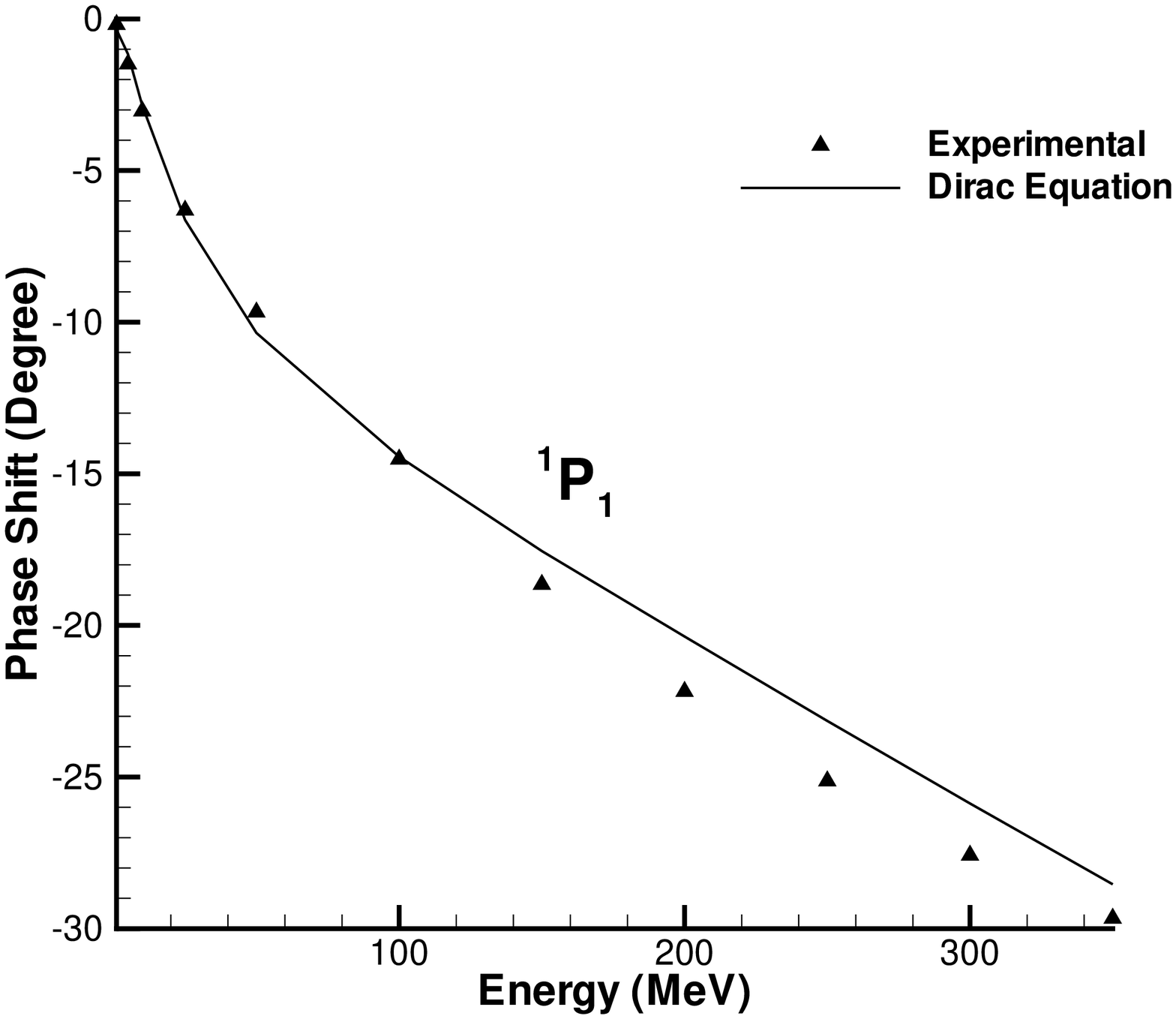}%
\caption{$np$ Scattering Phase Shift of $^{1}P_{1}$ State (Model 2)}%
\label{fig:p12}%
\end{center}
\end{figure}
\begin{figure}
[ptbptbptb]
\begin{center}
\includegraphics[
height=4.7202in,
width=5.3082in
]%
{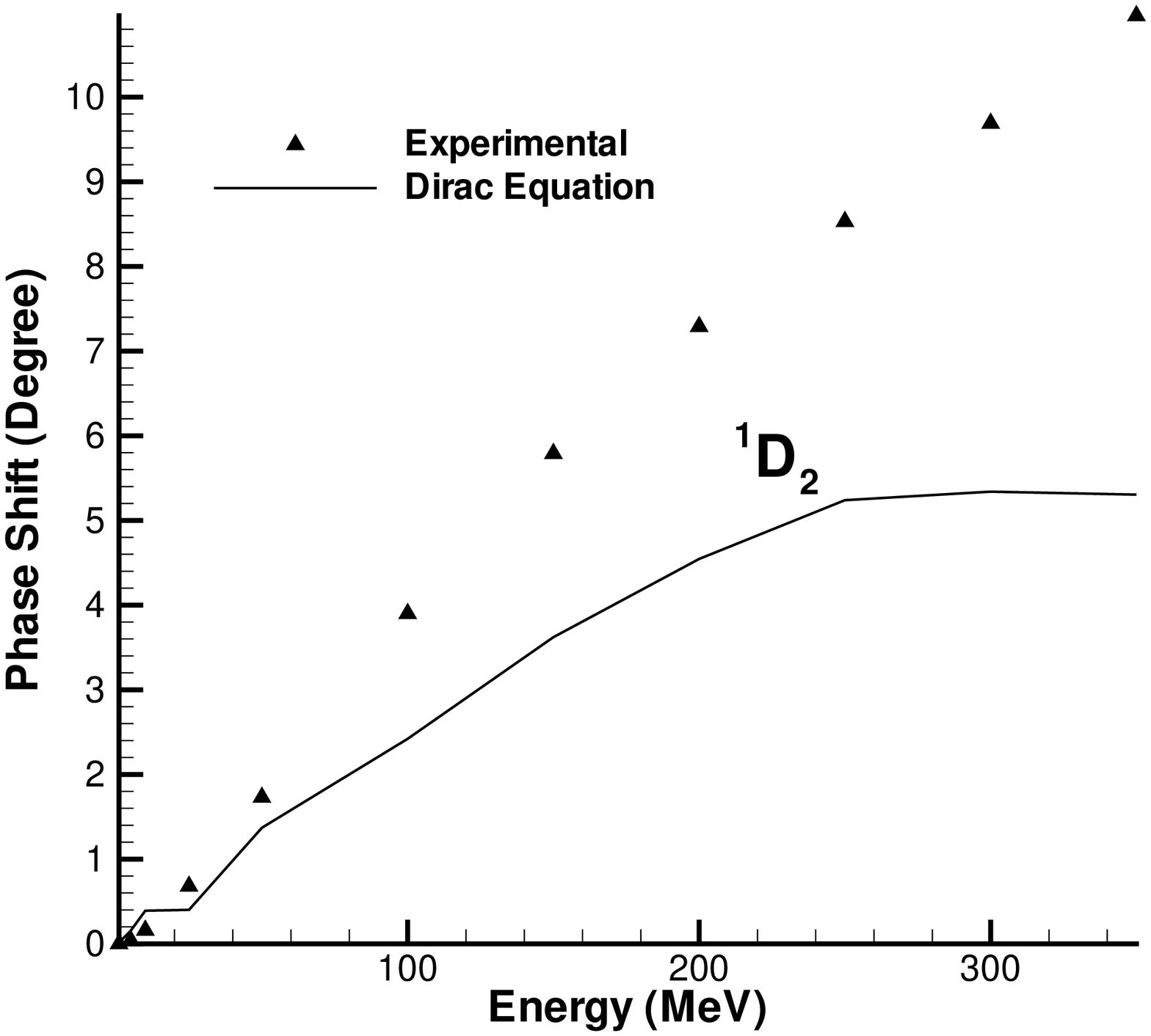}%
\caption{$np$ Scattering Phase Shift of $^{1}D_{2}$ State (Model 2)}%
\label{fig:d22}%
\end{center}
\end{figure}
\begin{figure}
[ptbptbptbptb]
\begin{center}
\includegraphics[
height=4.7202in,
width=5.3082in
]%
{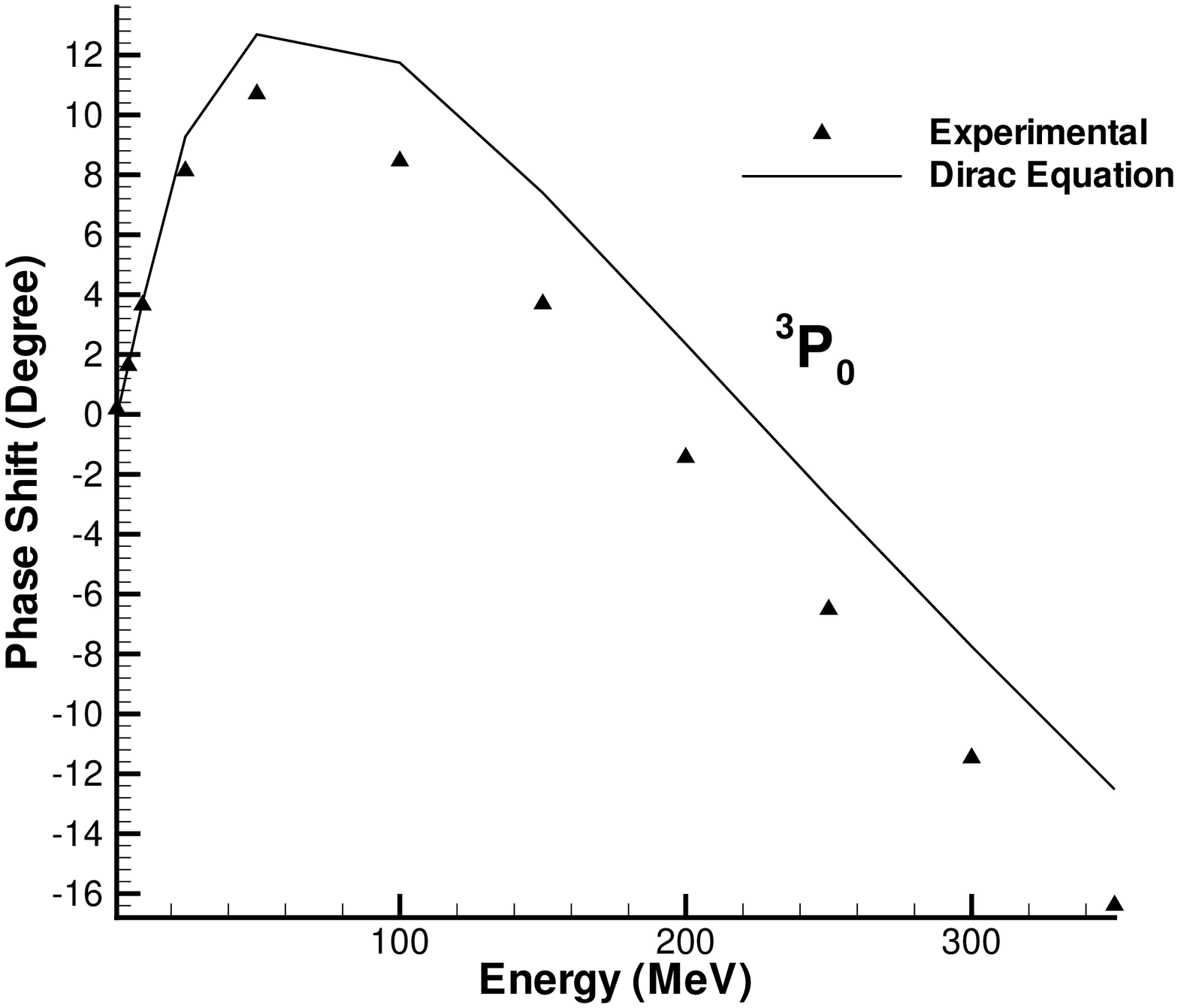}%
\caption{$np$ Scattering Phase Shift of $^{3}P_{0}$ State (Model 2)}%
\label{fig:p302}%
\end{center}
\end{figure}
\begin{figure}
[ptbptbptbptbptb]
\begin{center}
\includegraphics[
height=4.7202in,
width=5.3082in
]%
{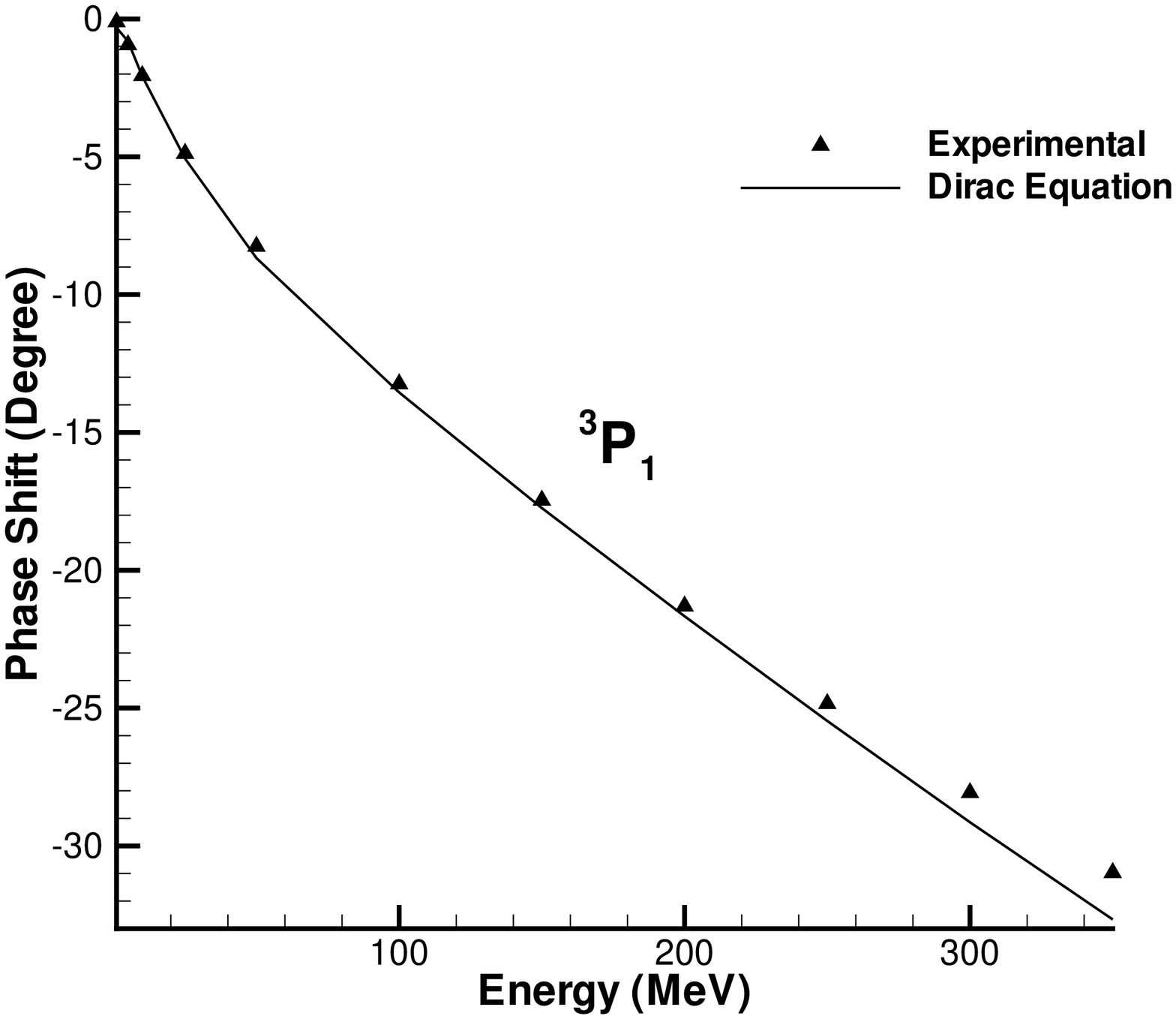}%
\caption{$np$ Scattering Phase Shift of $^{3}P_{1}$ State (Model 2)}%
\label{fig:p312}%
\end{center}
\end{figure}
\begin{figure}
[ptbptbptbptbptbptb]
\begin{center}
\includegraphics[
height=4.7202in,
width=5.3082in
]%
{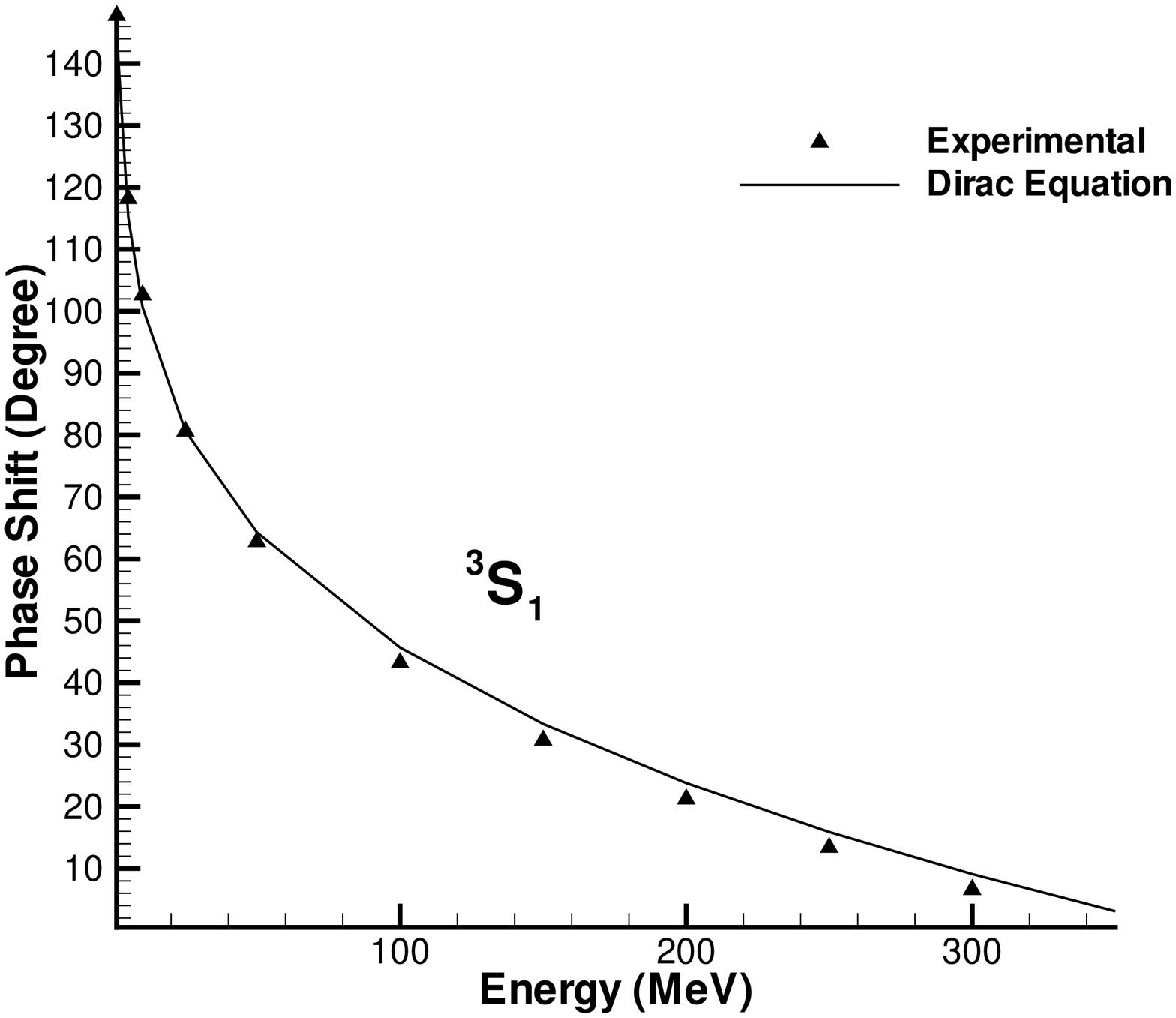}%
\caption{$np$ Scattering Phase Shift of $^{3}S_{1}$ State (Model 2)}%
\label{fig:s312}%
\end{center}
\end{figure}
\begin{figure}
[ptbptbptbptbptbptbptbptbptb]
\begin{center}
\includegraphics[
height=4.7202in,
width=5.3082in
]%
{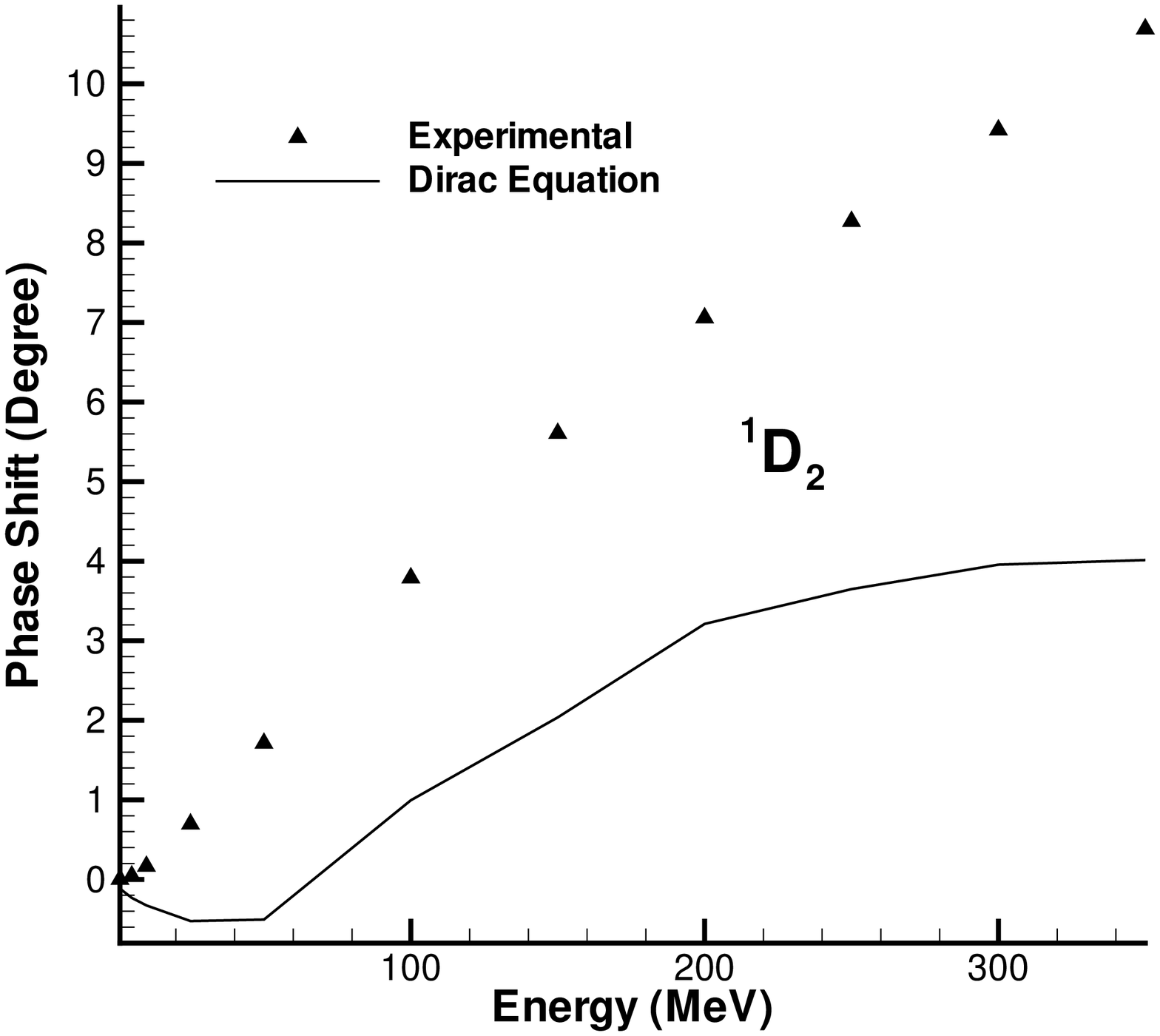}%
\caption{$pp$ Scattering Phase Shift of $^{1}D_{2}$ State (Model 2)}%
\label{fig:d2pp2}%
\end{center}
\end{figure}
\begin{figure}
[ptbptbptbptbptbptbptbptbptbptb]
\begin{center}
\includegraphics[
height=4.7202in,
width=5.3082in
]%
{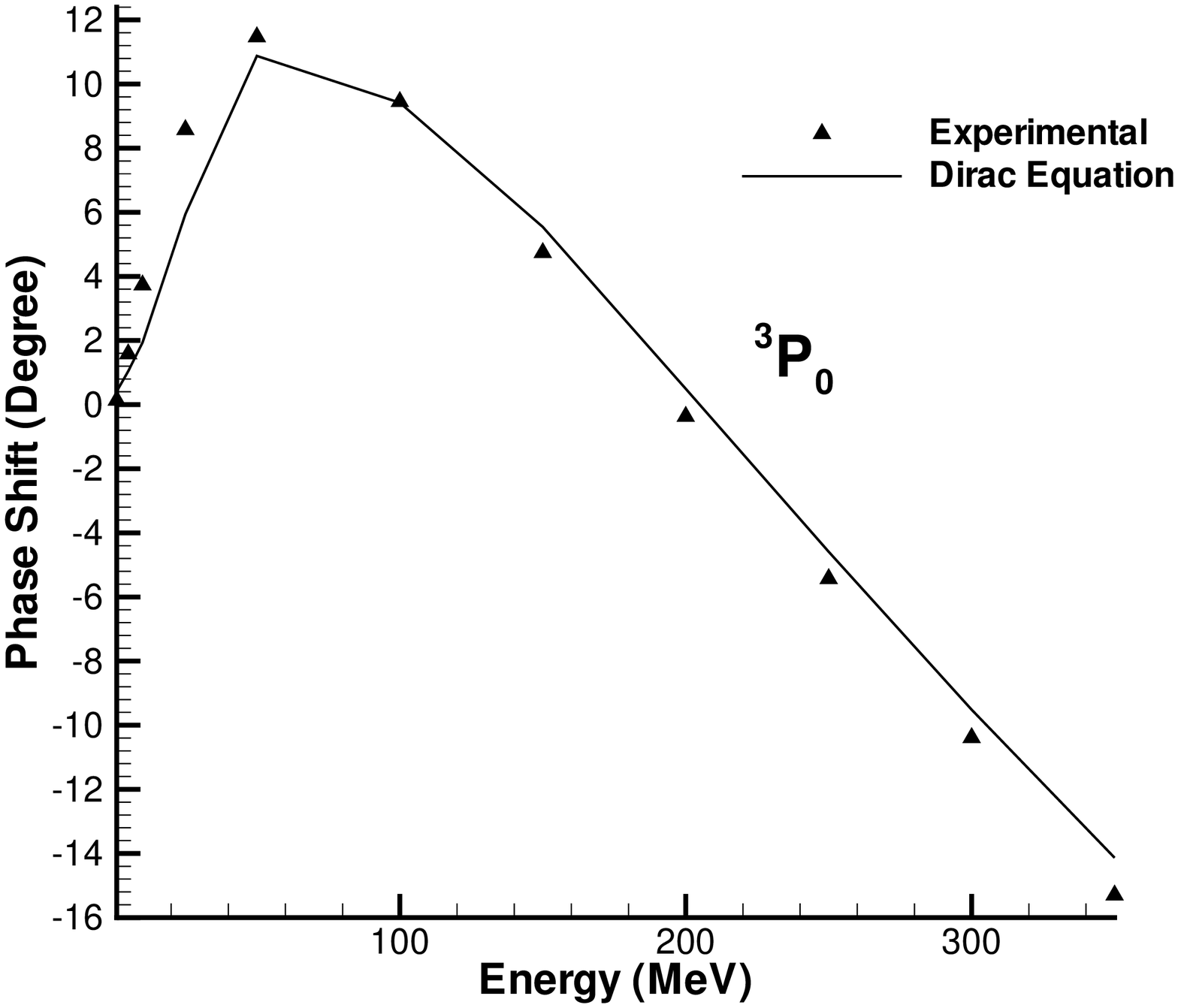}%
\caption{$pp$ Scattering Phase Shift of $^{3}P_{0}$ State (Model 2)}%
\label{fig:p30pp2}%
\end{center}
\end{figure}

\section{Conclusion}

The two-body Dirac equations of constraint dynamics constitutes the first
fully covariant treatment of the relativistic two-body problem that

a) includes constituent spin, b) regulates the relative time in a covariant
manner, c) provides an exact reduction to 4 decoupled 4-component wave
equations, d) includes non-perturbative recoil effects in a natural way that
eliminates the need for singularity-softening parameters or finite particle
size in semiphenomenological applications to QCD, e) is canonically equivalent
in the semi-relativistic approximation to the Fermi-Breit approximation to the
Bethe-Salpeter equation, f) unlike the Bethe-Salpeter equation and most other
relativisitc approaches has a local momentum structure as simple as that of
the nonrelativistic Schr\"{o}dinger equation. g) is well-defined for zero-mass
constituents (hence, permits investigation of the chiral symmetry limit) h)
possesses spin structure that yields an exact solution for singlet
positronium, i) has static limits that are relativistic, reducing to the
ordinary single-particle Dirac equation in the limit that either particle
becomes infinitely heavy. j) possesses a great variety of equivalent forms
that are rearrangements of its two coupled Dirac equations (hence is directly
related to many previously-known quantum descriptions of the relativistic
two-body system). \ These structures play an essential role in the success of
this approach to both QCD and QED bound states. What is noteworthy in the
latter appllication is that one need only identify the non-relativistic parts,
i.e. the lowest order forms of $\mathcal{A}$ and $S$. The spin-dependent and
covariant structure of the two-body Dirac formalism then automatically stamps
out the correct semirelativistic spin dependent and spin independent
corrections and provides well defined higher order relativistic corrections as
well. \ In addition the constraint formalism, although rooted in classical
mechanics, has close connections to the Bethe-Salpeter equation of quantum
field theory \cite{saz4} and with Wigner's formulation of relativistic quantum
mechanics as a symmetry of quantum theory \cite{plyz}.

In this paper we have shown that these two-body Dirac equations may provide a
reasonable account of the nucleon-nucleon scattering data when combined with
the meson exchange model. \ What makes this result important is that it is
accomplished with a local and covariant formulation of the two-body problem.
\ What makes this unique is that this approach has been thoroughly tested in a
nonperturbative context for both QED and QCD bound states. It is not a given
that success in one or even both areas would imply that the formalism would do
well in another. \ In particular, the fits could have easily been disastrous
given that the minimal coupling idea we have used (based in part on the
earlier work on the quasipotential approach of Todorov). \ The reason for some
doubt is that these minimal coupling forms (generalized to the scalar
interactions as well as the vector) lead to the scalar and vector potentials
appearing squared. \ Because of the size of the coupling constants, the
deviation from the standard effective potentials could have been considerable
in all cases. There are other nonperturbative structures that appear in the
Pauli reduction of our equations to Schr\"{o}dinger-like form (typical of what
appears in the Pauli reduction of the one-body Dirac equation) that could also
have prevented any reasonable results. So the general agreement we obtained
with the data is very encouraging that this approach could be extended to
include more general interactions.

An important step in our reduction was that we put the equation in a form for
which we can apply the techniques which have been already developed for the
Schr\"{o}dinger-like system in nonrelativistic quantum mechanics. \ This
required that we get rid of first derivative terms. For the uncoupled states,
it is pretty straightforward. For the coupled states we used a different
spin-matirx approach that works for both the uncoupled and coupled states simultaneously.

We then tested several models by using the variable phase methods. We found it
most \ convenient to put all the angular momentum barrier terms in the
potentials, and change all the phase shift equations to the form of $S$
state-like phase shift equations (see Eqs.(\ref{mes},\ref{psa},\ref{psb}%
,\ref{psc})).

After several models and several methods to minimize our $\chi^{2}$ tested, we
found two models which can lead us to a fairly good fit to the experimental
phase shift data.

The most important equation used in our phase shift analysis for
nucleon-nucleon scattering is Eq.(\ref{master_eq}) . It is a coupled
Schr\"{o}dinger-like equation derived from two-body Dirac equations with no
approximations. All of our radial wave equations for any specific angular
momentum state are obtained from this equation.

We use nine mesons in our fit. We summarize the meson-nucleon interactions we
used by writing the quantum field theory Lagrange function for their effective
interactions
\begin{align}
L_{I}  &  =g_{\sigma}\overline{\psi}\psi\sigma+g_{f_{0}}\overline{\psi}\psi
f_{0}+g_{a_{0}}\overline{\psi}\mathbf{\tau}\psi\cdot\mathbf{a}_{0}\nonumber\\
&  +g_{\rho}\overline{\psi}\gamma^{\mu}\mathbf{\tau}\psi\cdot\mathbf{\rho
}_{\mu}+g_{\omega}\overline{\psi}\gamma^{\mu}\psi\omega_{\mu}+g_{\phi
}\overline{\psi}\gamma^{\mu}\psi\phi_{\mu}\nonumber\\
&  +g_{\pi}\overline{\psi}\gamma^{5}\mathbf{\tau}\psi\cdot\mathbf{\pi}%
+g_{\eta}\overline{\psi}\gamma^{5}\psi\eta+g_{\eta^{\prime}}\overline{\psi
}\gamma^{5}\psi\eta^{\prime} \label{effect}%
\end{align}
where $\psi$ represent the nucleon field, $\sigma$, $f_{0}$, ... represent the
meson fields.

Several models have been tested by using the variable phase methods, two
models can lead us to a fairly good fit to the experimental phase shift data.
We use the parameters which gives good fits to the $np$ scattering data to
predict the phase shifts for the $pp$ scattering. \ These lead to a good
prediction for the $pp$ scattering based on the parameters we obtained (with
noted exceptions). This means that our work has shown a promising result. The
following are some suggestions to improve our work in the future.

\section{Suggestions For Future Work}

\subsection{Other Model Tests}

More model testing is absolutely necessary in the future. By model we mean the
way we place the perturbative interactions that arise from Eq.(\ref{effect})
into the nonperturbative forms we need for $L$, $C$ and $\mathcal{G}$. During
our fits, we found that our final result are sensitive to the model we chose
ranging from very bad fits to the fits presented here. Changing the way to
modify the interactions and the way of mesons enter into the two-body Dirac
equations may provide a new opportunity to improve our fit.

\subsection{Including World Tensor Interactions}

We have included just scalar, pseudoscalar and vector interactions in our
potentials through the invariant forms like $L$, $C$ and $\mathcal{G}$.
Treating two-body Dirac equations with tensor interactions of the vector meson
may improve our fit. These tensor interacting were discussed earlier (see
Eq.(\ref{tensor})) and correspond to non-minimal coupling of spin one-half
particle not present in QED but which can not be ruled out in massive vector
meson-nucleon interactions. The corresponding field theory interaction is%

\begin{equation}
\Delta L_{I}=g_{\rho}^{\prime}\overline{\psi}\sigma^{\mu\nu}\mathbf{\tau}%
\psi\mathbf{\rho}_{\mu\nu}+g_{\omega}^{\prime}\overline{\psi}\sigma^{\mu\nu
}\psi\omega_{\mu\nu}+g_{\phi}^{\prime}\overline{\psi}\sigma^{\mu\nu}\psi
\phi_{\mu\nu}%
\end{equation}
and would correspond to relaxing the free field equation assumption made in
Eq.(\ref{q-sqrt0})).

\subsection{Include Pseudovector Interactions}

Another option is to allow the pseudoscalar mesons($\pi$, $\eta$, and
$\eta^{\prime}$) to interact with the nucleon not only by the pseudoscalar
interaction(as in Eq.(\ref{effect})) but also by the way of the pseudovector
interactions as below%

\begin{equation}
\Delta L_{I}=g_{\pi}^{\prime}\overline{\psi}\gamma^{\mu}\gamma^{5}%
\mathbf{\tau}\psi\partial_{\mu}\mathbf{\pi}+g_{\eta}^{\prime}\overline{\psi
}\gamma^{\mu}\gamma^{5}\psi\partial_{\mu}\eta+g_{\eta^{\prime}}^{\prime
}\overline{\psi}\gamma^{\mu}\gamma^{5}\psi\partial_{\mu}\eta^{\prime}%
\end{equation}

\subsection{Include Full Massive Spin-One Propagator}

We have ignored a portion of the massive spin-one propagator in our fit which
is zero for particles on the mass shell. To include this portion of massive
spin-one propagator we would have to change the vector propagator as below%

\begin{equation}
\frac{\eta^{\mu\nu}}{q^{2}+m_{\rho}^{2}-i\varepsilon}\longrightarrow\frac
{\eta^{\mu\nu}+\frac{q^{\mu}q^{\nu}}{m_{\rho}^{2}}}{q^{2}+m_{\rho}%
^{2}-i\varepsilon}.
\end{equation}

Among all the four suggestions, the first one would be technically easiest
once we finds models more general than the two we have presented here . The
last three suggestions would involve corresponding additions to the
interaction that appear in the two-body Dirac equations. Because the above
interactions all involve derivative couplings we will \ have to examine the
CTBDE for the corresponding invariant $\Delta^{\prime}s$. These would include
not only the eight invariants listed earlier (see Eq.(\ref{interL}) to
Eq.(\ref{interA})) but also four additional ones corresponding to
$\Delta=R\theta_{1}\cdot\hat{x}_{\perp}\theta_{2}\cdot\hat{x}_{\perp},$
$~2S\theta_{51}\theta_{52}\theta_{1}\cdot\hat{x}_{\perp}\theta_{2}\cdot\hat
{x}_{\perp}\ ,~$\ $~2T\theta_{1}\cdot\hat{P}\theta_{1}\cdot\hat{P}\theta
_{1}\cdot\hat{x}_{\perp}\theta_{2}\cdot\hat{x}_{\perp}$ and $4U\theta
_{51}\theta_{52}\theta_{1}\cdot\hat{P}\theta_{1}\cdot\hat{P}\theta_{1}%
\cdot\hat{x}_{\perp}\theta_{2}\cdot\hat{x}_{\perp}.$ \ The four functions
$R,S,T,U$ are each functions of $x_{\perp}$ and they represent space-like
interactions paralleling those corresponding to $\mathcal{G},I,Y,$ and
$\mathcal{F}$ respectively given earlier. \ To include all 12 covariant matrix
interactions will involve a significant modification of our basic equation
Eq.(\ref{master_eq}) as well as the two-body Dirac equations given in
Eqs(\ref{e1},\ref{e2}). \ 

\subsection{Extentions to the N-Body Problem}

Can the constraint formalism be extended to $N$-bodies? \ There is no solution
to the compatibility condition%
\begin{equation}
\lbrack\mathcal{H}_{i},\mathcal{H}_{j}]|\psi\rangle=0;~~i,j=1,..,N
\label{ncmp}%
\end{equation}
of generalized mass-shell constraints (or their Dirac counterparts) that has
the simplicity of \ the \textquotedblleft third law\textquotedblright\ and
tranversality conditions given in (\ref{8}) and (\ref{9}). \ The difficulty
involves satisfying Eq.(\ref{ncmp}) and cluster separability (needed to
describe scattering states) at the same time. \ Rohrlich has shown that this
necessarily involves the introduction of $N-$body forces \cite{rohr}. If one
is willing to limit $N-$body considerations to bound states (so that cluster
considerations are not important) then Ref. \cite{saz6} provides a constraint
formalism in which a single dynamical wave equation (as in the two-body case)
determines the bound state energies. Ref. \cite{luca} (and references
contained therein) provides an $N-$body constraint formalism that involves
particles and fields leading in the end to directly inteacting particles by
elimination of the field degrees of freedom by second class contraints.

\appendix\renewcommand{\theequation}{\Alph{section}.\arabic{equation}}

\section{Pauli-Form of Two-Body Dirac Equations}

\bigskip We rewrite Eqs.({\ref{f1},\ref{f2}}) by multiplying the first by
$\sqrt{2}i\beta_{1}$ and the second by $\sqrt{2}i\beta_{2}$ yielding
\cite{long1}%

\[
\lbrack T_{1}(\beta_{1}\beta_{2})+U_{1}(\beta_{1}\beta_{2})\gamma_{51}%
\gamma_{52}]|\psi\rangle=(E_{1}+M_{1}\beta_{1})\gamma_{51}|\psi\rangle
\]%
\begin{equation}
-[T_{2}(\beta_{1}\beta_{2})+U_{2}(\beta_{1}\beta_{2})\gamma_{51}\gamma
_{52}]|\psi\rangle=(E_{2}+M_{2}\beta_{2})\gamma_{52}|\psi\rangle\label{a12}%
\end{equation}
in which the kinetic and recoil terms are
\[
T_{1}(\beta_{1}\beta_{2})=\exp(\mathcal{G})[\mathbf{\Sigma}_{1}\cdot
p-{\frac{i}{2}}\beta_{1}\beta_{2}(\mathbf{\Sigma}_{2}\cdot\partial
(-C+\mathcal{G}\beta_{1}\beta_{2}\mathbf{\Sigma}_{1}\cdot\mathbf{\Sigma}%
_{2})]
\]%
\begin{equation}
T_{2}(\beta_{1}\beta_{2})=\exp(\mathcal{G})[\mathbf{\Sigma}_{2}\cdot
p-{\frac{i}{2}}\beta_{1}\beta_{2}(\mathbf{\Sigma}_{1}\cdot\partial
(-C+\mathcal{G}\beta_{1}\beta_{2}\mathbf{\Sigma}_{1}\cdot\mathbf{\Sigma}_{2})]
\end{equation}%
\[
U_{1}(\beta_{1}\beta_{2})=\exp(\mathcal{G})[-{\frac{i}{2}}\beta_{1}\beta
_{2}\mathbf{\Sigma}_{2}\cdot\partial(J\beta_{1}\beta_{2}-L)]
\]%
\begin{equation}
U_{2}(\beta_{1}\beta_{2})=\exp(\mathcal{G})[-{\frac{i}{2}}\beta_{1}\beta
_{2}\mathbf{\Sigma}_{1}\cdot\partial(J\beta_{1}\beta_{2}-L)]
\end{equation}
while the timelike and scalar potentials $E_{i},M_{i}$ are given above in
Eqs.(\ref{m}) and (\ref{e})

The final result of the matrix multiplication in Eqs.({\ref{a12}}) is a set of
eight simultaneous equations for the Dirac spinors $|\psi\rangle_{1}%
,|\psi\rangle_{2},|\psi\rangle_{3},|\psi\rangle_{4}$. In an arbitrary frame,
the result of the matrix calculation produces the eight simultaneous equations
($\sigma_{i}^{\mu}|\psi\rangle\rightarrow\Sigma_{i}^{\mu}|\psi\rangle
_{1,2,3,4}$)\cite{long1}. One then reduces the eight equations to a second
order Schr\"{o}dinger-like equation by a process of substitution and
elimination using the combinations of the four Dirac-spinors given below
\cite{long1}:%

\[
|\phi_{\pm}\rangle\equiv|\psi\rangle_{1}\pm|\psi\rangle_{4},
\]%
\begin{equation}
|\chi_{\pm}\rangle\equiv|\psi\rangle_{2}\pm|\psi\rangle_{3}.
\end{equation}
We display all the general spin dependent structures in $\Phi(\mathbf{r}%
,\mathbf{p},{\mbox{\boldmath $\sigma$}}_{1},{\mbox{\boldmath $\sigma$}}%
_{2},w)$ explicitly, very similar to what appears in nonrelativistic
formalisms such as seen in the Hamada-Johnson and Yale group models (as well
as the nonrelativistic limit of Gross's equation). We do this by expressing it
explicitly in terms of its matrix (${\mbox{\boldmath $\sigma$}}_{1}%
,{\mbox{\boldmath $\sigma$}}_{2}$), and operator $\mathbf{p}$ structure in the
CM system ($\hat{P}=(1,\mathbf{0})$ ). We are working in the CM frame (i.e.
$x_{\perp}=(\mathbf{r},0)$), so all the interaction functions ($L(x_{\perp
}),J(x_{\perp}),C(x_{\perp}),\mathcal{G}(x_{\perp})$) are functions of
$r=\sqrt{x_{\perp}^{2}}=\left\vert \mathbf{r}\right\vert $, $F=F(r)$

Ref. \cite{long1} finds the reduction
\[
h[E_{1}[{\mbox{\boldmath $\sigma$}}_{1}\cdot\mathbf{p}%
-i{\mbox{\boldmath $\sigma$}}_{2}\cdot(\mathbf{d+k}{\mbox{\boldmath $\sigma$}}%
_{1}\cdot{\mbox{\boldmath $\sigma$}}_{2})]hF_{1}[{\mbox{\boldmath $\sigma$}}%
_{1}\cdot\mathbf{p}-i{{\mbox{\boldmath $\sigma$}}_{2}\cdot}(\mathbf{z+k}%
{\mbox{\boldmath $\sigma$}}_{1}\cdot{\mbox{\boldmath $\sigma$}}_{2})]|\phi
_{+}\rangle\ \ \ \ \ \ \ \ \ \ \ \ \ \ \ \
\]%
\[
+h[M_{1}[{\mbox{\boldmath $\sigma$}}_{1}\cdot\mathbf{p}%
-i{{\mbox{\boldmath $\sigma$}}_{2}\cdot}(\mathbf{o+k}%
{\mbox{\boldmath $\sigma$}}_{1}\cdot{\mbox{\boldmath $\sigma$}}_{2}%
)]hF_{3}[{\mbox{\boldmath $\sigma$}}_{1}\cdot\mathbf{p}%
-i{{\mbox{\boldmath $\sigma$}}_{2}\cdot}(\mathbf{z+k}%
{\mbox{\boldmath $\sigma$}}_{1}\cdot{\mbox{\boldmath $\sigma$}}_{2})]|\phi
_{+}\rangle\ \ \ \ \ \ \ \ \ \ \ \ \ \ \ \
\]%
\[
-h[E_{1}[{\mbox{\boldmath $\sigma$}}_{1}\cdot\mathbf{p}%
-i{{\mbox{\boldmath $\sigma$}}_{2}\cdot}(\mathbf{d+k}%
{\mbox{\boldmath $\sigma$}}_{1}\cdot{\mbox{\boldmath $\sigma$}}_{2}%
)]hF_{2}[{\mbox{\boldmath $\sigma$}}_{2}\cdot\mathbf{p}%
-i{{\mbox{\boldmath $\sigma$}}_{1}\cdot}(\mathbf{z+k}%
{\mbox{\boldmath $\sigma$}}_{1}\cdot{\mbox{\boldmath $\sigma$}}_{2})]|\phi
_{+}\rangle\ \ \ \ \ \ \ \ \ \ \ \ \ \ \ \
\]%
\[
+h[M_{1}[{\mbox{\boldmath $\sigma$}}_{1}\cdot\mathbf{p}%
-i{{\mbox{\boldmath $\sigma$}}_{2}\cdot}(\mathbf{o+k}%
{\mbox{\boldmath $\sigma$}}_{1}\cdot{\mbox{\boldmath $\sigma$}}_{2}%
)]hF_{4}[{\mbox{\boldmath $\sigma$}}_{2}\cdot\mathbf{p}%
-i{{\mbox{\boldmath $\sigma$}}_{1}\cdot}(\mathbf{z+k}%
{\mbox{\boldmath $\sigma$}}_{1}\cdot{\mbox{\boldmath $\sigma$}}_{2})]|\phi
_{+}\rangle\ \ \ \ \ \ \ \ \ \ \ \ \ \ \ \
\]%
\begin{equation}
=\mathcal{B}^{2}|\phi_{+}\rangle. \label{main}%
\end{equation}
in which%
\[
\mathcal{B}^{2}=E_{1}{}^{2}-M_{1}{}^{2}=E_{2}{}^{2}-M_{2}{}^{2}%
\]%
\begin{align}
&  =b^{2}(w)+(\epsilon_{1}^{2}+\epsilon_{2}^{2})\mathrm{\sinh}^{2}%
(J)+2\epsilon_{1}\epsilon_{2}\sinh(J)\cosh(J)\\
&  -(m_{1}^{2}+m_{2}^{2})\mathrm{\sinh}^{2}(L)-2m_{1}m_{2}\sinh(L)\cosh
(L).\nonumber
\end{align}
and%
\begin{align}
h  &  \equiv\exp(\mathcal{G}),\qquad\nonumber\\
\mathbf{k}  &  \equiv{\frac{1}{2}}\nabla\log(h),\nonumber\\
\mathbf{z}  &  \equiv{\frac{1}{2}}\nabla(-C+J-L)\nonumber\\
\mathbf{d}  &  \equiv{\frac{1}{2}}\nabla(C+J+L)\nonumber\\
\mathbf{o}  &  \equiv{\frac{1}{2}}\nabla(C-J-L).
\end{align}
with%
\[
F_{1}\equiv{\frac{M_{2}}{\mathcal{D}}}%
\]%
\[
F_{2}\equiv{\frac{M_{1}}{\mathcal{D}}}%
\]%
\[
F_{3}\equiv{\frac{E_{2}}{\mathcal{D}}}%
\]%
\begin{equation}
F_{4}\equiv{\frac{E_{1}}{\mathcal{D}}}%
\end{equation}%
\begin{equation}
\mathcal{D}\equiv E_{1}M_{2}+E_{2}M_{1}.
\end{equation}
Eq.({\ref{main}}) is a second-order Schr\"{o}dinger-like eigenvalue equation
for the newly defined wavefunction $|\phi_{+}\rangle$ in the form.
\begin{equation}
(p_{\perp}^{2}+\Phi(r,\mbox{\boldmath $\sigma$}_{1}%
,\mbox{\boldmath $\sigma$}_{2},w))|\phi_{+}\rangle=b^{2}(w)|\phi_{+}\rangle.
\label{s-like}%
\end{equation}
Eq.({\ref{spnin}}) for $\mathcal{B}^{2}$ provide us with the primary spin
independent part of $\Phi$, the quasipotential. Note that in the CM system
$p_{\perp}^{2}=\mathbf{p}^{2}$, $\sigma=(0,{\mbox{\boldmath $\sigma$}}).$For
future reference we will refer to the four sets of terms on the left hand side
as the {Eq.(\ref{main})} (a),(b),(c),(d) term.

Now we proceed with a different derivation than Long and Crater's derivation
\cite{long1}. The aim is to produce a Schr\"{o}dinger like form like in
Eq.(\ref{s-like}) involving the Pauli matrices for both particles.

Substitute $\mathbf{d}$, $h$, $F_{1}$, $\mathbf{z}$, $\mathbf{k}$'s
expressions to (a) term of {Eq.(\ref{main}), we obtain}
\[
\text{(a)\ \ term}=\exp(\mathcal{G})E_{1}\{[{\mbox{\boldmath $\sigma$}}%
_{1}\cdot\mathbf{p}-\frac{i}{2}{\mbox{\boldmath $\sigma$}}_{2}\cdot
\nabla(C+J+L)-\frac{i}{2}\nabla\mathcal{G}\cdot({\mbox{\boldmath $\sigma$}}%
_{1}+i{\mbox{\boldmath $\sigma$}}_{1}\times{\mbox{\boldmath $\sigma$}}_{2})]
\]%
\begin{equation}
\times\exp(\mathcal{G}){\frac{M_{2}}{\mathcal{D}}[}{\mbox{\boldmath $\sigma$}}%
_{1}\cdot\mathbf{p}-\frac{i}{2}{\mbox{\boldmath $\sigma$}}_{2}\cdot
\nabla(-C+J-L)-\frac{i}{2}\nabla\mathcal{G}\cdot({\mbox{\boldmath $\sigma$}}%
_{1}+i{\mbox{\boldmath $\sigma$}}_{1}\times{\mbox{\boldmath $\sigma$}}_{2})]\}
\end{equation}
working out the commutation relation of ${\mbox{\boldmath $\sigma$}}_{1}%
\cdot\mathbf{p} $ in above expression, we can find the (a) term is

(a) term$=\exp(\mathcal{G})E_{1}\times$
\[
\{\exp(\mathcal{G}){\frac{M_{2}}{\mathcal{D}}[\mathbf{p}}^{2}-\frac{i}%
{2}{\mbox{\boldmath $\sigma$}}_{2}\cdot\nabla
(-C+J-L)({\mbox{\boldmath $\sigma$}}_{1}\cdot\mathbf{p})-\frac{i}{2}%
\nabla\mathcal{G}\cdot\lbrack(\mathbf{p}+i({\mbox{\boldmath $\sigma$}}%
_{1}\times\mathbf{p})-({\mbox{\boldmath $\sigma$}}_{1}\cdot
{\mbox{\boldmath $\sigma$}}_{2})\mathbf{p}+{\mbox{\boldmath $\sigma$}}%
_{1}({\mbox{\boldmath $\sigma$}}_{2}\cdot\mathbf{p}%
)-i({\mbox{\boldmath $\sigma$}}_{2}\times\mathbf{p})]]
\]%
\[
+\frac{1}{i}{\mbox{\boldmath $\sigma$}}_{1}\cdot\partial\lbrack\exp
(\mathcal{G}){\frac{M_{2}}{\mathcal{D}}[}{\mbox{\boldmath $\sigma$}}_{1}%
\cdot\mathbf{p}-\frac{i}{2}{\mbox{\boldmath $\sigma$}}_{2}\cdot\nabla
(-C+J-L)-\frac{i}{2}\nabla\mathcal{G}\cdot({\mbox{\boldmath $\sigma$}}%
_{1}+i{\mbox{\boldmath $\sigma$}}_{1}\times{\mbox{\boldmath $\sigma$}}%
_{2})]]-
\]%
\[
\frac{i}{2}[{\mbox{\boldmath $\sigma$}}_{2}\cdot\nabla(C+J+L)+\nabla
\mathcal{G}\cdot({\mbox{\boldmath $\sigma$}}_{1}+i{\mbox{\boldmath $\sigma$}}%
_{1}\times{\mbox{\boldmath $\sigma$}}_{2})]\exp(\mathcal{G}){\frac{M_{2}%
}{\mathcal{D}}[}{\mbox{\boldmath $\sigma$}}_{1}\cdot\mathbf{p}-\frac{i}%
{2}{\mbox{\boldmath $\sigma$}}_{2}\cdot\nabla(-C+J-L)-\frac{i}{2}%
\nabla\mathcal{G}\cdot({\mbox{\boldmath $\sigma$}}_{1}%
+i{\mbox{\boldmath $\sigma$}}_{1}\times{\mbox{\boldmath $\sigma$}}_{2})]\}
\]
Likewise we can the find (b),(c),(d) terms.

(b) term$=\exp(\mathcal{G})M_{1}\times$%

\[
\{\exp(\mathcal{G}){\frac{E_{2}}{\mathcal{D}}[\mathbf{p}}^{2}-\frac{i}%
{2}{\mbox{\boldmath $\sigma$}}_{2}\cdot\nabla
(-C+J-L)({\mbox{\boldmath $\sigma$}}_{1}\cdot\mathbf{p})-\frac{i}{2}%
\nabla\mathcal{G}\cdot\lbrack(\mathbf{p}+i({\mbox{\boldmath $\sigma$}}%
_{1}\times\mathbf{p})-({\mbox{\boldmath $\sigma$}}_{1}\cdot
{\mbox{\boldmath $\sigma$}}_{2})\mathbf{p}+{\mbox{\boldmath $\sigma$}}%
_{1}({\mbox{\boldmath $\sigma$}}_{2}\cdot\mathbf{p}%
)-i({\mbox{\boldmath $\sigma$}}_{2}\times\mathbf{p})]]
\]

\[
+\frac{1}{i}{\mbox{\boldmath $\sigma$}}_{1}\cdot\partial\lbrack\exp
(\mathcal{G}){\frac{E_{2}}{\mathcal{D}}[}{\mbox{\boldmath $\sigma$}}_{1}%
\cdot\mathbf{p}-\frac{i}{2}{\mbox{\boldmath $\sigma$}}_{2}\cdot\nabla
(-C+J-L)-\frac{i}{2}\nabla\mathcal{G}\cdot({\mbox{\boldmath $\sigma$}}%
_{1}+i{\mbox{\boldmath $\sigma$}}_{1}\times{\mbox{\boldmath $\sigma$}}%
_{2})]]-
\]

\[
\frac{i}{2}[{\mbox{\boldmath $\sigma$}}_{2}\cdot\nabla(C-J-L)+\nabla
\mathcal{G}\cdot({\mbox{\boldmath $\sigma$}}_{1}+i{\mbox{\boldmath $\sigma$}}%
_{1}\times{\mbox{\boldmath $\sigma$}}_{2})]\exp(\mathcal{G}){\frac{E_{2}%
}{\mathcal{D}}[}{\mbox{\boldmath $\sigma$}}_{1}\cdot\mathbf{p}-\frac{i}%
{2}{\mbox{\boldmath $\sigma$}}_{2}\cdot\nabla(-C+J-L)-\frac{i}{2}%
\nabla\mathcal{G}\cdot({\mbox{\boldmath $\sigma$}}_{1}%
+i{\mbox{\boldmath $\sigma$}}_{1}\times{\mbox{\boldmath $\sigma$}}_{2})]\}
\]

(c) term$=-\exp(\mathcal{G})E_{1}\times$%

\[
\{\exp(\mathcal{G}){\frac{M_{1}}{\mathcal{D}}[}({\mbox{\boldmath $\sigma$}}%
_{2}\cdot\mathbf{p})({\mbox{\boldmath $\sigma$}}_{1}\cdot\mathbf{p})-\frac
{i}{2}{\mbox{\boldmath $\sigma$}}_{1}\cdot\nabla
(-C+J-L)({\mbox{\boldmath $\sigma$}}_{1}\cdot\mathbf{p})-\frac{i}{2}%
\nabla\mathcal{G}\cdot\lbrack({\mbox{\boldmath $\sigma$}}_{2}%
({\mbox{\boldmath $\sigma$} }_{1}\cdot\mathbf{p})-({\mbox{\boldmath $\sigma$}}%
_{1}\cdot{\mbox{\boldmath $\sigma$}}_{2})\mathbf{p}%
+{\mbox{\boldmath $\sigma$}}_{1}({\mbox{\boldmath $\sigma$}}_{2}%
\cdot\mathbf{p})
\]

\[
+i({\mbox{\boldmath $\sigma$}}_{2}\times\mathbf{p})]]+\frac{1}{i}%
{\mbox{\boldmath $\sigma$}}_{1}\cdot\partial\lbrack\exp(\mathcal{G}%
){\frac{M_{1}}{\mathcal{D}}[}{\mbox{\boldmath $\sigma$}}_{2}\cdot
\mathbf{p}-\frac{i}{2}{\mbox{\boldmath $\sigma$}}_{1}\cdot\nabla
(-C+J-L)-\frac{i}{2}\nabla\mathcal{G}\cdot({\mbox{\boldmath $\sigma$}}%
_{2}+i{\mbox{\boldmath $\sigma$}}_{2}\times{\mbox{\boldmath $\sigma$}}%
_{1})]]-
\]

\[
\frac{i}{2}[{\mbox{\boldmath $\sigma$}}_{2}\cdot\nabla(C+J+L)+\nabla
\mathcal{G}\cdot({\mbox{\boldmath $\sigma$}}_{1}+i{\mbox{\boldmath $\sigma$}}%
_{1}\times{\mbox{\boldmath $\sigma$}}_{2})]\exp(\mathcal{G}){\frac{M_{1}%
}{\mathcal{D}}[}{\mbox{\boldmath $\sigma$}}_{2}\cdot\mathbf{p}-\frac{i}%
{2}{\mbox{\boldmath $\sigma$}}_{1}\cdot\nabla(-C+J-L)-\frac{i}{2}%
\nabla\mathcal{G}\cdot({\mbox{\boldmath $\sigma$}}_{2}%
+i{\mbox{\boldmath $\sigma$}}_{2}\times{\mbox{\boldmath $\sigma$}}_{1})]\}
\]

(d) term$=\exp(\mathcal{G})M_{1}\times$%

\[
\{\exp(\mathcal{G}){\frac{E_{1}}{\mathcal{D}}[}({\mbox{\boldmath $\sigma$}}%
_{2}\cdot\mathbf{p})({\mbox{\boldmath $\sigma$}}_{1}\cdot\mathbf{p})-\frac
{i}{2}{\mbox{\boldmath $\sigma$}}_{1}\cdot\nabla
(-C+J-L)({\mbox{\boldmath $\sigma$}}_{1}\cdot\mathbf{p})-\frac{i}{2}%
\nabla\mathcal{G}\cdot\lbrack({\mbox{\boldmath $\sigma$}}_{2}%
({\mbox{\boldmath $\sigma$} }_{1}\cdot\mathbf{p})-({\mbox{\boldmath $\sigma$}}%
_{1}\cdot{\mbox{\boldmath $\sigma$}}_{2})\mathbf{p}%
+{\mbox{\boldmath $\sigma$}}_{1}({\mbox{\boldmath $\sigma$}}_{2}%
\cdot\mathbf{p})
\]

\[
+i({\mbox{\boldmath $\sigma$}}_{2}\times\mathbf{p})]]+\frac{1}{i}%
{\mbox{\boldmath $\sigma$}}_{1}\cdot\partial\lbrack\exp(\mathcal{G}%
){\frac{E_{1}}{\mathcal{D}}[}{\mbox{\boldmath $\sigma$}}_{2}\cdot
\mathbf{p}-\frac{i}{2}{\mbox{\boldmath $\sigma$}}_{1}\cdot\nabla
(-C+J-L)-\frac{i}{2}\nabla\mathcal{G}\cdot({\mbox{\boldmath $\sigma$}}%
_{2}+i{\mbox{\boldmath $\sigma$}}_{2}\times{\mbox{\boldmath $\sigma$}}%
_{1})]]-
\]

\[
\frac{i}{2}[{\mbox{\boldmath $\sigma$}}_{2}\cdot\nabla(C-J-L)+\nabla
\mathcal{G}\cdot({\mbox{\boldmath $\sigma$}}_{1}+i{\mbox{\boldmath $\sigma$}}%
_{1}\times{\mbox{\boldmath $\sigma$}}_{2})]\exp(\mathcal{G}){\frac{E_{1}%
}{\mathcal{D}}[}{\mbox{\boldmath $\sigma$}}_{2}\cdot\mathbf{p}-\frac{i}%
{2}{\mbox{\boldmath $\sigma$}}_{1}\cdot\nabla(-C+J-L)-\frac{i}{2}%
\nabla\mathcal{G}\cdot({\mbox{\boldmath $\sigma$}}_{2}%
+i{\mbox{\boldmath $\sigma$}}_{2}\times{\mbox{\boldmath $\sigma$}}_{1})]\}
\]

We simplify the above expressions by using identities involving
${\mbox{\boldmath $\sigma$}}_{1}$ and ${\mbox{\boldmath $\sigma$}}_{2}$ and
group above equations by the $\mathbf{p}^{2}$ term , Darwin term
$(\hat{\mathbf{r}}\cdot\mathbf{p)}$, spin-orbit angular momentum term
$\mathbf{L}\cdot({\mbox{\boldmath $\sigma$}}_{1}+{\mbox{\boldmath $\sigma$}}%
_{2})$, spin-orbit angular momentum difference term $\mathbf{L}\cdot
({\mbox{\boldmath $\sigma$}}_{1}-{\mbox{\boldmath $\sigma$}}_{2})$, spin-spin
term $({\mbox{\boldmath $\sigma$}}_{1}\cdot{\mbox{\boldmath $\sigma$}}_{2})$,
tensor term $({\mbox{\boldmath $\sigma$} }_{1}\cdot\hat{\mathbf{r}%
})({\mbox{\boldmath $\sigma$}}_{2}\cdot\hat{\mathbf{r}})$, additional spin
dependent terms $\mathbf{L}\cdot({\mbox{\boldmath $\sigma$}}_{1}%
\times{\mbox{\boldmath $\sigma$}}_{2})$ and $({\mbox{\boldmath $\sigma$}}%
_{1}\cdot\hat{\mathbf{r}})({\mbox{\boldmath $\sigma$}}_{2}\cdot\mathbf{p}%
)+({\mbox{\boldmath $\sigma$}}_{2}\cdot\hat{\mathbf{r}}%
)({\mbox{\boldmath $\sigma$}}_{1}\cdot\mathbf{p}),$ and spin independent
terms. Collecting all terms for the (a) + (b) + (c) + (d) terms our
Eq.(\ref{main}) becomes Eq.(\ref{main1}).

\section{Radial Equations \ \ }

The following are radial eigenvalue equations corresponding to Eq.(\ref{main1}%
) after getting rid of the first derivative terms for singlet states
$^{1}S_{0}$, $^{1}P_{1}$, $^{1}D_{2}$( a general singlet $^{1}J_{j}$), triplet
states $^{3}P_{1}$( a general let $^{3}J_{j}$), a general $s=1,$ $j=l+1$ (
$^{3}P_{0}$,$^{3}S_{1}$ states $)$, and a general $s=1,$ $j=l+1$ $(^{3}D_{1}$ state).

$^{1}S_{0}$, $^{1}P_{1}$, $^{1}D_{2}$ ( a general singlet $^{1}J_{j}$)
$\mathbf{L}\cdot({\mbox{\boldmath $\sigma$}}_{1}+{\mbox{\boldmath $\sigma$}}%
_{2})=0$, ${\mbox{\boldmath $\sigma$} }_{1}\cdot{\mbox{\boldmath $\sigma$}}%
_{2}=-3$, ${\mbox{\boldmath $\sigma$}}_{1}\cdot\mathbf{\hat{r}}%
\mbox{\boldmath $\sigma$}_{2}\cdot\mathbf{\hat{r}=-}1.$%

\begin{equation}
\{-\frac{d^{2}}{dr^{2}}+\frac{j(j+1)}{r^{2}}+\frac{g^{\prime2}}{4}+h^{\prime
2}+\frac{g^{\prime\prime}}{2}+\frac{g^{\prime}}{r}\mathbf{-}3k-j-g^{\prime
}h^{\prime}-h^{\prime\prime}-\frac{2h^{\prime}}{r}+m\}v=\mathcal{B}^{2}%
\exp(-2\mathcal{G})v
\end{equation}
$^{3}P_{1}$( a general triplet $^{3}J_{j}$) $\mathbf{L}\cdot
({\mbox{\boldmath $\sigma$} }_{1}+{\mbox{\boldmath $\sigma$}}_{2})=-2$,
${\mbox{\boldmath $\sigma$}}_{1}\cdot{\mbox{\boldmath $\sigma$}}_{2}=1$,
${\mbox{\boldmath $\sigma$}}_{1}\cdot\mathbf{\hat{r}}%
\mbox{\boldmath $\sigma$}_{2}\cdot\mathbf{\hat{r}=}1.$%

\begin{equation}
\{-\frac{d^{2}}{dr^{2}}+\frac{j(j+1)}{r^{2}}+\frac{g^{\prime2}}{4}+h^{\prime
2}+\frac{g^{\prime\prime}}{2}+k+n+g^{\prime}h^{\prime}+h^{\prime\prime
}+m\}v=\mathcal{B}^{2}\exp(-2\mathcal{G})v,
\end{equation}
$s=1,$ $j=l+1$ ( $^{3}S_{1}$ states $)$

$\mathbf{L}\cdot({\mbox{\boldmath $\sigma$}}_{1}+{\mbox{\boldmath $\sigma$}}%
_{2})=2(j-1)$, ${\mbox{\boldmath $\sigma$}}_{1}\cdot
{\mbox{\boldmath $\sigma$}}_{2}=1$, ${\mbox{\boldmath $\sigma$}}_{1}%
\cdot\mathbf{\hat{r}}\mbox{\boldmath $\sigma$}_{2}\cdot\mathbf{\hat{r}=}%
\frac{1}{2j+1}$ (diagonal term), and ${\mbox{\boldmath $\sigma$}}_{1}%
\cdot\mathbf{\hat{r}}\mbox{\boldmath $\sigma$}_{2}\cdot\mathbf{\hat{r}=}%
\frac{2\sqrt{j(j+1)}}{2j+1}$(off diagonal term).%

\[
\{-\frac{d^{2}}{dr^{2}}+\frac{j(j-1)}{r^{2}}{+}\frac{3g^{\prime}\sinh^{2}h}%
{r}+6h^{\prime}\frac{\cosh h\sinh h}{r}-\frac{6\sinh^{2}h}{r^{2}}%
-\frac{g^{\prime}\cosh h\sinh h}{r}%
\]

\[
-2h^{\prime}\frac{\sinh^{2}h}{r}+2\frac{\cosh h\sinh h}{r^{2}}+\frac
{g^{\prime2}}{4}+h^{\prime2}+\frac{g^{\prime\prime}}{2}+\frac{g^{\prime}}%
{r}+k
\]

\[
+2(j-1)[\frac{g^{\prime}}{2r}+\frac{g^{\prime}\sinh^{2}h}{r}-2\frac{\sinh
^{2}h}{r^{2}}+2h^{\prime}\frac{\cosh h\sinh h}{r}]
\]

\[
+\frac{2(j-1)}{2j+1}[2h^{\prime}\frac{\sinh^{2}h}{r}-2\frac{\cosh h\sinh
h}{r^{2}}+\frac{h^{\prime}}{r}+\frac{g^{\prime}\cosh h\sinh h}{r}]
\]

\[
+\frac{1}{2j+1}[\frac{3g^{\prime}\cosh h\sinh h}{r}-\frac{g^{\prime}\sinh
^{2}h}{r}-2h^{\prime}\frac{\cosh h\sinh h}{r}+6h^{\prime}\frac{\sinh^{2}h}{r}%
\]

\[
-6\frac{\cosh h\sinh h}{r^{2}}+2\frac{\sinh^{2}h}{r^{2}}+n+g^{\prime}%
h^{\prime}+h^{\prime\prime}+\frac{2h^{\prime}}{r}]+m\}u_{+}%
\]

\[
+\frac{2\sqrt{j(j+1)}}{2j+1}\{\frac{3g^{\prime}\cosh h\sinh h}{r}%
-\frac{g^{\prime}\sinh^{2}h}{r}-2h^{\prime}\frac{\cosh h\sinh h}{r}%
+6h^{\prime}\frac{\sinh^{2}h}{r}%
\]

\[
-6\frac{\cosh h\sinh h}{r^{2}}+2\frac{\sinh^{2}h}{r^{2}}+n+g^{\prime}%
h^{\prime}+h^{\prime\prime}+\frac{2h^{\prime}}{r}%
\]

\begin{equation}
+2(j-1)[\frac{2h^{\prime}\sinh^{2}(h)}{r}-\frac{2\cosh(h)\sinh(h)}{r^{2}%
}+\frac{h^{\prime}}{r}+\frac{g^{\prime}\cosh(h)\sinh(h)}{r}]\}u_{-}%
=\mathcal{B}^{2}\exp(-2\mathcal{G})u_{+},
\end{equation}

$s=1,$ $j=l-1$ ( $^{3}P_{0}$,$^{3}D_{1}$ states $)$

$\mathbf{L}\cdot({\mbox{\boldmath $\sigma$}}_{1}+{\mbox{\boldmath $\sigma$}}%
_{2})=-2(j+2)$, ${\mbox{\boldmath $\sigma$}}_{1}\cdot
{\mbox{\boldmath $\sigma$}}_{2}=1$, ${\mbox{\boldmath $\sigma$}}_{1}%
\cdot\mathbf{\hat{r}}\mbox{\boldmath $\sigma$}_{2}\cdot\mathbf{\hat{r}=-}%
\frac{1}{2j+1}$(diagonal term), and ${\mbox{\boldmath $\sigma$}}_{1}%
\cdot\mathbf{\hat{r}}\mbox{\boldmath $\sigma$}_{2}\cdot\mathbf{\hat{r}=}%
\frac{2\sqrt{j(j+1)}}{2j+1}$(off diagonal term).%

\[
\{-\frac{d^{2}}{dr^{2}}+\frac{(j+1)(j+2)}{r^{2}}{+}\frac{3g^{\prime}\sinh
^{2}h}{r}+6h^{\prime}\frac{\cosh h\sinh h}{r}-\frac{6\sinh^{2}h}{r^{2}}%
-\frac{g^{\prime}\cosh h\sinh h}{r}%
\]

\[
-2h^{\prime}\frac{\sinh^{2}h}{r}+2\frac{\cosh h\sinh h}{r^{2}}+\frac
{g^{\prime2}}{4}+h^{\prime2}+\frac{g^{\prime\prime}}{2}+\frac{g^{\prime}}%
{r}+k
\]

\[
+2(j+2)[\frac{g^{\prime}}{2r}+\frac{g^{\prime}\sinh^{2}h}{r}-2\frac{\sinh
^{2}h}{r^{2}}+2h^{\prime}\frac{\cosh h\sinh h}{r}]
\]

\[
+\frac{2(j-1)}{2j+1}[2h^{\prime}\frac{\sinh^{2}h}{r}-2\frac{\cosh h\sinh
h}{r^{2}}+\frac{h^{\prime}}{r}+\frac{g^{\prime}\cosh h\sinh h}{r}]
\]

\[
-\frac{1}{2j+1}[\frac{3g^{\prime}\cosh h\sinh h}{r}-\frac{g^{\prime}\sinh
^{2}h}{r}-2h^{\prime}\frac{\cosh h\sinh h}{r}+6h^{\prime}\frac{\sinh^{2}h}{r}%
\]

\[
-6\frac{\cosh h\sinh h}{r^{2}}+2\frac{\sinh^{2}h}{r^{2}}+n+g^{\prime}%
h^{\prime}+h^{\prime\prime}+\frac{2h^{\prime}}{r}]+m\}u_{-}%
\]

\[
+\frac{2\sqrt{j(j+1)}}{2j+1}\{\frac{3g^{\prime}\cosh h\sinh h}{r}%
-\frac{g^{\prime}\sinh^{2}h}{r}-2h^{\prime}\frac{\cosh h\sinh h}{r}%
+6h^{\prime}\frac{\sinh^{2}h}{r}%
\]

\[
-6\frac{\cosh h\sinh h}{r^{2}}+2\frac{\sinh^{2}h}{r^{2}}+n+g^{\prime}%
h^{\prime}+h^{\prime\prime}+\frac{2h^{\prime}}{r}%
\]

\begin{equation}
-2(j+2)[\frac{2h^{\prime}\sinh^{2}(h)}{r}-\frac{2\cosh(h)\sinh(h)}{r^{2}%
}+\frac{h^{\prime}}{r}+\frac{g^{\prime}\cosh(h)\sinh(h)}{r}]\}u_{+}%
=\mathcal{B}^{2}\exp(-2\mathcal{G})u_{-},
\end{equation}

\bigskip Substituting for $g^{\prime},h^{\prime},m,n,k$ we obtain the radial
equations and potentials $\Phi$ given in the text.

\section{Derivation of Coupled Phase Shift Equations}

We have found that we can use the Messiah ansatz \cite{quant}.%
\begin{align}
u  &  =A\sin(br+\delta)\nonumber\\
u^{\prime}  &  =bA\cos(br+\delta)
\end{align}
for the solution of
\begin{equation}
(-\frac{d^{2}}{dr^{2}}+\Phi_{L}(r))u=b^{2}u(r)
\end{equation}
to yield%
\begin{equation}
\delta^{\prime}=-\frac{\Phi_{L}}{b}\sin^{2}(br+\delta)
\end{equation}

Next we see how this can be worked out in the case of coupled radial equations
of the form
\begin{equation}
-\left(
\begin{array}
[c]{c}%
u_{-}\\
u_{+}%
\end{array}
\right)  ^{\prime\prime}+\Phi_{L}\left(
\begin{array}
[c]{c}%
u_{-}\\
u_{+}%
\end{array}
\right)  =b^{2}\left(
\begin{array}
[c]{c}%
u_{-}\\
u_{+}%
\end{array}
\right)  .
\end{equation}
where $\Phi_{L}$ is a two by two matrix. This equation will have solutions
that are $S-$wave dominant and $D-$wave dominant. \ Form them together into a
$2\times2$ matrix $U$ so that the above equation becomes
\begin{equation}
-U^{\prime\prime}+\Phi_{L}U=b^{2}U.
\end{equation}
\ Then take its transpose and add the two. \ One obtains%
\begin{equation}
-(U^{\prime\prime}+U^{\prime^{\prime}T})+(\Phi_{L}U+U^{T}\Phi_{L}^{T}%
)=b^{2}(U+U^{T}) \label{seq}%
\end{equation}
In analogy to the uncoupled case we assume%
\begin{equation}
U=A(r)\sin(br+\mathcal{D(}r)\mathbb{)}%
\end{equation}
where \
\begin{align}
\mathcal{D}  &  \mathbb{=}\delta(r)+\mathbf{D(r)}\cdot
{\mbox{\boldmath $\sigma$}}\\
A  &  \mathbb{=}a(r)+\mathbf{A(r)}\cdot{\mbox{\boldmath $\sigma$}.}\nonumber
\end{align}
Let $R$ be a matrix that diagonalizes the phase shift matrix function
$\mathcal{D}(r)$ to the form $\delta(r)+D(r)\sigma_{3},$%
\begin{equation}
\tilde{U}\mathbf{=}RUR^{-1}=\tilde{A}\sin(br+\delta+D\sigma_{3})
\end{equation}
where%
\begin{equation}
\tilde{A}=RAR^{-1}%
\end{equation}
Continuing the analogy we let%
\begin{equation}
U^{\prime}=bA\cos(br+\mathcal{D}\mathbb{)}.
\end{equation}
Then%
\begin{align}
RU^{\prime}R^{-1}  &  =b\tilde{A}\cos(br+\delta+D\sigma_{3})=R(R^{-1}\tilde
{U}R)^{\prime}R^{-1}\\
&  =RR^{-1\prime}\tilde{A}\sin(br+\delta+D\sigma_{3})+\tilde{A}^{\prime}%
\sin(br+\delta+D\sigma_{3})\nonumber\\
&  +\tilde{A}(b+\delta^{\prime}+D^{\prime}\sigma_{3})\cos(br+\delta
+D\sigma_{3})+\tilde{A}\sin(br+\delta+D\sigma_{3})R^{\prime}R^{-1}\nonumber
\end{align}
But $RR^{-1\prime}=-R^{\prime}R^{-1}$ so that we obtain the condition%
\begin{equation}
\tilde{A}^{\prime}\sin(br+\delta+D\sigma_{3})+\tilde{A}(\delta^{\prime
}+D^{\prime}\sigma_{3})\cos(br+\delta+D\sigma_{3})+[\tilde{A}\sin
(br+\delta+D\sigma_{3}),R^{\prime}R^{-1}]_{-}=0. \label{wf}%
\end{equation}
In general we would take%
\begin{align}
\tilde{A}  &  \equiv a+\tilde{A}_{3}\sigma_{3}+\mathbf{\tilde{A}}_{\perp}%
\cdot{\mbox{\boldmath $\sigma$}}\\
&  \equiv\tilde{A}_{||}+\mathbf{\tilde{A}}_{\perp}\cdot
{\mbox{\boldmath $\sigma$} }\nonumber
\end{align}
and decompose Eq.(\ref{wf}) and Eq.(\ref{seq}) into two sets of \ four coupled equations.

Give $R$ the following general form:%
\begin{align}
R  &  =\exp(i\varepsilon(r)\sigma_{2})\exp(\eta(r)\sigma_{1})\nonumber\\
R^{-1}  &  =\exp(-\eta\sigma_{1})\exp(-i\varepsilon\sigma_{2})\nonumber\\
R^{\prime}  &  =i\varepsilon^{\prime}\sigma_{2}\exp(i\varepsilon\sigma
_{2})\exp(\eta\sigma_{1})+\exp(i\varepsilon\sigma_{2})\exp(\eta\sigma_{1}%
)\eta^{\prime}\sigma_{1}\nonumber\\
R^{\prime}R^{-1}  &  =i\varepsilon^{\prime}\sigma_{2}+\exp(i\varepsilon
\sigma_{2})\eta^{\prime}\sigma_{1}\exp(-i\varepsilon\sigma_{2})\nonumber\\
&  =i\varepsilon^{\prime}\sigma_{2}+\eta^{\prime}\cos(2\varepsilon)\sigma
_{1}+\eta^{\prime}\sin(2\varepsilon)\sigma_{3}%
\end{align}
We consider the case in which $\Phi_{L}$ is a symmtric matrix and furthermore
that as a result $\mathcal{D=D}^{T}.$ \ In that case our matrix is $R$ is
orthogonal ($\eta=0$).\ 

Next we examine the three terms of Eq.(\ref{seq}). \ We assume that
$\mathbf{\tilde{A}}_{\perp}$ is symmetric so that $\tilde{A}_{2}=0$ and
\begin{align}
b^{2}R(U+U^{T})R^{-1}  &  =b^{2}[(\tilde{A}_{||}+\mathbf{\tilde{A}}_{\perp
}\cdot{\mbox{\boldmath $\sigma$})}\sin(br+\delta+D\sigma_{3})+\sin
(br+\delta+D\sigma_{3})(\tilde{A}_{||}+\mathbf{\tilde{A}}_{\perp}%
\cdot{\mbox{\boldmath $\sigma$}}]\mathbf{)}\nonumber\\
&  =b^{2}[2\tilde{A}_{||}\sin(br+\delta+D\sigma_{3})+2(\tilde{A}_{1}%
\sin(br+\delta)\cos D)\sigma_{1}]
\end{align}
and%
\begin{align}
&  R(\Phi U+U^{T}\Phi^{T})R^{-1}\nonumber\\
&  =(\tilde{\Phi}_{||}+\mathbf{\tilde{\Phi}}_{\perp}{\cdot
\mbox{\boldmath $\sigma$} )}(\tilde{A}_{||}+\mathbf{\tilde{A}}_{\perp}%
\cdot{\mbox{\boldmath $\sigma$})}\sin(br+\delta+D\sigma_{3})+(\text{Transpose}%
)\nonumber\\
&  =(\tilde{\Phi}_{||}\tilde{A}_{||}+\mathbf{\tilde{\Phi}}_{\perp}%
\cdot\mathbf{\tilde{A}}_{\perp}\mathbf{)}\sin(br+\delta+D\sigma_{3}%
)\nonumber\\
&  +(\tilde{\Phi}_{||}\mathbf{\tilde{A}}_{\perp}\cdot
{\mbox{\boldmath $\sigma$} }+\mathbf{\tilde{\Phi}}_{\perp}\cdot
{\mbox{\boldmath $\sigma$}}\tilde{A}_{||}+i\mathbf{\tilde{\Phi}}_{\perp}%
\times\mathbf{\tilde{A}}_{\perp}\cdot{\mbox{\boldmath $\sigma$}}%
)\sin(br+\delta+D\sigma_{3})\nonumber\\
&  +\sin(br+\delta+D\sigma_{3})(\tilde{\Phi}_{||}\tilde{A}_{||}+\mathbf{\tilde
{\Phi}}_{\perp}\cdot\mathbf{\tilde{A}}_{\perp}\mathbf{)}\nonumber\\
&  +\sin(br+\delta+D\sigma_{3})(\mathbf{\tilde{A}}_{\perp}\cdot
{\mbox{\boldmath $\sigma$} }\tilde{\Phi}_{||}+\tilde{A}_{||}\mathbf{\tilde
{\Phi}}_{\perp}\cdot{\mbox{\boldmath $\sigma$}}+i\mathbf{\tilde{\Phi}}_{\perp
}\times\mathbf{\tilde{A}}_{\perp}\cdot{\mbox{\boldmath $\sigma$}})
\end{align}
The term $\mathbf{\tilde{\Phi}}_{\perp}\times\mathbf{\tilde{A}}_{\perp}%
\cdot{\mbox{\boldmath $\sigma$}}$ is zero since $\tilde{A}_{2}=0=\tilde{\Phi
}_{2}.$ The second derivative term is%
\begin{align}
&  R(U^{\prime\prime}+U^{\prime\prime T})R^{T}\nonumber\\
&  =R(R^{T}\tilde{U}^{\prime}R)^{\prime}R^{-1}+(\text{Transpose})\nonumber\\
&  =bR(R^{T}\tilde{A}\cos(br+\delta+D\sigma_{3})R)^{\prime}R^{T}%
+(\text{Transpose})\nonumber\\
&  =b\{[\tilde{A}\cos(br+\delta+D\sigma_{3}),R^{\prime}R^{T}]_{-}+\tilde
{A}^{\prime}\cos(br+\delta+D\sigma_{3})\nonumber\\
&  -\tilde{A}(b+\delta^{\prime}+D^{\prime}\sigma_{3})\sin(br+\delta
+D\sigma_{3})\}+(\text{Transpose})\nonumber\\
&  =b\{i\varepsilon^{\prime}[(\tilde{A}_{||}+\mathbf{\tilde{A}}_{\perp}%
\cdot{\mbox{\boldmath $\sigma$})}\cos(br+\delta+D\sigma_{3}),\sigma_{2}%
]_{-}+(\tilde{A}_{||}^{\prime}+\mathbf{\tilde{A}}_{\perp}^{\prime}%
\cdot{\mbox{\boldmath $\sigma$})}\cos(br+\delta+D\sigma_{3})\nonumber\\
&  -(\tilde{A}_{||}+\mathbf{\tilde{A}}_{\perp}\cdot
{\mbox{\boldmath $\sigma$})}(b+\delta^{\prime}+D^{\prime}\sigma_{3}%
)\sin(br+\delta+D\sigma_{3})\}+(\text{Transpose})
\end{align}
Using properties of the Pauli matrices and dividing Eq.(\ref{wf}) and
Eq.(\ref{seq}) into $||$ and $\perp$ components we obtain the four equations%
\begin{align}
&  -\tilde{A}_{||}(\delta^{\prime}+D^{\prime}\sigma_{3})\sin(br+\delta
+D\sigma_{3})\nonumber\\
&  +\tilde{A}_{||}^{\prime}\cos(br+\delta+D\sigma_{3})-2b\varepsilon^{\prime
}\sigma_{3}\tilde{A}_{1}\cos(br+\delta)\cos D\nonumber\\
&  =\frac{1}{b}(\tilde{\Phi}_{||}\tilde{A}_{||}+\tilde{\Phi}_{1}\tilde{A}%
_{1}\mathbf{)}\sin(br+\delta+D\sigma_{3}), \label{d1}%
\end{align}%
\begin{align}
&  \tilde{A}_{||}^{\prime}\sin(br+\delta+D\sigma_{3})+\tilde{A}_{||}%
(\delta^{\prime}+D^{\prime}\sigma_{3})\cos(br+\delta+D\sigma_{3}%
)-2\varepsilon^{\prime}\sigma_{3}\tilde{A}_{1}\sin(br+\delta)\cos D\nonumber\\
&  =0, \label{2}%
\end{align}%
\begin{align}
&  \cos(br+\delta)\cos D\tilde{A}_{1}^{\prime}-(\delta^{\prime}\sin
(br+\delta)\cos D+D^{\prime}\cos(br+\delta)\sin D)\tilde{A}_{1}\nonumber\\
&  +2\varepsilon^{\prime}(\tilde{A}_{3}\cos(br+\delta)\cos D-a\sin
(br+\delta)\sin D)\nonumber\\
&  =\frac{1}{b}[\phi\sin(br+\delta)\cos D\tilde{A}_{1}-\tilde{\Phi}_{3}%
\cos(br+\delta)\sin D\tilde{A}_{1}\nonumber\\
&  +(a\sin(br+\delta)\cos D+\tilde{A}_{3}(\cos(br+\delta)\sin D)\tilde{\Phi
}_{1}], \label{d3}%
\end{align}%
\begin{align}
&  \tilde{A}_{1}^{\prime}\sin(br+\delta)\cos D+\tilde{A}_{1}(\delta^{\prime
}\cos(br+\delta)\cos D-D^{\prime}\sin(br+\delta)\sin D)\nonumber\\
&  +2\varepsilon^{\prime}(a\cos(br+\delta)\sin D+\tilde{A}_{3}\sin
(br+\delta)\cos D)\nonumber\\
&  =0. \label{4}%
\end{align}
Combining Eq.(\ref{d1}) and Eq.(\ref{2}) we obtain%
\begin{align}
&  -\tilde{A}_{||}(\delta^{\prime}+D^{\prime}\sigma_{3})-2\varepsilon^{\prime
}\tilde{A}_{1}\sin D\cos D\nonumber\\
&  =\frac{1}{b}(\tilde{\Phi}_{||}\tilde{A}_{||}+\tilde{\Phi}_{1}\tilde{A}%
_{1}\mathbf{)}\sin^{2}(br+\delta+D\sigma_{3}). \label{d5}%
\end{align}
Combining Eq.(\ref{d3}) and Eq.(\ref{4}) gives%
\begin{align}
&  \tilde{A}_{1}\delta^{\prime}\csc(br+\delta)\cos D-2\varepsilon^{\prime}%
\csc(br+\delta)\sin D\nonumber\\
&  =\frac{1}{b}[(\phi\sin(br+\delta)\cos D-\tilde{\Phi}_{3}\cos(br+\delta)\sin
D)\tilde{A}_{1}\nonumber\\
&  +(a\sin(br+\delta)\cos D+\tilde{A}_{3}(\cos(br+\delta)\sin D)\tilde{\Phi
}_{1}]. \label{6}%
\end{align}
Rewrite the above two equations as%
\begin{align}
&  \tilde{A}_{||}[(\delta^{\prime}+D^{\prime}\sigma_{3})+\frac{1}{b}%
\tilde{\Phi}_{||}\sin^{2}(br+\delta+D\sigma_{3})]\nonumber\\
&  +\tilde{A}_{1}(2\varepsilon^{\prime}\sin D\cos D+\frac{1}{b}\tilde{\Phi
}_{1}\sin^{2}(br+\delta+D\sigma_{3}))\nonumber\\
&  =0,\\
&  \frac{1}{b}(a\sin(br+\delta)\cos D+\tilde{A}_{3}\cos(br+\delta)\sin
D)\tilde{\Phi}_{1}+2\varepsilon^{\prime}a\csc(br+\delta)\sin D\nonumber\\
&  +\tilde{A}_{1}[\frac{1}{b}(\phi\sin(br+\delta)\cos D-\tilde{\Phi}_{3}%
\cos(br+\delta)\sin D)-\delta^{\prime}\csc(br+\delta)\cos D]\nonumber\\
&  =0. \label{7b}%
\end{align}
The first of these two equations is actually two equations%
\begin{align}
&  a[\delta^{\prime}+\frac{1}{2b}\phi(1-\cos2(br+\delta)\cos(2D))+\frac{1}%
{2b}\tilde{\Phi}_{3}\sin2(br+\delta)\sin(2D)]\nonumber\\
&  \tilde{A}_{3}[D^{\prime}+\frac{1}{2b}\tilde{\Phi}_{3}(1-\cos2(br+\delta
)\cos(2D))+\frac{1}{2b}\phi\sin2(br+\delta)\sin(2D)]\nonumber\\
&  +\tilde{A}_{1}((\varepsilon^{\prime}\sin2D+\frac{1}{2b}\tilde{\Phi}%
_{1}(1-\cos2(br+\delta)\cos2D))\nonumber\\
&  =0 \label{d8}%
\end{align}
and%
\begin{align}
&  a[D^{\prime}+\frac{1}{2b}\tilde{\Phi}_{3}(1-\cos2(br+\delta)\cos
(2D))+\frac{1}{2b}\phi\sin2(br+\delta)\sin(2D)]\nonumber\\
&  +\tilde{A}_{3}[\delta^{\prime}+\frac{1}{2b}\phi(1-\cos2(br+\delta
)\cos(2D))+\frac{1}{2b}\tilde{\Phi}_{3}\sin2(br+\delta)\sin(2D)]\nonumber\\
&  +\tilde{A}_{1}\frac{1}{2b}\tilde{\Phi}_{1}\sin2(br+\delta)\sin2D\nonumber\\
&  =0. \label{d9}%
\end{align}
So now together with Eq.(\ref{7b})%
\begin{align}
&  a[\frac{1}{b}\sin(br+\delta)\cos(D)\tilde{\Phi}_{1}+2\varepsilon^{\prime
}\csc(br+\delta)\sin D]\nonumber\\
&  +\tilde{A}_{3}[\frac{1}{b}\cos(br+\delta)\sin(D)\tilde{\Phi}_{1}%
]\nonumber\\
&  +\tilde{A}_{1}[\frac{1}{b}(\phi\sin(br+\delta)\cos D-\tilde{\Phi}_{3}%
\cos(br+\delta)\sin D)-\delta^{\prime}\csc(br+\delta)\cos D]\nonumber\\
&  =0 \label{10}%
\end{align}
we have three homogeneous equations in $a,\tilde{A}_{3},\tilde{A}_{1}$. \ We
simplify these equations further by assuming that $\tilde{A}_{1}=0.\ $
\begin{align}
&  =a[\delta^{\prime}+\frac{1}{2b}\phi(1-\cos2(br+\delta)\cos(2D))+\frac
{1}{2b}\tilde{\Phi}_{3}\sin2(br+\delta)\sin(2D)]\nonumber\\
&  \tilde{A}_{3}[D^{\prime}+\frac{1}{2b}\tilde{\Phi}_{3}(1-\cos2(br+\delta
)\cos(2D))+\frac{1}{2b}\phi\sin2(br+\delta)\sin(2D)]\nonumber\\
&  =0
\end{align}%
\begin{align}
&  =a[D^{\prime}+\frac{1}{2b}\tilde{\Phi}_{3}(1-\cos2(br+\delta)\cos
(2D))+\frac{1}{2b}\phi\sin2(br+\delta)\sin(2D)]\nonumber\\
&  +\tilde{A}_{3}[\delta^{\prime}+\frac{1}{2b}\phi(1-\cos2(br+\delta
)\cos(2D))+\frac{1}{2b}\tilde{\Phi}_{3}\sin2(br+\delta)\sin(2D)]\nonumber\\
&  =0
\end{align}%
\begin{align}
&  a[\frac{1}{b}\sin(br+\delta)\cos(D)\tilde{\Phi}_{1}+2\varepsilon^{\prime
}\csc(br+\delta)\sin D]\nonumber\\
&  +\tilde{A}_{3}[\frac{1}{b}\cos(br+\delta)\sin(D)\tilde{\Phi}_{1}%
]\nonumber\\
&  =0 \label{11}%
\end{align}
The solution we seek is
\begin{equation}
\delta^{\prime}+\frac{1}{2b}\phi(1-\cos2(br+\delta)\cos(2D))+\frac{1}%
{2b}\tilde{\Phi}_{3}\sin2(br+\delta)\sin(2D)=0
\end{equation}%
\begin{equation}
D^{\prime}+\frac{1}{2b}\tilde{\Phi}_{3}(1-\cos2(br+\delta)\cos(2D))+\frac
{1}{2b}\phi\sin2(br+\delta)\sin(2D)=0.
\end{equation}
Let%
\begin{align}
\delta &  =\frac{1}{2}(\delta_{1}+\delta_{2})\nonumber\\
D  &  =\frac{1}{2}(\delta_{1}-\delta_{2}),
\end{align}
and that leads to
\begin{equation}
\delta_{1}^{\prime}=-\frac{1}{b}(\phi+\tilde{\Phi}_{3})\sin^{2}(br+\delta_{1})
\label{12}%
\end{equation}%
\begin{equation}
\delta_{2}^{\prime}=-\frac{1}{b}(\phi-\tilde{\Phi}_{3})\sin^{2}(br+\delta_{2})
\label{13}%
\end{equation}
Returning to Eq.(\ref{4}) we find it reduces to
\begin{equation}
\tilde{A}_{3}=-a\cot(br+\delta)\tan D.
\end{equation}
and combining that with Eq.(\ref{11}) yields%
\begin{align*}
&  [\frac{1}{b}\sin(br+\delta)\cos(D)\tilde{\Phi}_{1}+2\varepsilon^{\prime
}\csc(br+\delta)\sin D]\\
&  -\cot(br+\delta)\tan D[\frac{1}{b}\cos(br+\delta)\sin(D)\tilde{\Phi}_{1}]\\
&  =0
\end{align*}
so that%
\begin{align}
\varepsilon^{\prime}  &  =\frac{1}{2b}\tilde{\Phi}_{1}(\tan D\cos
^{2}(br+\delta)-\sin^{2}(br+\delta)\cot(D))\nonumber\\
&  =\frac{1}{2\sin D\cos(D)b}\tilde{\Phi}_{1}(\sin^{2}D\cos^{2}(br+\delta
)-\sin^{2}(br+\delta)\cos^{2}(D))\nonumber\\
&  =\frac{1}{b\sin2D}\tilde{\Phi}_{1}\sin(br+\delta+D)\sin(br+\delta-D)
\label{14}%
\end{align}
\ From the definintion of $\tilde{\Phi}$ we see that%
\begin{align}
&  \left(
\begin{array}
[c]{cc}%
\cos\varepsilon & \sin\varepsilon\\
-\sin\varepsilon & \cos\varepsilon
\end{array}
\right)  \left(
\begin{array}
[c]{cc}%
\Phi_{3} & \Phi_{1}\\
\Phi_{1} & -\Phi_{3}%
\end{array}
\right)  \left(
\begin{array}
[c]{cc}%
\cos\varepsilon & -\sin\varepsilon\\
\sin\varepsilon & \cos\varepsilon
\end{array}
\right) \label{ej}\\
&  =\left(
\begin{array}
[c]{cc}%
\Phi_{3}\cos2\varepsilon+\Phi_{1}\sin2\varepsilon & -\Phi_{3}\sin
2\varepsilon+\Phi_{1}\cos2\varepsilon\\
-\Phi_{3}\sin2\varepsilon+\Phi_{1}\cos2\varepsilon & -\Phi_{3}\cos
2\varepsilon-\Phi_{1}\sin2\varepsilon
\end{array}
\right)  =\left(
\begin{array}
[c]{cc}%
\tilde{\Phi}_{3} & \tilde{\Phi}_{1}\\
\tilde{\Phi}_{1} & -\tilde{\Phi}_{3}%
\end{array}
\right) \nonumber
\end{align}
So from this and Eqs.(\ref{12}) and Eq.(\ref{13}) we obtain the phase shift
equations Eqs.(\ref{psa},\ref{psb}) given in the text while Eq.(\ref{14} )
gives us Eq.(\ref{psc}).

\section{\bigskip\ Phase Shift Equation With The Coulomb Potential}

We review here the necessary modification of our phase equations when we
consider $pp$ scattering \cite{c6,cal}. When we study $pp$ scattering, we must
consider the influence of the Coulomb potential. The general form of the
uncoupled Schr\"{o}dinger-like equation with Coulomb potential is \cite{c6}%

\begin{equation}
\left[  -\frac{d^{2}}{dr^{2}}+\frac{l(l+1)}{r^{2}}-\frac{2\epsilon_{w}\alpha
}{r}+\Delta\Phi\right]  u(r)=b^{2}u(r), \label{Cou}%
\end{equation}
where $\Delta\Phi$ consists of the short range parts of the effective
potential, $\alpha$ is the fine structure constant. (Compare the Coulomb term
with the first term on the right hand sides of Eqs.(\ref{faiaa},\ref{faiaa1}%
)). Due to the long range behavior of the potential in above equation, the
asymptotic behavior of the wave function is%

\begin{equation}
u(r)\overset{r\rightarrow\infty}{\longrightarrow}const\cdot\sin(br-\eta
\ln2br+\Delta),
\end{equation}
in which
\begin{equation}
\Delta=\delta_{l}+\sigma_{l}-\frac{l\pi}{2},
\end{equation}
where $\sigma_{l}=\arg\Gamma(l+1+i\eta)$ is the Coulomb phase shift, here
$\eta=-\frac{\epsilon_{w}\alpha}{b}.$

We describe here the variable phase method to calculate the phase shift with
the Coulomb potential. Consider the two differential equations%

\begin{equation}
u^{\prime\prime}+(b^{2}-W-\overline{W})u=0,
\end{equation}
and
\begin{equation}
\overline{u}^{\prime\prime}+(b^{2}-\overline{W})\overline{u}=0,\hspace
{0.2in}\hspace{0.2in}i=1,2
\end{equation}
in which $u(0)=\overline{u}_{1}(0)=0.$ Let%

\begin{align}
\overline{W}(r)  &  =-\frac{2\epsilon_{w}\alpha}{r},\\
W(r)  &  =\frac{l(l+1)}{r^{2}}+\Delta\Phi,\nonumber
\end{align}
so that
\[
\overline{u}_{1}(r)\overset{r\rightarrow\infty}{\longrightarrow}const\cdot
\sin(br-\eta\ln2br+\overline{\Delta}),
\]%
\begin{equation}
\overline{u}_{2}(r)\overset{r\rightarrow\infty}{\longrightarrow}const\cdot
\cos(br-\eta\ln2br+\overline{\Delta}),
\end{equation}
where $\overline{\Delta}=$ $\sigma_{0}.$

Just as in the variable phase method, we obtain a nonlinear first order
differential equation for the phase shift function $\delta_{l}(r)$ such that
$\delta_{l}(\infty)=\delta_{l},$ and $\delta_{l}(0)=0.$ This is done by
rewriting $u(r)$ as%

\begin{equation}
u(r)=\alpha(r)[\cos\gamma(r)\overline{u}_{1}(r)+\sin\gamma(r)\overline{u}%
_{2}(r)]
\end{equation}
so that
\begin{equation}
\Delta=\overline{\Delta}+\gamma(\infty).
\end{equation}
Since we have rewritten $u(r)$ in two arbitrary functions, we are free to
impose a condition on $u(r)$%

\begin{equation}
u^{\prime}(r)=\alpha^{\prime}(r)[\cos\gamma(r)\overline{u}_{1}^{\prime
}(r)+\sin\gamma(r)\overline{u}_{2}^{\prime}(r)].
\end{equation}
Combining $u(r)$ and $u^{\prime}(r)$ leads to%

\begin{equation}
\gamma(r)=-\tan^{-1}[\frac{u(r)\overline{u}_{1}^{\prime}(r)-u^{\prime
}(r)\overline{u}_{1}(r)}{u(r)\overline{u}_{2}^{\prime}(r)-u^{\prime
}(r)\overline{u}_{2}(r)}]
\end{equation}
where $\gamma(0)=0$, and $\overline{u}_{1}(r)=F_{0}(\eta,br)$ and
$\overline{u}_{2}(r)=G_{0}(\eta,br)$ are the two Coulomb wave functions$.$
With the Wronskian $F_{0}G_{0}^{\prime}-$ $F_{0}^{\prime}G_{0}=b$, we obtain,
by differentiating, the differential equation%

\begin{equation}
\gamma^{\prime}(r)=-W(r)[\cos\gamma(r)F_{0}(\eta,br)+\sin\gamma(r)G_{0}%
(\eta,br)]^{2}/b.
\end{equation}
Note that for%

\[
W(r)\overset{r\rightarrow0}{\longrightarrow}\frac{\lambda(\lambda+1)}{r^{2}%
},\hspace{0.5in}\frac{\lambda(\lambda+1)}{r^{2}}=\frac{l(l+1)}{r^{2}}%
-\frac{\alpha^{2}}{r^{2}},
\]

\[
F_{0}(\eta,br)\overset{r\rightarrow0}{\longrightarrow}C_{0}br,
\]

\begin{equation}
G_{0}(\eta,br)\overset{r\rightarrow0}{\longrightarrow}\frac{1}{C_{0}},
\end{equation}
we obtain the relation%

\begin{equation}
\gamma^{\prime}(0)=-\frac{C_{0}^{2}b\lambda}{\lambda(\lambda+1)}.
\end{equation}
Letting%

\begin{equation}
\gamma(r)=\beta(r)+\eta(r),
\end{equation}
where $\beta(r)$ is defined as%

\begin{equation}
\beta^{\prime}(r)=-\frac{l(l+1)}{r^{2}}[\cos\gamma(r)F_{0}(\eta,br)+\sin
\gamma(r)G_{0}(\eta,br)]^{2}/b
\end{equation}
$\beta(r)$ has the exact solution%

\begin{equation}
\gamma(r)=-\tan^{-1}[\frac{F_{l}(\eta,br)F_{0}^{\prime}(\eta,br)-F_{l}%
^{\prime}(\eta,br)F_{0}(\eta,br)}{F_{l}(\eta,br)G_{0}^{\prime}(\eta
,br)-F_{l}^{\prime}(\eta,br)G_{0}(\eta,br)}]
\end{equation}
with $\beta(0)=0$ and $\beta^{\prime}(0)=-\frac{C_{0}^{2}bl}{l(l+1)}$ and
$\beta(\infty)=\sigma_{l}-\frac{l\pi}{2}-\sigma_{0},$ lead to%

\begin{equation}
\delta_{l}=\eta(\infty).
\end{equation}
Thus , if we solve
\begin{align}
\eta^{\prime}(r)  &  =[-\frac{l(l+1)}{r^{2}}+\Delta\Phi][\cos(\beta
(r)+\eta(r))F_{0}(\eta,br)+\sin(\beta(r)+\eta(r))G_{0}(\eta,br)]^{2}%
/b\nonumber\\
&  +\frac{l(l+1)}{r^{2}}[\cos\beta(r)F_{0}(\eta,br)+\sin\beta(r)G_{0}%
(\eta,br)]^{2}/b
\end{align}
with the condition $\eta(0)=0$, we obtain the additional phase shift(above the
Coulomb phase shift ) by integration to $\eta(\infty).$

There is no Coulomb scattering for the coupled triplet $^{3}S_{1}$ and
$^{3}D_{1}$ states as a consideration of Pauli principal would show.

\section*{\bigskip\textbf{Acknowledgement}}

\bigskip

The authors wish to thank J. Lewis, C. Parigger, L. Davis, and H.V. von Geramb
for valuable suggestions.


\begin{thebibliography}{99}                                                                                               %
\bibitem {liu}This paper is based on the Ph.D. dissertation "Two-Body Dirac
Equations and Nucleon-Nucleon Scattering Phase Shift Analysis" of Bin Liu, The
University of Tennessee (2001) -http://etd.utk.edu:/2001/LiuBin.pdf

\bibitem {van1}P. Van Alstine and H. W. Crater, J. Math. Phys. \textbf{23}, 1997(1982)

\bibitem {c1}H. W. Crater, P. Van Alstine Ann. Phys. \textbf{148}, 57(1983)

\bibitem {c2}H. W. Crater, P. Van Alstine Phys. Rev. Lett. \textbf{53}, 1527(1984)

\bibitem {c3}H. W. Crater, P. Van Alstine Phys. Rev. \textbf{D36} 3007(1987)

\bibitem {c4}H. W. Crater, P. Van Alstine Phys. Rev. \textbf{D37}, 1982(1988)

\bibitem {Saz1}H. Sazdjian, Phys. Rev. \textbf{D33}, 3401(1988)

\bibitem {long1}P. Long and H. Crater, J. of Math. Phys. \textbf{39}, 124 (1998)

\bibitem {1}J. M. B. Kellogg, I. I. Rabi, N. F. Ramsey, Jr., and J. R.
Zacharias, Phys. Rev. \textbf{55}, 318(1939) and Phys. Rev. \textbf{57}, 677(1940)

\bibitem {2}O. Chamberlain, E. Segre, RDD. D. Tripp, et al Phys. Rev.
\textbf{105}, 288(1957)

\bibitem {3}H. P. Stapp, et al, Phys. Rev. \textbf{105}, 302(1957)

\bibitem {4}T. Hamada and I. D. Johnston, Nucl. Phys. \textbf{34}, 382(1962)

\bibitem {5}K. E. Lassila, M. H. Hull , et al Phys. Rev. \textbf{126}, 881(1962)

\bibitem {6}Roderick V. Reid Jr., Annals of Physics, Vol. \textbf{50}, 411-448 (1968)

\bibitem {7}K. E. Lassila, M. H. Hull , Jr., H. M. Ruppel, F. A. MacDonald and
G. Breit, Phys. Rev. \textbf{126}(1962)881

\bibitem {cal}Variable Phase Approach to Potential Scattering. F. Calogero,
Academic Press, 1967

\bibitem {dega}A. Degasperis, Nuovo Cimento, Vol. XXXIV, N. 6 (1964).

\bibitem {gross1}Franz Gross, Phys. Rev. \textbf{D 10}, 223(1974)

\bibitem {gross2}Franz Gross, et al, Phys. Rev. \textbf{C45}, 2094(1992)

\bibitem {stoks}V. G. J. Stoks, et al, Phys. Rev. \textbf{C48}, 792(1993)

\bibitem {mac}J.W. Negele and Erich Vogt (eds.) Advances in Nuclear Physics.-
R. Machleidt, - Chapter 2, The Meson Theory of Nuclear Forces and Nuclear
Structure- See also R. Machleidt, Phys. Rev. \textbf{C63, }024001 (2001)

\bibitem {c6}H. W. Crater, R. L. Becker, C.Y. Wong and P. Van Alstine Phys.
Rev. \textbf{D46} 5117(1992)

\bibitem {c7}H. W. Crater, P. Van Alstine Found. of Phys. \textbf{24}, 297(1994)

\bibitem {c15}H. Crater, P. Van Alstine, hep-ph/0208186 , submitted for
publication to Phys. Rev. D1.

\bibitem {Breit1}G. Breit, Phys. Rev. \textbf{34}, 553(1929)

\bibitem {Breit2}G. Breit, Phys. Rev. \textbf{36}, 383(1930)

\bibitem {Breit3}G. Breit, Phys. Rev. \textbf{39}, 616(1932)

\bibitem {Dirac}P. A. M. Dirac , Proc. R. Soc. London \textbf{A 117}, 610(1928)

\bibitem {bse}H.A. Bethe and E. E. Salpeter, \textit{Quantum Mechanics of One
and} \noindent\textit{Two Electron Atoms} (Springer, Berlin, 1957).

\bibitem {c10}H. W. Crater, C. W. Wong, and C. Y. Wong, Intl. J. Mod. Phys.
\textbf{E 5,} 589(1996)

\bibitem {c9}H. W. Crater, P. Van Alstine, Phys. Lett. \textbf{100B} 166, (1981)

\bibitem {c8}H. W. Crater, P. Van Alstine, Phys. Rev. \textbf{D30,} 2585(1984)

\bibitem {Tod1}I. T. Todorov, in \textquotedblright\ Properties of the
Fundamental Interactions\textquotedblright, ed by A. Zichichi, Vol. 9, Part C,
pp. 953-979; Phys. Rev. \textbf{D3} 2351(1971)

\bibitem {dirac2}P.A.M. Dirac, \underline{Lectures on Quantum Mechanics}
(Yeshiva University, Hew York, 1964).

\bibitem {c12}H. W. Crater and P. Van Alstine, J. Math. Phys. \textbf{31}, 1998(1990)

\bibitem {saz4}H. Sazdjian has shown how the Bethe-Salpeter equation for two
particles may be algebraically transformed into the two generalized mass shell
constraints equations (\ref{1}) and their spin-one half versions. H. Sazdjian
J. Math. Phys. \textbf{28,} 2618 (1987), Extended Objects and Bound Systems,
Proc. of the Karuizawa International Symposium, 1992, edited by O. Hara, S.
Ishida and S. Naka (World Scientific, Singapore, 1992), p. 117.

\bibitem {Wong1}C. Y. Wong, H. Crater, Phys. Rev \textbf{C63} 044907 (2001) ( nucl-th/0010003)

\bibitem {Saz2}J. Mourad, H. Sazdjian, J. of Phys. \textbf{G 21}, 267(1995)

\bibitem {c13}H. W. Crater and D. Yang, J. Math. Phys. \textbf{32}, 2374(1991)

\bibitem {c14}H. W. Crater and P. Van Alstine, Phys. Rev. \textbf{D46.} 766(1992)

\bibitem {saz3}H. Jallouli and H. Sazdjian, Annals of Physics, \textbf{253}
,376 (1997)

\bibitem {wig}E. Wigner, Ann. Math. \textbf{40,} 149 1939

\bibitem {plyz}W. N. Polyzou, Phys. Rev. \textbf{32}, 995 (1985)

\bibitem {c16}Since the CTBDE are the counterparts of the one-body Dirac
equation with eliminated field they possess an analogous inherited dynamical
\ \textquotedblleft gauge invariance\textquotedblright\ with Eq.(\ref{D-a})
and Eq.(\ref{D-b}) invariant under any gauge transformation of the form
$A_{i}^{\mu}\rightarrow A_{i}^{\mu}+\partial_{i}^{\mu}\chi(x_{\perp})$ (
$\chi$ is the phase change of the wave function).

\bibitem {cn}Review of Particle Physics, D. E. Groom et al., The European
Physical Journal \textbf{C15} (2000) 1.

\bibitem {quant}A. Messiah, Quantum Mechanics, John Wiley, 1963, Chapter 3,
Sec 9

\bibitem {smlr}See $\Phi_{11}$ and $\Phi_{12}$ in Eq.(\ref{phi1}) and
$\Phi_{21}$ and $\Phi_{22}$ in Eq.(\ref{phi2}). Here we let $\Phi_{12}%
,\Phi_{21}\rightarrow\frac{1}{2}(\Phi_{12}+\Phi_{21})$. Strictly speaking to
accommodate nonsymmetric potentials the variables phase method must be
modified. \ A possible route to this is indicated in Appendix C. In our
present case $\eta_{-}=\lim(r\rightarrow0)r^{2}\Phi_{11},~\eta_{+}%
=\lim(r\rightarrow0)r^{2}(\Phi_{22}+6/r^{2}),$ and we take $\eta_{0}%
=\lim(r\rightarrow0)r^{2}\frac{1}{2}(\Phi_{12}+\Phi_{21})$.

\bibitem {koch}R. Koch and E. Pietarinen, Nucl. Phys. \textbf{A336}, 331(1980)

\bibitem {arndt}R. A. Arndt, Z. Li, Phys. Rev. Lett. \textbf{65}, 157(1990);R.
A. Arndt and R. L. Workman, Phys. Rev. C43, 2436(1991)

\bibitem {berger}J. Bergervoet, et al., Phys. Rev. Lett.\textbf{ 59},
2255(1987). J. Bergervoet, et al., Phys. Rev. C41, 1435(1990)

\bibitem {9}J. J. Sakurai, Modern Quantum Mechanics, 1994 by Addison-Wesley
Publishing Company, ISBN 0-201-53929-2.

\bibitem {rohr}F. Rohrlich, Phys. Rev. \textbf{D23} , 1305 (1981)

\bibitem {saz6}H. Sazdjian, Ann. Phys. \textbf{191}, 52 (1989)

\bibitem {luca}H. Crater and L. Lusanna, Ann. Phys. \textbf{289}, 87 (2001)
\end{thebibliography}
\end{document}